%% file: Winnberg-etal-jpg.tex
\documentclass{aa}
\usepackage{txfonts}
\usepackage[english]{babel}  
\usepackage{graphicx}  
\usepackage[utf8]{inputenc}  
\usepackage{microtype}  
\usepackage{booktabs}  
\usepackage{rotating}  
\usepackage{lscape}
\usepackage[squaren]{SIunits}
\usepackage{natbib}
\usepackage{color}
\usepackage{amsmath}
\usepackage{amssymb}
\newcommand{\water}{H$_2$O}

\newcommand{\recta}[4]{$\alpha=#1^{\rm{h}}\,#2^{\rm{m}}\,#3\fs#4$}
\newcommand{\dec}[3]{$\delta=#1\degr\,#2\arcmin\,#3\arcsec$}

\newcommand{\powerten}[1]{$10^{#1}$}

\newcommand{\Myr}{$M_{\sun}$\,yr$^{-1}$}
\newcommand{\mdot}{$\dot{M}$}

\newcommand{\Lup}{$L_{\rm H_2O}^{\rm up}$}
\newcommand{\gm}{$\;\;\;\;\;$}

\newcommand{\kms}{km\,s$^{-1}$}

\newcommand{\pad}{.\hskip-2pt$^\circ$}
\newcommand{\pam}{.\hskip-2pt$^{\prime}$}
\newcommand{\pas}{.\hskip-2pt$^{\prime\prime}$}

\newcommand{\gsim}{\;\lower.6ex\hbox{$\sim$}\kern-7.75pt\raise.65ex\hbox{$>$}\;}
\newcommand{\lsim}{\;\lower.6ex\hbox{$\sim$}\kern-7.75pt\raise.65ex\hbox{$<$}\;}
\newcommand{\Done}{D$^\prime$}
\newcommand{\Dtwo}{D$^{\prime\prime}$}
\newcommand{\Gone}{G$^\prime$}
\newcommand{\Gtwo}{G$^{\prime\prime}$}

\begin{document} 

\title{Water vapour masers in long-period variable stars}
\subtitle{III. Mira variables
\object{U\,Her} and \object{RR\,Aql}\thanks{The maser spectra and the VLA data cubes are available via anonymous ftp to cdsarc.u-strasbg.fr (130.79.128.5) or via http://cdsarc.u-strasbg.fr/cgi-bin/qcat?J/A+A/}}

\author {A.~Winnberg \inst{1}
        \and J.~Brand \inst{2}
        \and D.~Engels \inst{3}}

\offprints{J. Brand or D. Engels,\\
\email{brand@ira.inaf.it, dengels@hs.uni-hamburg.de}
}

\institute{Onsala Rymdobservatorium, Observatoriev\"{a}gen,
           S--43992 Onsala, Sweden
      \and INAF - Istituto di Radioastronomia \& Italian ALMA Regional Centre, Via P. Gobetti 101,
           I--40129 Bologna, Italy
      \and Hamburger Sternwarte, Universit\"{a}t Hamburg, Gojenbergsweg 112,
           D--21029 Hamburg, Germany
           }

\date{Received date; accepted date: 12/2/24}

\abstract {Water maser emission is often found in the circumstellar envelopes of evolved stars, i.e. asymptotic-giant-branch stars and red supergiants with oxygen-rich chemistry. The \water\ emission shows strong variability in evolved stars of all these types.}
{We wish to understand the reasons for the strong variability of water masers emitted at 22 GHz. In this paper we study U Her and RR Aql as representatives of Mira variable stars.}
{We monitored U\,Her and RR\,Aql in the 22-GHz maser line of water vapour with single-dish telescopes. The monitoring period covered about two decades between 1990 and 2011, with a gap between 1997 and 2000 in the case of RR\,Aql. Observations were taken also in 1987 and 2015 before and after the period of contiguous monitoring. In addition, maps were obtained in the period 1990--1992 of U\,Her with the Very Large Array.}
{We find that the strongest emission in U\,Her is located in a shell with boundaries 11 -- 25 AU. The gas crossing time is 8.5 years. We derive lifetimes for individual maser clouds of $\le$4 years, based on the absence of detectable line-of-sight velocity drifts of the maser emission. The shell is not evenly filled, and its structure is maintained on timescales much longer than those of individual maser clouds. Both stars show brightness variability on several timescales. The prevalent variation is periodic, following the optical variability of the stars with a lag of 2--3 months. Superposed are irregular fluctuations, of a few months' duration, of increased or decreased excitation at particular locations, and long-term systematic variations on timescales of a decade or more.}
{The properties of the maser emission are governed by those of the stellar wind while traversing the \water\ maser shell. Inhomogeneities in the wind affecting the excitation conditions and prevalent beaming directions likely cause the variations seen on timescales longer than the stellar pulsation period. We propose the existence of long-living regions in the shells, which maintain favourable excitation conditions on timescales of the wind crossing times through the shells or orbital periods of (sub-)stellar companions. The \water\ maser properties in these two Mira variables are remarkably similar to those in the semiregular variables studied in our previous papers, regarding shell location, outflow velocities, and lifetimes. The only difference is the regular brightness variations of the Mira variables caused by the periodic pulsation of the stars.}


\maketitle


\section{\label{intro} Introduction}
Maser emission of SiO, \water\ and OH is frequently found in the circumstellar shells or envelopes (CSEs) of oxygen-rich stars on the asymptotic giant branch (AGB) and in several red supergiants (RSGs). Within the CSEs, where conditions are favourable for the excitation of one or another of these masers depends on local density, temperature and dynamics and thus in practice on distance to the stellar surface. In the case of Mira variables the \water\ masers are typically found at radii of 5 to 50 AU \citep{bowers93, bowers94, colomer00, bains03, imai03, xu22}.

Early observing programs to monitor water masers found strong variability in their spectra \citep[and references therein]{schwartz74, berulis83, habing96} particularly noticeable in the integrated maser emission \citep{berulis98}. Depending on the type of star observed and the duration of the monitoring, several types of variability can be recognised. The first and often most evident is a variation in delayed sync with the light variations of the central star (same period but with an offset in phase); superposed on this regular variation there often is an erratic variability, occurring on shorter timescales, including bursts of individual maser lines lasting weeks to months. If the monitoring takes place over long periods of time, variability in overall brightness of the maser emission may be detected, lasting many years \citep{brand20} and may have repetitive patterns ('superperiods';  \citealt{rudnitskii05}).

\begin{table*}
  \caption{Basic information on the two Mira variables monitored in the period 1990--2011.}
\label{centralcoords} 

\resizebox{\textwidth}{!}{%
\begin{tabular}{rlllrrrclcc}
\hline\noalign{\smallskip}
\multicolumn{1}{c}{Name} &  \multicolumn{2}{c}{$\alpha$\,\,\,\, (J2000)\,\,\,\, 
 $\delta$} & 
 \multicolumn{1}{c}{$D^{\rm a}$} &
  \multicolumn{1}{c}{$V_{\ast}^{\rm b}$} & 
 \multicolumn{1}{c}{$V_{\rm exp}^{\rm b}$} & 
 \multicolumn{1}{c}{$V_{\rm b}, V_{\rm r}$} & 
 \multicolumn{1}{c}{$P_{\rm opt}$} &
 \multicolumn{1}{c}{TJD$_{max}$} &
\multicolumn{1}{c}{$P_{\rm rad}$} &
\multicolumn{1}{c}{$\phi_{\rm lag}$} \\
\multicolumn{1}{c}{} &
\multicolumn{1}{l}{\, h\,\,  m\, \,   s}
 & \multicolumn{1}{l}{\, \,  $\circ$\,\,\, $\prime$\, \,  $\prime\prime$}
 & \multicolumn{1}{c}{pc} 
 & \multicolumn{1}{c}{\kms}
 & \multicolumn{1}{c}{\kms}
 & \multicolumn{1}{c}{\kms}
 & \multicolumn{1}{c}{days}
 & \multicolumn{1}{c}{days}
 & \multicolumn{1}{c}{days} 
 & \multicolumn{1}{c}{} \\
\hline\noalign{\smallskip}
U~Her & 16:25:47.5 & +18:53:33&266$^{+32}_{-18}$ &$-$15.0&13.1 &$-$23.3,  $-$7.1 & 405 &6668$\pm$3 &407$\pm$11 &0.16 \\[0.1cm]
RR~Aql& 19:57:36.1 & $-$01:53:11&410$^{+12}_{-11}$&28.5&9.0 &23.2, 31.9 & 400&6487$\pm$5 &400$\pm$5 &0.21 \\[0.1cm]
\noalign{\smallskip}
\hline
\end{tabular}}
\\[0.1cm]
References. $^{\rm (a)}$ For distances: 
U Her: \cite{vlemmings07};  
RR Aql: \cite{sun22}.
$^{\rm (b)}$ For stellar systemic and expansion velocities.
U Her:~\cite{gon-alfonso98}; RR Aql:~\cite{danilovich15}.
\end{table*}

Maps made from interferometric observations taken many months apart show that also the distribution of the maser emission sites in the CSEs changes considerably \citep{johnston85}. The masers are thought to reside in clouds of size 2--5 AU \citep{bains03, richards11} embedded in the stellar wind, which in the case of Semiregular and Mira variables are identifiable for at most a few years \citep{bains03, winnberg08}. The crossing times through the \water\ maser shells, located within $\sim$50 stellar radii, have timescales of decades, so that the disappearance of the emission of particular maser features after few years would indicate that the clouds either dissipate or change their beaming direction \citep{bains03, richards12}.

Besides brightness variations, also variations of the velocities of the maser lines were studied. Velocity drifts attributed to the passage of shocks in the \water\ maser shell were reported for several stars \citep{shintani08}. The monitoring of the velocity variations through high-resolution interferometry, make it possible to trace the structure of the stellar wind passing through the shell, as shown recently by \cite{xu22} for the Mira variable BX\,Cam (IRC+70066). 

In order to improve the understanding of the properties of \water\ maser variability for different types of late-type stars, we started in 1987 the Medicina/Effelsberg monitoring program of several such stars using the Medicina 32-m and Effelsberg 100-m radio telescopes. With data covering 20--30 years, we expect to elucidate the changes of maser excitation conditions within the \water\ maser shells, which in AGB stars are crossed by the stellar wind on timescales of the same order. The sample included Semi-regular Variables (SRV), Mira variables, OH/IR stars and RSGs. For each class of stars we added several interferometric observations of a prototypical object using the Very Large Array (VLA), to study the development of the emission pattern in the maps and the response of the single-dish spectra to it.

In our first two papers we presented the results for the SRVs in our sample: RX~Boo and SV~Peg (\citealt{winnberg08}; hereafter Paper I), and R~Crt and RT~Vir (\citealt{brand20}; hereafter paper II). In the period 1990--1992 the \water\ maser emission of RX~Boo, taken as representative of the class, was found in an incomplete ring with an inner radius of 15~AU and a shell thickness of 22~AU. The variability of \water\ masers in RX~Boo, as well as in SV~Peg, R~Crt and RT~Vir, is due to the emergence and disappearance of maser clouds with lifetimes of $\sim$1 year. The maser emission regions do not evenly fill the shell of RX~Boo, as indicated by the asymmetry in the spatial distribution, which persists at least an order of magnitude longer. An exception to the generally short lifetime of individual maser clouds is the "11~\kms\ feature" in RT~Vir, originating in a cloud with an estimated lifetime of $>$ 7.5 years \citep{brand20}. 

In this paper we present the results for approximately two decades of monitoring of the Mira variables U~Her and RR~Aql. We chose U~Her as the representative star of the class of Mira variables.  Interferometric maps were taken for this star between 1990 and 1992. Preliminary results of the U~Her observations were reported in \cite{engels99} and \cite{winnberg11}. In addition, here we will use also other interferometric maps from the literature made in the same period as the single-dish monitoring program. The results of the Mira-like variable stars IK~Tau, and of R~Cas, R~Leo and $\chi$~Cyg 
will be the subject of  separate papers. The results for the RSGs will be presented in a forthcoming paper.

In Table \ref{centralcoords} we present some basic information. It gives the name of the object in column (Col.) 1; the coordinates are in Cols. 2 and 3; in Col. 4 we show the distance, the references for which are given in the footnote. 
All linear sizes in this paper are scaled to these distances. The radial velocity of the star,  $V_{\ast}$, and the final expansion velocity in the CSE, $V_{\rm exp}$, are given in Cols. 5 and 6. These velocities are our best estimates using the data obtained from observations of molecular emission (mostly CO) by the references listed in the footnote. In Col.~7 we give the (blue and red) boundaries $V_{\rm b}$, $V_{\rm r}$ of the range in velocity, over which \water\ emission was found during the monitoring period. Col. 8 gives the optical pulsation period $P_{\rm opt}$, and Col. 9 the date TJD$_{max}$\footnote{Truncated Julian Date, TJD=JD-2440000.5} of the last optical maximum before the monitoring started. Col. 10 shows the radio pulsation period $P_{\rm rad}$, and in Col. 11 we give the lag $\phi_{\rm lag}$ of the phase of the radio light curve with respect to the optical phase. The entries for Cols. 7--11 for the individual stars are taken from the sub-sections in this paper, where the \water\ maser properties are analysed individually. 
 
This paper is organised as follows: in Sect.~\ref{sec:observations} we describe the observations, and in Sect.~\ref{sec:presdata} we explain methods and definitions to present the data. In Sect.~\ref{sec:uher} we analyse the single dish and interferometric data of U\,Her, and present the model and 3-dimensional structure of the circumstellar envelope of U\,Her. The single dish data for RR\,Aql are discussed in Sect.~\ref{sec:rraql}. The properties of the \water\ maser emission in the circumstellar envelopes of Mira variables are discussed in Sect.~\ref{sec:miraprop}, while our findings are summarised in Sect.~\ref{sec:conclusions}.

\section{Observations  \label{sec:observations}}
Single dish observations of the \water\ maser line at 22235.08~MHz were made with the Medicina 32-m and Effelsberg 100-m telescopes at typical intervals of a few months. Initial observations began in 1987 with the Medicina telescope, and the regular monitoring for the stars discussed here was performed between 1990 and 2011. Some additional spectra were taken in 2015. For both stars 
one spectrum taken between 1987 and 1989 has been published before, by \cite{comoretto90}. 
The Effelsberg telescope participated in the monitoring program between 1990 and 1999, and in the case of U~Her until 2002. 
VLA observations of U~Her were made on four occasions in the period 1990--1992.

\subsection{Medicina}
Between March 1987 and March 2011, and again in 2015, we searched for H$_2$O($6_{16}-5_{23}$) (22.2350798~GHz) maser emission with the Medicina 32-m telescope\footnote{The Medicina 32--m VLBI radiotelescope is operated by INAF--Istituto di Radioastronomia.} towards the stars listed in Table~\ref{centralcoords}. We used a digital autocorrelator backend with a bandwidth of 10~MHz and 1024 channels, resulting in a resolution of 9.76~kHz (0.132 \kms); the half-power beam width (HPBW) at 22~GHz was $\sim$1\pam 9. During this period the sample was observed four to five times per year in separate sessions. For more information on the changes in the system during these years, see Paper~I.

The telescope pointing model was typically updated a few times per year, and quickly checked every few weeks by observing strong maser sources (e.g. W3~OH, Orion-KL, W49~N, Sgr~B2, and W51). The pointing accuracy was always better than 25\arcsec; the rms residuals from the pointing model were of the order of 8\arcsec--10\arcsec. 

Observations were taken in total power mode, with both ON and OFF scans of 5~min duration. The OFF position was taken \mbox{1\pad25 E} of the source position to rescan the same path as the ON scan. Usually two ON/OFF pairs were taken at each position. Only the left-hand circular (LHC) polarisation output from the receiver was registered\footnote{In Paper I this was erroneously reported as only RHC (right-hand circular).}. In 2015 both polarisations were recorded (and averaged during data reduction).
The observations were embedded in a larger program. We could thus determine the antenna gain as a function of elevation by observing several times during the day the continuum source DR~21 (for which we assumed a flux density of 16.4 Jy after scaling the value of 17.04~Jy given by \cite{ott94} for the ratio of the source size to the Medicina beam) at a range of elevations. Antenna temperatures were derived from total power measurements in position switching mode. The integration time at each position was 10 sec with 400~MHz bandwidth. 

The daily gain curve was determined by fitting a polynomial curve to the DR~21 data; this was then used to convert antenna temperature to flux density for all spectra taken that day. From the dispersion of the single measurements around the curve, we found the typical calibration uncertainty to be 20\%.

\begin{table}
\caption[]{\label{tab:VLAspec} VLA map specifications for U~Her}
\begin{flushleft}
\begin{tabular}{lrrrrr}
\hline\noalign{\smallskip}
Date &  \multicolumn{3}{c}{HPBW} & rms & $S/N$ \\
 & maj.a.  & min.a. & p. a.  &  & \\
 & (\arcsec) & (\arcsec) & (\degr) & (Jy/b.) & \\
\noalign{\smallskip} \hline\noalign{\smallskip}
1990 Feb. 26 & 0.091 & 0.081 & 71.31 & 0.012 & 1080 \\
1990 June 03 & 0.130 & 0.106 & 54.79 & 0.015 & 500 \\
1991 Oct. 20 & 0.409 & 0.106 & $-65.62$ & 0.021 & 380 \\
1992 Dec. 28 & 0.075 & 0.072 & $-24.46$ & 0.017 & 420 \\
\noalign{\smallskip} \hline
\end{tabular}
\end{flushleft}
\emph{maj.a.}: half-power beam width (HPBW) for the major axis of
the best-fit three-dimensional Gaussian component to the synthesised beam \\
\emph{min.a.}: HPBW for the minor axis \\
\emph{p.a.}: position angle of the major axis (E of N)\\
\emph{rms}: the root-mean-square noise fluctuations in signal-free
channels in units of Jansky per beam area \\
\emph{S/N}: signal-to-noise ratio or `dynamic range' in the
channel with strongest signal
\end{table}

\subsection{Effelsberg}
Between 1990 and 1999 we observed the sources 
with the Effelsberg 100-m antenna\footnote{The Effelsberg 100-m radiotelescope is operated by the Max-Planck-Institut f\"ur Radioastronomie, Bonn}. 
To observe the $6_{16}\rightarrow 5_{23}$ transition of the water molecule, 18--26~GHz receivers with cooled masers as pre-amplifier were used until 1999. Only one polarisation direction, the LHC, was recorded, as circumstellar water masers were found to be unpolarised to limits of a few percent (\citealt{barvainis89}). U~Her was observed also in 2002 using the 1.3cm prime-focus receiver, which measured two linear polarisations averaged during post-processing. At 1.3~cm wavelength the beam width is $\sim$40\arcsec\ (HPBW).  We observed in total power mode integrating ON and OFF the source in general for 3--10~min each. 
`ON-source' the telescope was positioned on the coordinates given in Table \ref{centralcoords}, while the `OFF-source' position was displaced 3\arcmin\ to the east of the source. 

Until 1999 the backend consisted of a 1024 channel autocorrelator, while in 2002 an 8192 channel autocorrelator was used (4096 channels per polarisation). Observations were made with a bandwidth of 6.25~MHz (5 MHz in 2002), centred on the stellar radial velocity. The velocity coverage was $70$ or $80$~\kms\ and the velocity resolution 0.08~\kms\ (0.016 \kms\ in 2002). 
For procedures to reduce the spectra and for the calibration we refer to Paper I. We estimate that the flux densitiy values are not reliable to better than 30~\%. 

\subsection{VLA observations}
U~Her was observed with the Very Large Array (VLA)\footnote{The VLA is operated by the National Radio Astronomy Observatory, which is a facility of the National Science Foundation operated under cooperative agreement by Associated Universities, Inc.
} on four occasions between February 1990 and December 1992. All 27 antennas were used yielding synthesised beamwidths down to $\sim$70 mas (Table \ref{tab:VLAspec}). For three of the four epochs we used the largest extent ("A" configuration), while the October 1991 observations were carried out with a hybrid configuration ("BnA"). We chose a backend bandwidth of 3.125~MHz to obtain a total velocity range of 42~\kms\ and the bandwidth was split into 64 channels, yielding a velocity resolution of 0.66~\kms. Data from the right and left circular polarization modes were averaged. Typical integration times were 30~min on the star and 12~min on the phase calibrator J1608+1029 with a sampling time of 30~s. Flux calibration was obtained relative to 3C286 that was assumed to have a flux density of 2.55~Jy and 3C84 was used to correct for the bandpass shapes.

\begin{figure*}
\resizebox{18cm}{!}{
\includegraphics{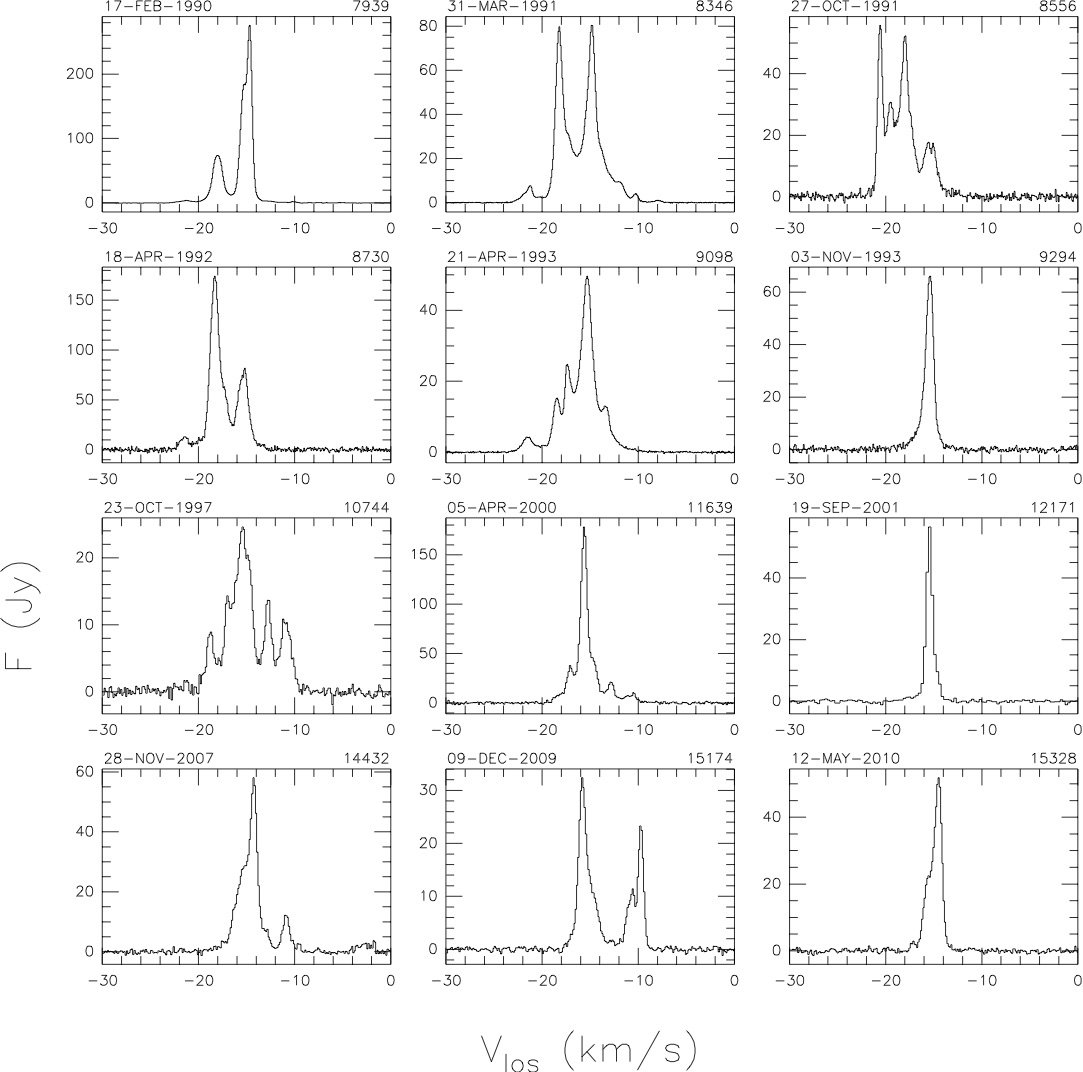}}
\caption{Selected H$_2$O maser spectra of U~Her. The calendar date of the observation is indicated on the top left above each panel, the TJD (JD-2440000.5), 
on the top right.}
\label{fig:uher_sel}
\end{figure*}

\section{Presentation of the data \label{sec:presdata}}
Before we present and discuss the data on the stars in our sample, we need to describe the tools and define the parameters we used in our analysis. For each star we also show a selection of the spectra taken over the years, in the sub-sections where they are presented. See Fig. \ref{fig:uher_sel} as an example. All maser spectra for the stars are presented in Fig. \ref{fig:uher_all} and \ref{fig:rraql_all} (Appendix C).

\subsection{Diagnostic plots} \label{diagplots}

For each star we show a number of plots that summarise the behaviour of the water maser emission in time, intensity and velocity-range. We give a brief description of these diagnostic plots, and refer to Paper II for more details. 

\begin{figure*}
\begin{minipage}[t]{17cm}
 \begin{minipage}[t]{8.5cm}
  \begin{flushleft}
   {\includegraphics*[width=8cm,angle=0]{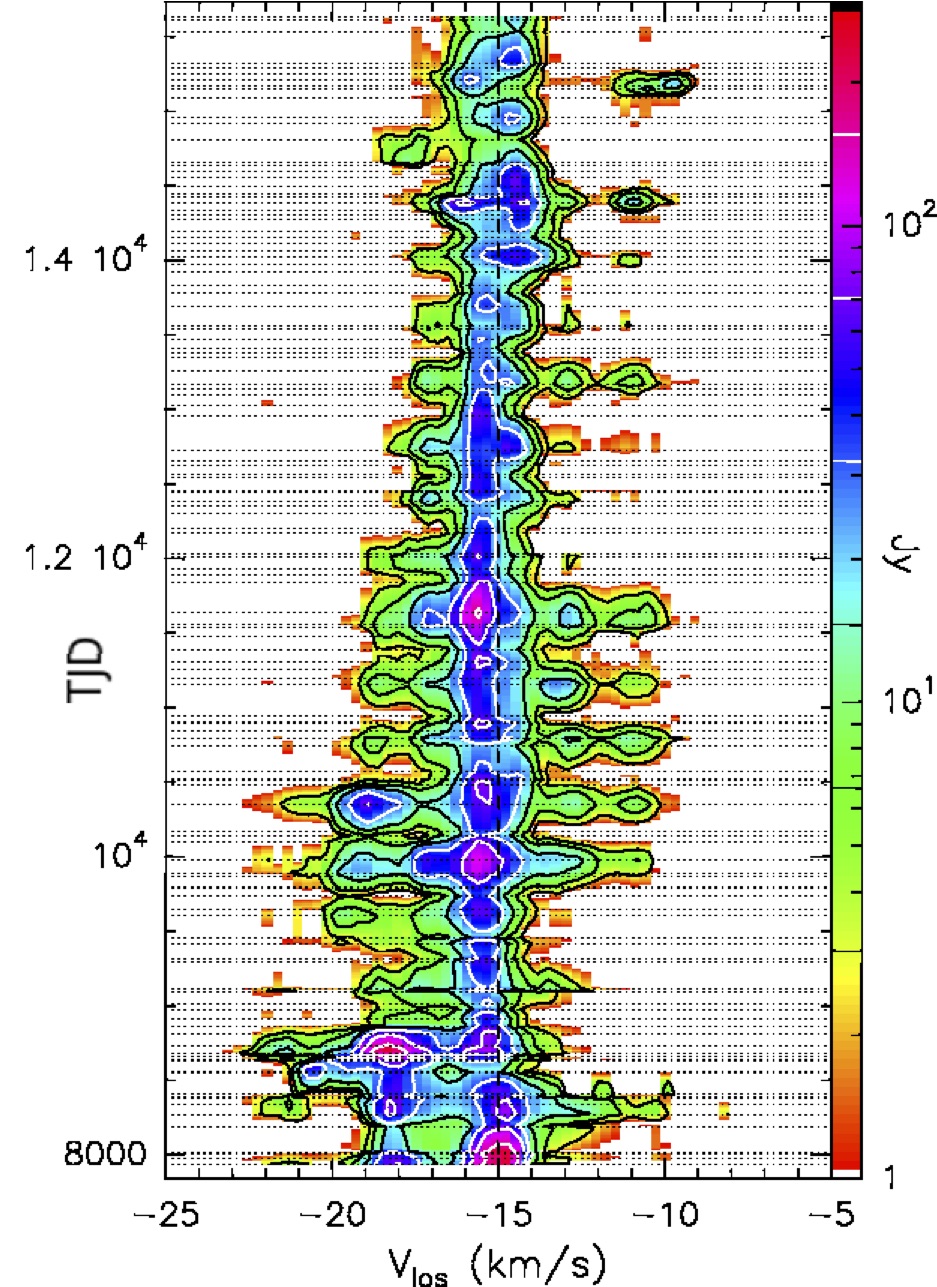}}
  \end{flushleft}
 \end{minipage}
 \begin{minipage}[t]{8.5cm}
  \begin{flushright}
        {\includegraphics*[width=8cm,angle=0]{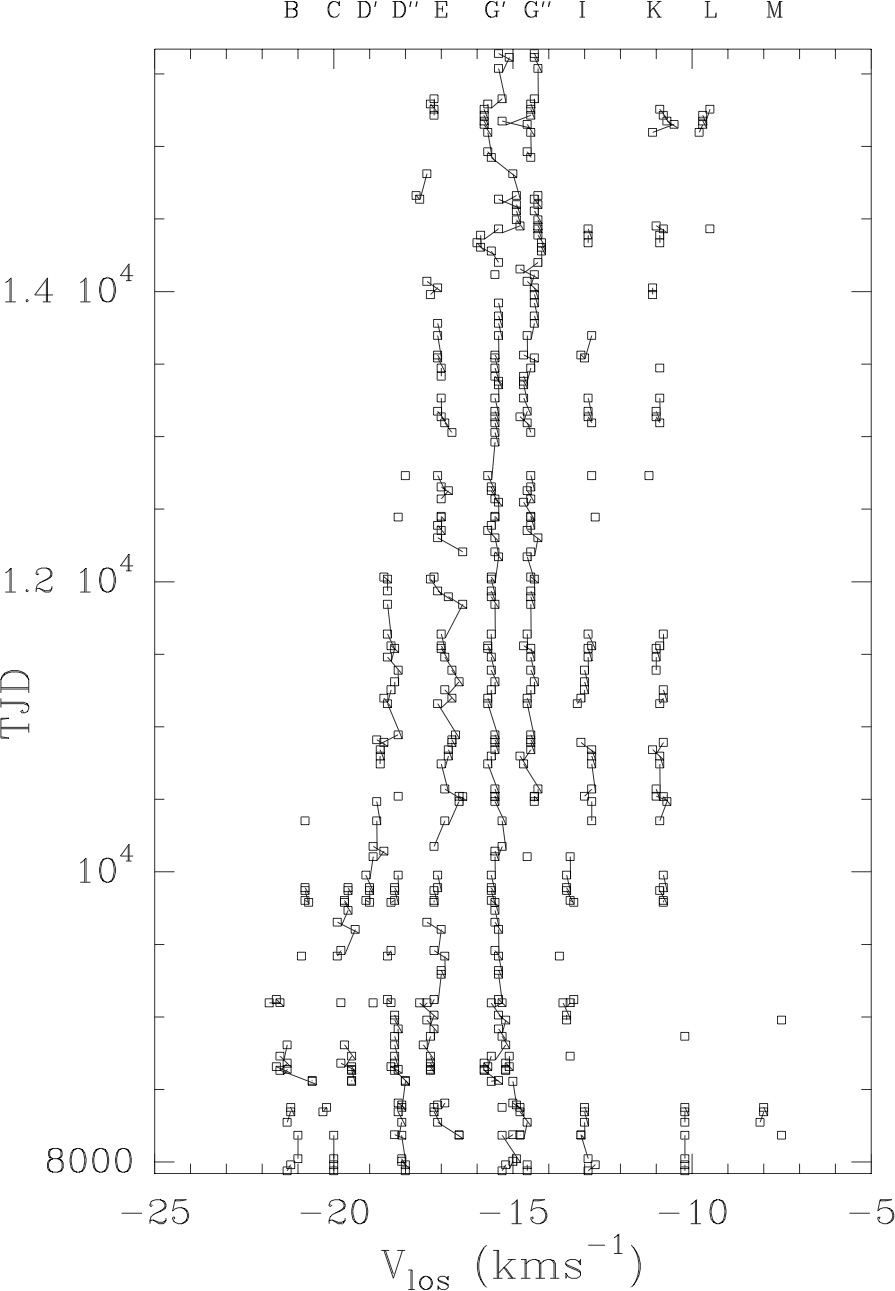}}
  \end{flushright}
 \end{minipage}
\end{minipage}
\caption{{\it Left:} Flux density versus line-of-sight as a function of time (FVt)-plot for U~Her. Each horizontal dotted line indicates an observation (spectra taken within 4 days from each other were averaged). Data are resampled to a resolution of 0.3~\kms\ and only emission at levels $\geq 3\sigma$ and $\geq$ 1~Jy is shown. The first spectrum in this plot was taken on 16 February 1990; JD = 2447938.5, TJD = 7938. Last spectrum shown is for 20 March 2011. \\{\it Right:} Spectral components identified by the component fit of the single-dish spectra as listed in  Tables~\ref{tab:compUHerB-E} and \ref{tab:compUHerG-M} (see also Sect.~\ref{sdd_LineProfAna}). The component designations are given above the plot. 
Features which have been detected in adjacent spectra, are connected by solid lines.}
\label{fig:uher-fvt}
\end{figure*}

\smallskip
{\it FVt-plot:}\ The time variation of the maser emission is visualised by plotting the flux density versus time and line-of-sight (los) velocity, $V_{\rm los}$, in a so-called FVt-diagram (cf. \citealt{felli07}). An example is shown in Fig.~\ref{fig:uher-fvt}.
Between consecutive observations linear interpolation was applied; when there is a long time-interval between two consecutive observations this produces an apparent persistence or increase in the lifetime of a feature.
Although we also took 5 spectra in 2015, the last spectra used in the FVt-plots are from March 2011, to avoid a 4-year gap. 

\smallskip
{\it Upper envelope spectrum:}\ this was obtained by assigning to each velocity channel the maximum (if  ${>} 3\sigma$, after resampling to a resolution of 0.3 \kms) signal detected during our observations (including spectra taken before and after the monitoring period 1990 -- 2011.
This 'envelope' represents the maser spectrum if all velocity components were to emit at their maximum level and at the same time.
See Fig.~\ref{fig:uher-upenv} for an example. 

\smallskip
{\it Lower envelope spectrum}:\ as the upper envelope, but obtained by finding the minimum flux density in each velocity channel, setting it to zero, unless it is ${>} 3\sigma$ (after resampling to a resolution of 0.3 \kms). 
An example is shown in Fig.~\ref{fig:uher-loenv}. 

\smallskip
{\it Detection-rate histogram:}\ this shows the rate-of-occurrence of maser emission above the 3$\sigma$ noise level for each velocity channel (for 0.3 \kms\ resolution), both in absolute numbers (left axis) as in percentage (right axis). This simply counts for each channel the number of times the flux density in the channel is greater than the 3$\sigma$ noise level of the spectrum.
An example is shown in Fig.~\ref{fig:uher-histo}. 

\smallskip
{\it Radio (maser) light curves}\ are obtained by plotting integrated flux densities versus TJD or versus optical phase. 
The integrated flux density $S(\rm tot)$  is determined over a fixed velocity interval encompassing all velocities $V_{\rm los}$ at which maser emission was detected. 
The optical phase $\varphi_{\rm s}$ is obtained from a fit with a sine-function to the optical light curve (for details see Sect.~\ref{opt-radio-variability}).

\subsection{Velocities and velocity ranges \label{velodef}}

\noindent
In the following we define velocities and velocity ranges that we shall use in the analysis of the spectra. Only a short description is given here; for a detailed definition we refer to 
Paper II.
The observed velocity ranges of the \water\ maser emission are analysed in the frame of the 'standard model' for CSEs in evolved stars \citep{hoefner18}. This model assumes that the stars have radially symmetric outflowing winds, which form a spherical shell of dust and gas around them. 
The winds are accelerated so that the outflow velocity $V_{\rm out}$ is increasing with radial distance from the star before it reaches 
the final expansion velocity $V_{\rm exp}$.

The velocity range over which \water\ maser emission can be expected is 
constrained by the velocity ranges given by the OH maser and CO thermal emission. Both species are found beyond the typical \water\ maser shells in regions where the wind acceleration has already ceased and the outflow velocity is constant \citep{hoefner18}. 
Then $V_{\rm out} \le V_{\rm exp}$
and the observed \water\ maser velocities $V_{\rm los}$  
are expected in the range 
$V_{\ast}-V_{\rm exp} \le V_{\rm los} \le V_{\ast}+V_{\rm exp}$.

We call the blue and red extremes of 
the observed \water\ maser velocity range  $V_{\rm b}$ and $V_{\rm r}$, respectively. We use the  detection-rate histogram for the determination of the observed maximum extent of the \water\ maser velocity range $\Delta V_{\rm los} = V_{\rm r} - V_{\rm b}$ (hereafter 'maximum velocity range') valid for the period of observations. 
This method gives accurate values ($\approx 0.15$ \kms) 
for the maximum velocity range $\Delta V_{\rm los}$. In the case of spherical symmetry, we expect that the centre of the \water\ maser velocity range is 
$(V_{\rm b}+V_{\rm r})/2 = V_{\ast}$. 

One should note that the velocity range of individual observations and the maximum velocity range do vary with time because of two effects. First, for periods of time the outermost features may fall in brightness below the detection limit leading to an apparent variation of the observed velocity range. And second, maser emission might be excited out to larger/smaller distances 
for periods of time leading to a real increase/decrease of the maximum velocity range. 

\noindent

\section{U~Her \label{sec:uher}}
U\,Her is a long-period variable AGB star at a distance of $266^{+32}_{-18}$ pc (Table \ref{centralcoords}), based on the OH maser parallax measured by \cite{vlemmings07}. We prefer the distance obtained by radio interferometry, because there is a large difference between the distances measured by the two astrometric satellites Hipparcos ($235^{+58}_{-39}$ pc; \citealt{vanleeuwen07, vanleeuwen08}) and Gaia EDR3 ($424^{+14}_{-13}$ pc; \citealt{gaiacol20}), which may be caused by uncertainties introduced by stellar activity on optical parallaxes of nearby AGB stars \citep{chiavassa18}. Radial velocity determinations of U\,Her agree within $\sim$0.5 \kms\ centred on $V_{\ast} = -15.0$ \kms. The final expansion velocity $V_{\rm exp}$ in the CSE can be as high as 20 \kms\ \citep{gottlieb22}, but here we use a more conservative value $V_{\rm exp} = 13.1$ \kms\ (see Table \ref{centralcoords}). 

The \water\ maser of U~Her was first detected in 1969 by \cite{schwartz70b, schwartz70a} as a single feature at $-15$ \kms\ (their detection limit was $\approx$ 10~Jy). Until 1984 the maser was observed several times with detections in the velocity range $-24$ to $-7$ \kms. 
The strongest peak was found either at $-15$ or $-17$ \kms\ \citep[and references therein]{engels88}. Interferometric observations were made until the early time of our monitoring program with the VLA in 1983, 1988 and 1990
\citep{lane87, bowers94, colomer00} and with MERLIN in 1985 \citep{yates94}. They found the masers to be located in an unevenly filled ring-like structure with typical inner and outer radii of $\sim$10 and $\sim$20 AU, respectively.

\begin{figure}
\resizebox{9cm}{!}{\rotatebox{270}{
\includegraphics
{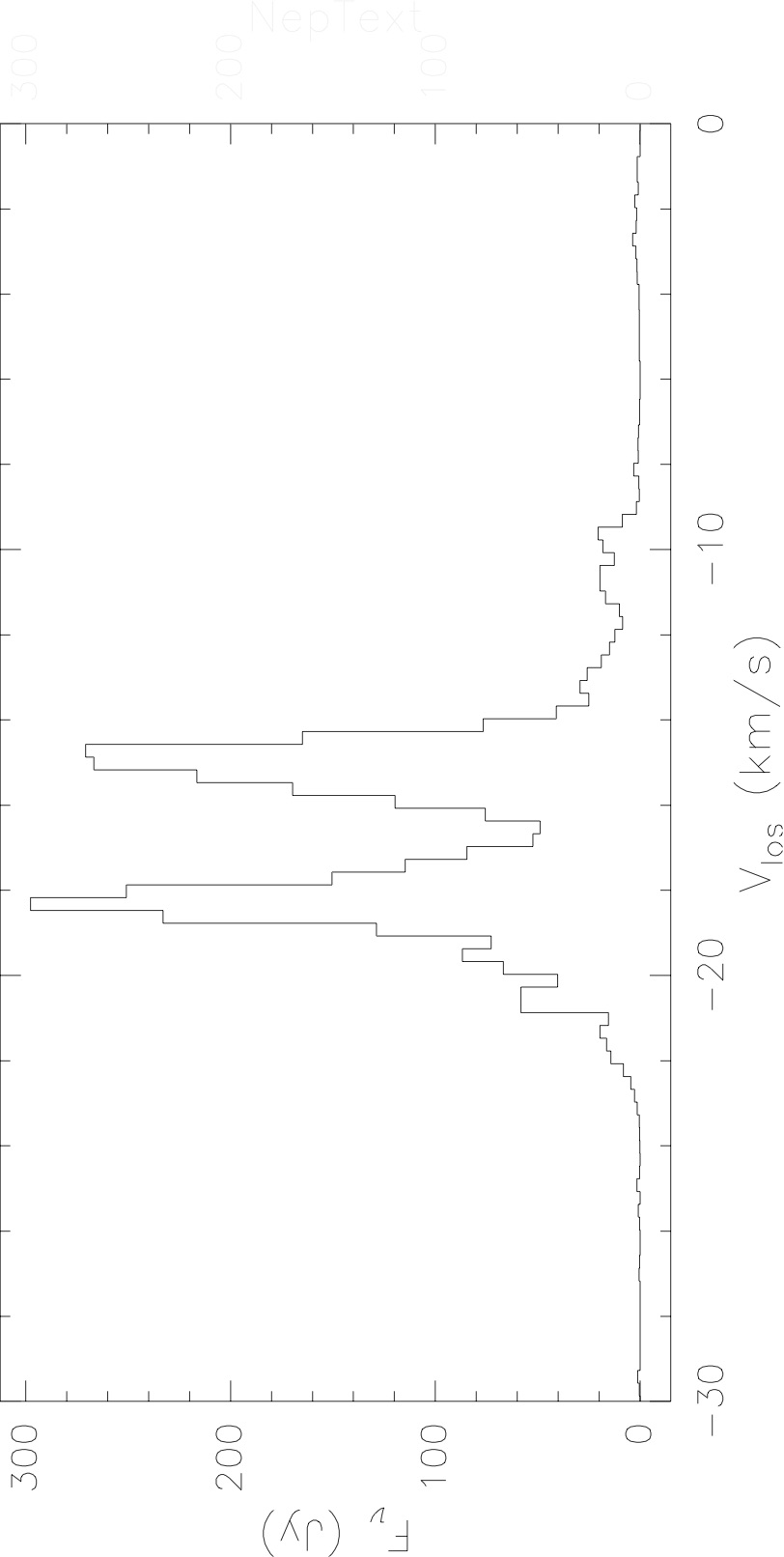}}}
\caption{Upper envelope spectrum for U~Her; 1987-2015.}
\label{fig:uher-upenv}
\end{figure}

\begin{figure}
\resizebox{9cm}{!}{\rotatebox{270}{
\includegraphics
{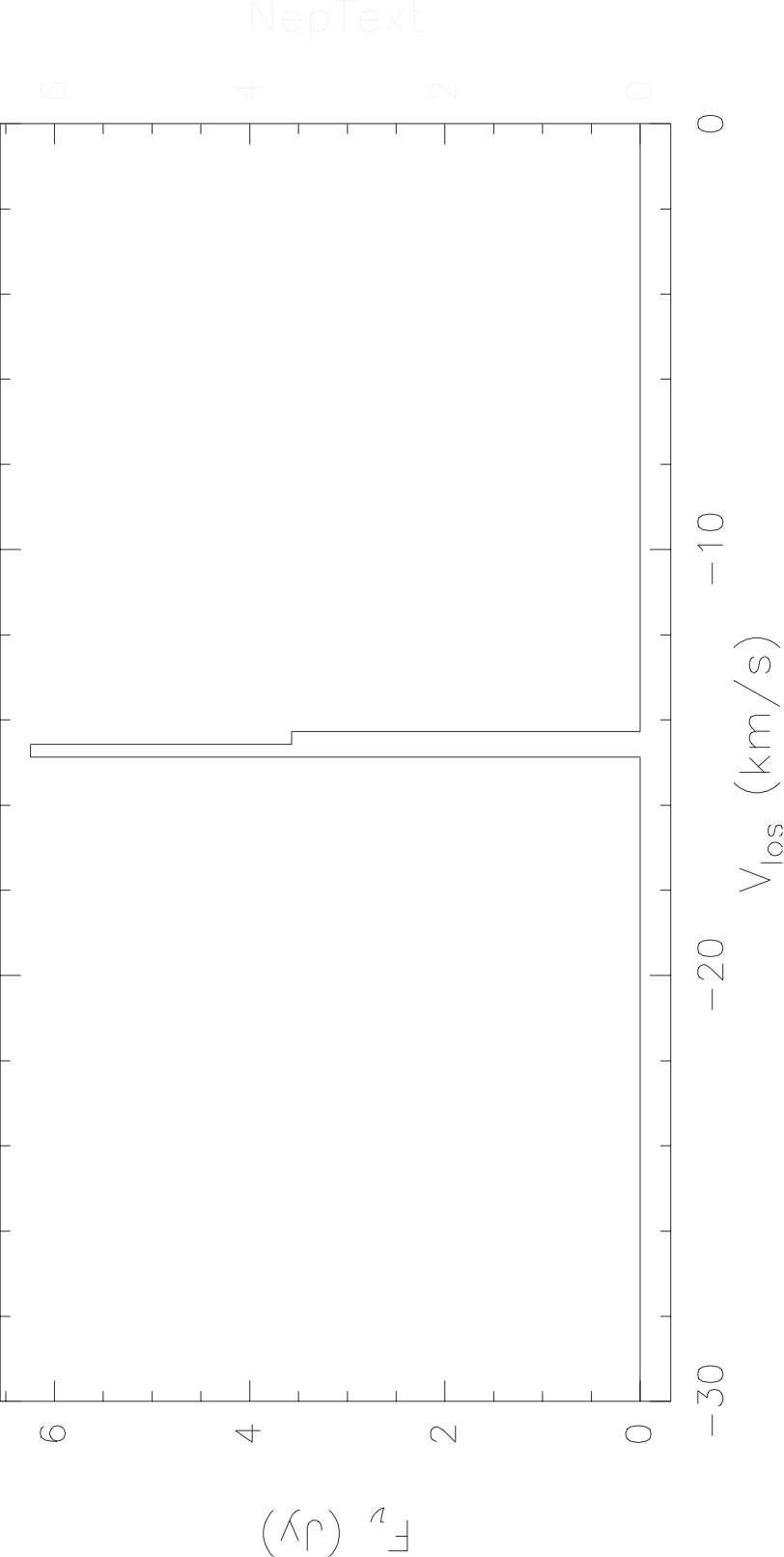}}}
\caption{Lower envelope spectrum for U~Her; 1987-2015.}
\label{fig:uher-loenv}
\end{figure}

\subsection{Single-dish data}
\subsubsection{\label{sdd_MaserSpec} Variations in brightness of the \water\ maser profile}
Our observations of U~Her cover more than 28 years, from March 1987 to October 2015. Contiguous monitoring was made between 1990 and 2011 with typically 5--6 observations per year. Depending on telescope (i.e. Effelsberg or Medicina), date of observation and integration time, the rms sensitivity of the observations was very inhomogeneous ranging from 0.1 to $\sim$4 Jy, depending on resolution and date. At the 0.3~\kms\ resolution used here, after mid-1991, with few exceptions all rms were $<$ 1.0~Jy. All 137 spectra taken are shown in the Appendix (Fig.~\ref{fig:uher_all}). Sample spectra showing typical profiles are given in Fig. \ref{fig:uher_sel}. 

A general view of the properties of the profile variations is given in the FVt plot (Fig. \ref{fig:uher-fvt}, left panel) covering the years 1990 -- 2011. The profile is usually dominated by emission  in the velocity range $-16 < V_{\rm los} < -14$ \kms\ close to the stellar radial velocity $V_\ast = -15.0$ \kms, while in the outer parts of the profile  ($V_{\rm los} < -18$ and $> -14$ \kms) the emission is much weaker and at $V_{\rm los} > -14$ \kms\ appeared more or less regularly only around the maximum of the periodic stellar light variations. 
In the velocity range $-18 < V_{\rm los} < -16$ \kms\ emission was generally present but never dominating the profile; the peak at TJD = 8681 (29 February 1992) is caused by a spectral component at $-18.2$ \kms, just outside this
range (component \Dtwo\ in Table \ref{tab:components}).
The plot clearly demonstrates that the maser emission is responding in strength to the periodic variability of the star. 
There is also an apparent broadening of the profile at regular time intervals, likewise connected to the pulsational period of the star. As will be shown in Section \ref{opt-radio-variability}, the radio emission varies with the optical period but is lagging behind the optical one by about three months. In 1987 and 2015 the profiles were similar to those in 1990--2011 but the emission in the outer parts of the profile was not detected (Fig. \ref{fig:uher_all}, Appendix C).

On top of the regular component of variability, non-regular flux density variations of individual maser features occurred, which led to strong profile variations over the years. This is exemplified by the upper envelope spectrum (Fig. \ref{fig:uher-upenv}), where the strongest feature is at $-18.3$ \kms. This feature was strong for about 18 months between January 1991 (TJD $\sim$8250) and July 1992 (TJD $\sim$8800) (Fig.~\ref{fig:uher_sel} and Fig.~\ref{fig:uher_all}; Appendix C). In the following this period will be referred to as the '1991/1992 peculiar phase'. The feature brightened again in autumn 1996 (TJD = 10352) for less than a year. No comparable brightenings were observed redwards of $-14$ \kms. In contrast, the second prominent feature in the upper envelope spectrum at $-15$ \kms\ was permanently present, even in 1987 and 2015 prior to and after the phase of contiguous observations (see the lower envelope spectrum, Fig. \ref{fig:uher-loenv}). 
The emission close to the borders of the velocity range at $V_{\rm los} < -20$ and $V_{\rm los} > -12$ \kms\ (cf. also Fig. \ref{fig:uher_sel}) is usually weak and becomes strong only occasionally. After 1996 (TJD $\ga$ 10500) the blue-shifted emission at velocities $V_{\rm los} < -18$ \kms\ faded away and after 2003 (TJD $\ga$13000) it was not detected anymore by us. The long-term brightness variations are reflected in the FVt-plot (Fig. \ref{fig:uher-fvt}) as prominent asymmetry in the observed velocity range over time.

\begin{figure}
\resizebox{9cm}{!}{\rotatebox{270}{
\includegraphics
{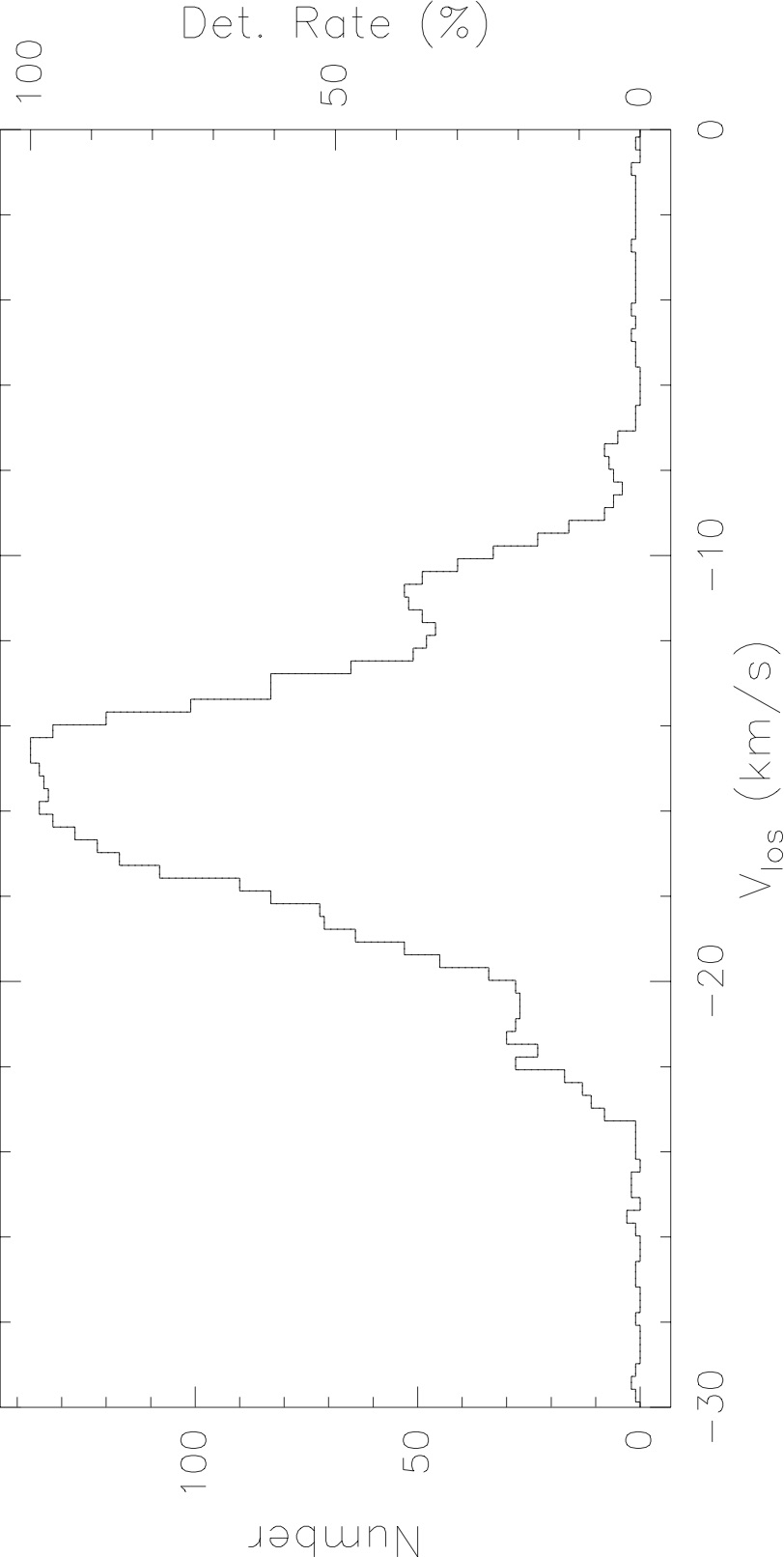}}}
\caption{Detection rate histogram for U~Her; 1987-2015.}
\label{fig:uher-histo}
\end{figure}

\subsubsection{\label{sdd_MaserVel} Variations of the \water\ maser velocity range}
The detection rate histogram (Fig. \ref{fig:uher-histo}) confirms that the dominant spectral features occurred between $-16$ and $-14$ \kms. It also shows that the total velocity range over which emission was detected is $-23.3 < V_{\rm los} <  -7.1$ \kms\ (Table \ref{centralcoords}), which is symmetric with respect to the stellar radial velocity. The FVt-plot shows also that the width of the observed velocity range is varying. This is caused by the  drop of the maser brightness at the weaker outer parts of the maser profile below the threshold of the FVt-plot ($\sim$1 Jy) during the faint part of the stellar variability cycle. 

The blue border of the \water\ maser profile of U~Her had been a point of discussion in the past, after  emission had been detected at velocities $\sim$2 \kms\ bluewards of the velocity range covered by the OH 1667 MHz maser emission and other molecular species \citep{engels88,bowers94}.  However, given the final expansion  velocity as obtained from more recent CO observations (see Table 1) the extreme blue \water\ maser velocities at $<-23$ \kms\ \citep{engels88}, seen before the start of our observations are not 'forbidden' by the 'standard model' anymore, and instead asymmetries in the OH maser shell could be responsible for the lack of OH maser emission at very blue velocities.

\subsubsection{Periodicity in the optical and radio light curves
\label{opt-radio-variability}}
As is evident from the FVt-plot (Fig. \ref{fig:uher-fvt}), the \water\ maser variations of U~Her show periodic behaviour, which is caused by the maser's strong response to the stellar brightness variations. 

We created the radio light curve of U~Her using the integrated flux density determined over a fixed velocity interval encompassing all velocities at which maser emission was detected. The optical data (V-band) were taken from AAVSO (Kafka, 2021\footnote{Observations from the AAVSO International Database, https://www.aavso.org}) for the years 1986--2015 encompassing the monitoring program and consisted of $>$2400 observations, while the radio data consisted of 137 observations. For both data sets a Fourier analysis was made to search for periodicity. The Lomb periodogram \citep{press92} of the optical data showed a well defined period $P_{\rm opt} = 405\pm2$ days, in agreement with the VizieR\footnote{Ochsenbein F., et al., The VizieR database of astronomical catalogues, DOI = 10.26093/cds/vizier; \cite{ochsenbein00}}
period of 406 days. The periodogram of the radio light curve confirmed the optical period ($P_{\rm rad} = 407$ days; Table \ref{centralcoords}), albeit with much larger uncertainties. 
To analyse the maser variations in relation to the optical variations of the star, we modelled in the following the optical and maser light curves by sine-waves with a common period and related the model light curves to each other. 

\subsubsection{The model for the optical light curve \label{sdd_OptModelLcurve} }
The \water\ maser variations are not in phase with the optical variations. It is well-established that for Mira variables they lag behind several weeks to months \citep{staley94, berulis98, shintani08}. To study this behaviour quantitatively we set the optical reference phase $\varphi_{\rm s} = 0$ at the maximum of the optical model sine curve. These maxima are delayed in the mean by $25\pm10$ days relative to the real optical maxima ($\Delta\varphi_{\rm s} = 0.06\pm0.025$ in units of phase). The delay is caused by the asymmetry of the optical light curve of U~Her with a steeper rise to the maximum and a slower decline to the minimum and the scatter is due to the varying time differences between two real optical maxima. We found $\Delta T = 406 \pm 15$ days as the average time difference between two consecutive maxima, with extreme time differences of 374 and 426 days. The choice to link the optical phase to the model sine curve is therefore the only way to define an optical reference phase independent from the details of the optical light curve or the choice of the time interval over which the light curve is analysed. Adopting this approach, radio-optical phase lags can be compared between stars having different quality of the sampling of their optical light curves. Using for example the {\it mean} time difference between the observed optical maxima as reference is an alternative way to determine the delay $\Delta\varphi_{\rm s}$, but this method would be restricted to stars where the optical maxima are well observed. Our optical model light curve has a period $P_{\rm opt} = 405$ days and a reference epoch for maxima TJD$_{max} = 6668 \pm 3$ days (Table~\ref{centralcoords}). 

\begin{figure}
\includegraphics[angle=-90,width=\columnwidth]
{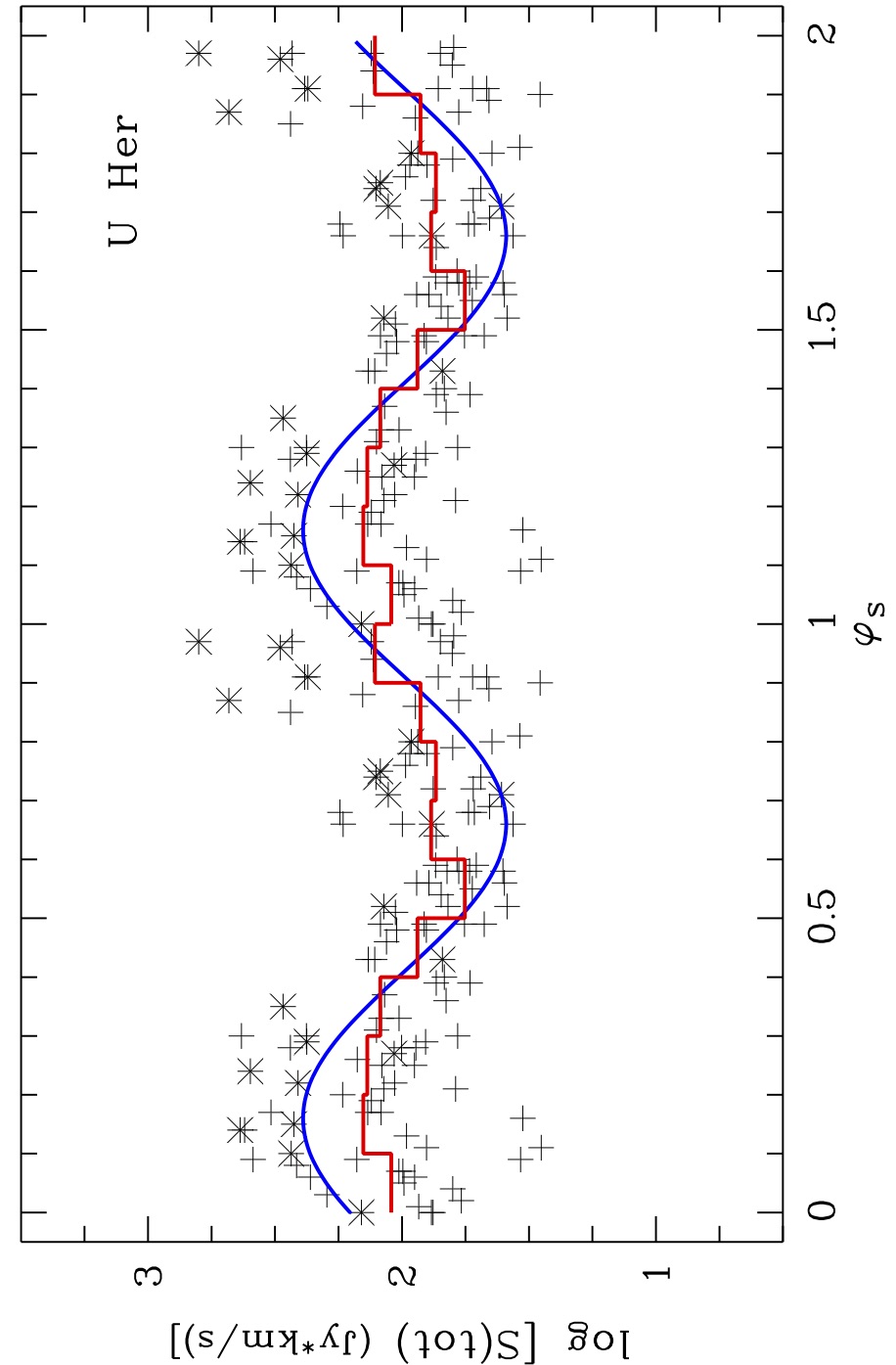}
\caption{U~Her \water\ maser light curve. Plotted are integrated fluxes $S(\rm tot)$ in the velocity range $-24 < V_{\rm los} <-6$ \kms\ in Jy \kms\ vs. optical phase   $\varphi_{\rm s}$. For better visualization the data are repeated for a second period. $\varphi_{\rm s} = 0$ is defined as the time of maximum optical brightness. Datapoints marked by an asterisk (*) are from Effelsberg, the plusses (+) are Medicina data. Overplotted are average integrated fluxes in phase bins of 0.1 (red), and a sine curve (blue) which was obtained by a fit to the 1990--2011 radio measurements with a period of $P_{\rm opt}=405$ days. The sine curve is delayed by $\phi_{\rm lag} = 0.16$, i.e. by 64 days with respect to the optical maximum.} 
\label{fig:uher-lcurve}
\end{figure}

\subsubsection{The phase lag between optical and radio light curve \label{sdd_phase-lag} }
\begin{figure}
\includegraphics[angle=-90,width=\columnwidth]
{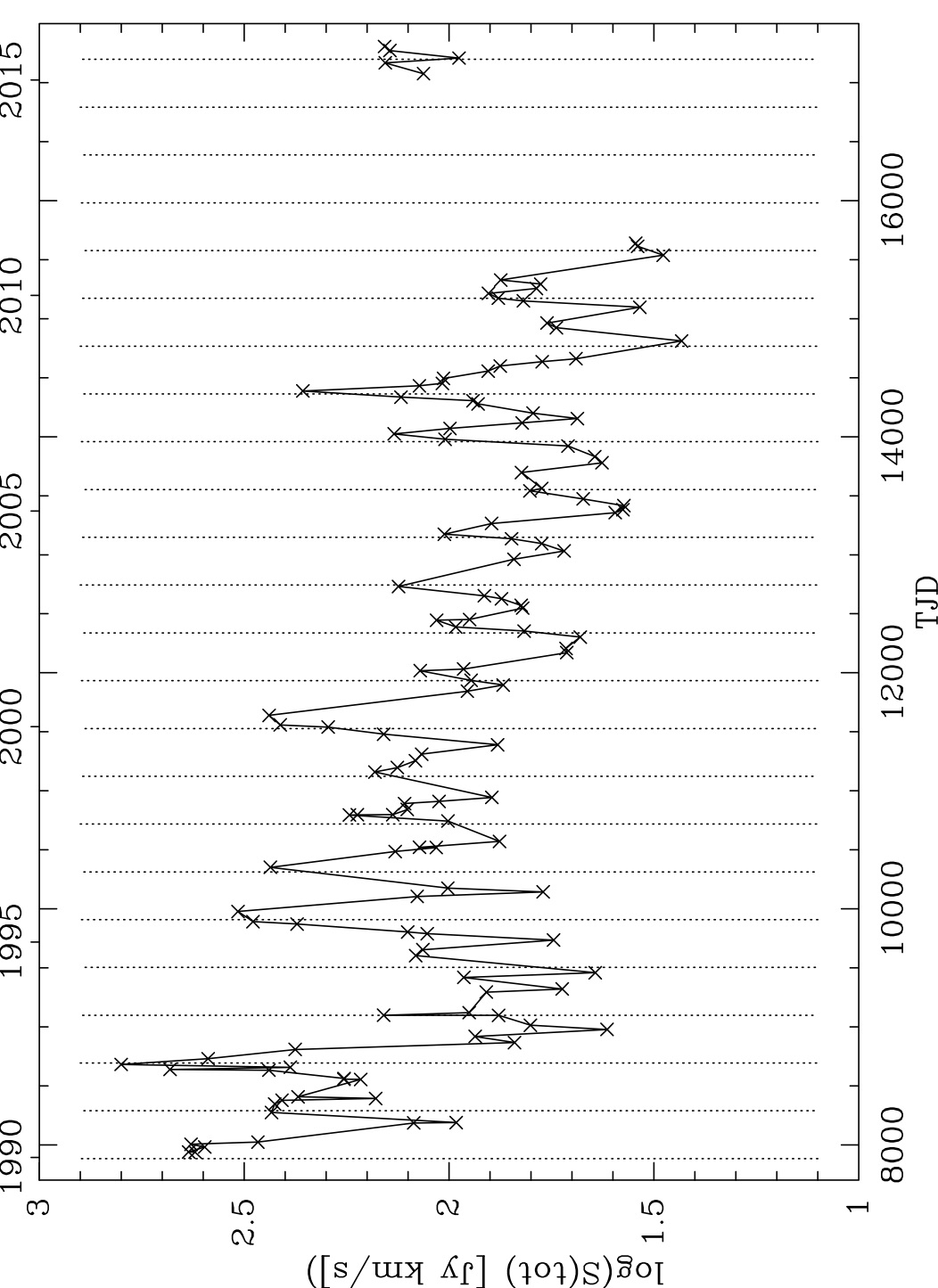}
\caption{U~Her \water\ maser light curve, showing the total flux (integrated between $V_{\rm los}$ $-24$~\kms\ and $-6$~\kms) as a function of TJD. The vertical dashed lines indicate the (modelled) optical maxima with $P = 405$~days.
} 
\label{fig:uher-stot-tjd}
\end{figure}

The lag of the radio light curve relative to the optical one was determined with the fit of a sine curve to the radio data using the optical period, and the amplitude as free parameter. Only radio observations during the continuous monitoring between 1990 and 2011 were used, and observations taken within 3 days were averaged. The final data-set to determine the radio light curve consisted of 125 maser spectra. The resulting lag of $\phi_{\rm lag} = 0.16$ (Table \ref{centralcoords}) is only weakly depending on the choice of the amplitude. The radio light curve is shown in Fig.~ \ref{fig:uher-lcurve} as a function of the optical phase $\varphi_{\rm s}$. It is immediately clear that the scatter in integrated fluxes $S$ is large for any particular phase, indicating that the luminosity variations of the star can explain only part of the maser variability seen. To visualize the periodic component of the variations we overplotted a binned light curve (average integrated fluxes in bins of 0.1 in phase) and the sine curve obtained from the fit using the optical period. In Fig.~ \ref{fig:uher-lcurve} the integrated flux densities $S(\rm tot)$ obtained with the Effelsberg telescope appear to be systematically brighter than those obtained with the Medicina telescope. This is a selection effect caused by the general brightness decrease of U\,Her's maser light curve (see Fig.~\ref{fig:uher-stot-tjd}) and the limitation of the Effelsberg observations to the first years. This leads to a fraction of observations during bright maser phases being significantly higher for the Effelsberg than for the Medicina radio telescope.

\begin{figure}
\includegraphics[angle=-90,width=\columnwidth]
{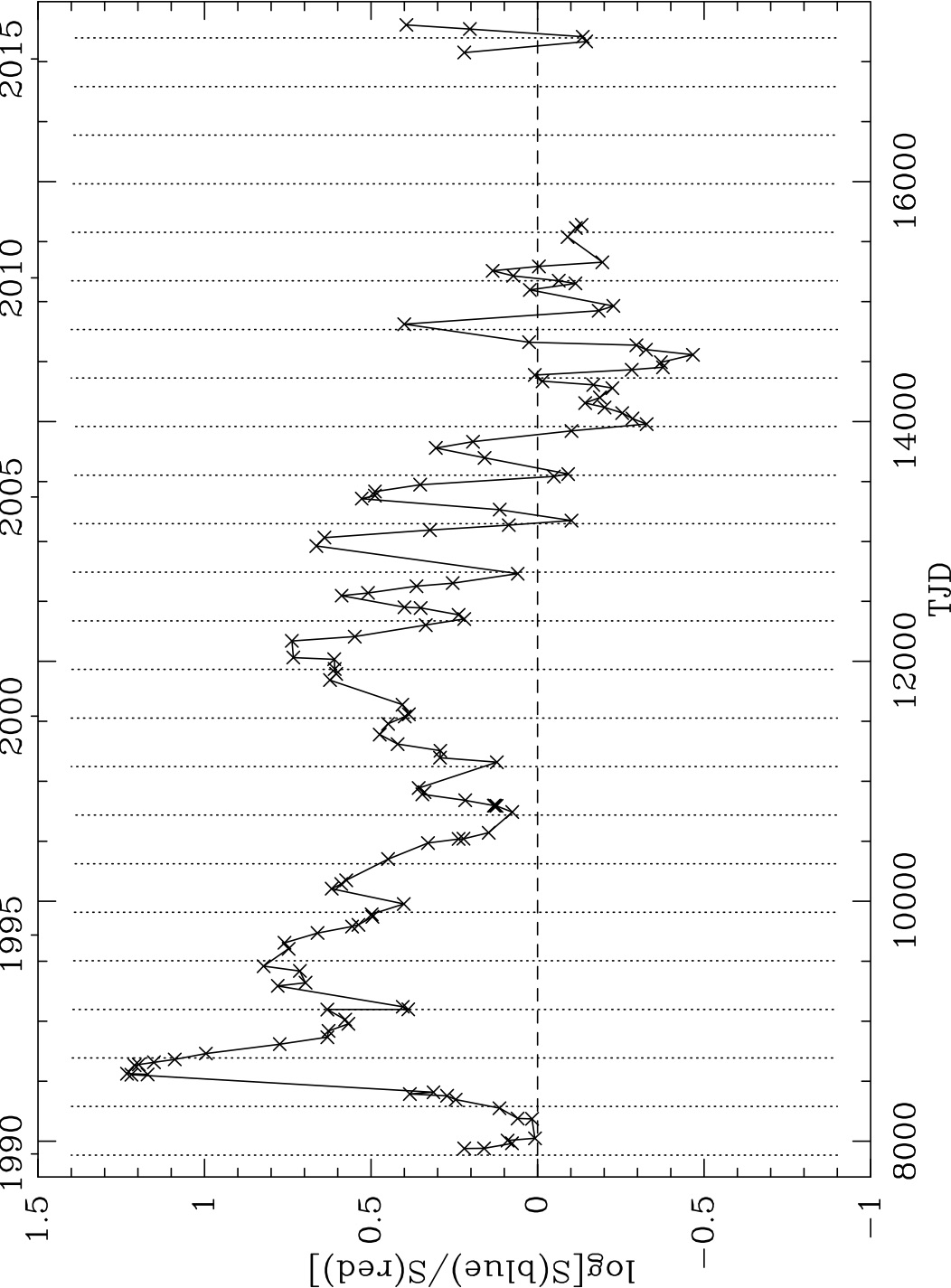}
\caption{The ratio $S(\rm blue)$/$S(\rm red)$ of the U~Her \water\ maser emission, of the $V_{\rm los} < -15$~\kms\ [$S(\rm blue)$] and $> -15$~\kms\ [$S(\rm red)$] part of the maser velocity range with respect to the stellar velocity, as a function of TJD. The vertical dashed lines indicate the (modelled) optical maxima with $P = 405$~days.
} 
\label{fig:uher-sratio-tjd}
\end{figure}

\subsubsection{\label{sdd_longterm_lc} The long-term radio light curve}
In addition to the periodic maser brightness variations, additional brightness changes are seen also on timescales shorter and longer than the stellar period. 

In Fig.~\ref{fig:uher-stot-tjd} we plot the total flux of the U~Her \water\ maser as a function of time between 1990 and 2015. As is evident also here, the variations in the maser emission follow the optical variations of the star, indicated by the dashed lines that mark the TJD of the (modelled) stellar maxima (see Sect.~\ref{sdd_OptModelLcurve}). Although the dominance of the emission in the $-16$ to $-14$ \kms\ velocity interval after 1992 suggests some long-term continuity, this continuity is restricted to velocities and not to brightness levels. The radio light curve shown in Fig.~\ref{fig:uher-stot-tjd} indicates a clear decrease by a factor of 4 of the average brightness level  between 1990 ($S(\rm tot) \sim 200$ Jy \kms) and 2011 ($S(\rm tot) \sim 50$ Jy \kms). In 2015 the brightness level had increased again to ($S(\rm tot) \sim 125$ Jy \kms), while in April 1984 the total flux was 185 Jy \kms\ \citep{engels88}. The strong emission in February 1992 during the ’1991/1992 peculiar phase’ (see Sect.~\ref{sdd_MaserSpec}) could have been a burst. 
After this phase emissions at $V_{\rm los} < -$17 \kms\ dropped sharply and the total flux went through a weak phase lasting until 1995, when a brightness increase of the $-15.5$ \kms\ feature brought the total flux back to a level following the long-term decline of the average brightness.\\

The 1990--2011 long-term brightness decrease is not uniform over the velocity range, but due to a systematic brightness decrease of the emission that is blue-shifted with respect to the stellar velocity ($V_*$ = $-$15 \kms). As shown in Figure~\ref{fig:uher-sratio-tjd} the ratio $S(\rm blue)$/$S(\rm red)$ between the blue- and red-shifted total flux is continuously decreasing between $\sim$1992 and $\sim$2007. During the ’1991/1992 peculiar phase’ $S(\rm blue)$ was $\sim$15 times stronger than $S(\rm red)$, while in 2007/2008 the ratio could be as small as $\sim$0.5. The strength of the red-shifted emission in the period 2007 -- 2011 (TJD $>$ 13500) appears to be due to the shift of the peak emission in the  $-16$ to $-14$ \kms\ velocity interval by $< 1$~\kms\ to the red (see the FVt-diagram,  Fig.~\ref{fig:uher-fvt}, left). The choice of the stellar radial velocity influences the ratio quantitatively but its trend remains for any radial velocity within the dominant $-16$  to $-14$ \kms\ interval. The $S(\rm blue)$/$S(\rm red)$ ratio and its variation indicate an asymmetry of the excitation conditions in the front part of the \water\ maser shell of U~Her, where the blue-shifted emission comes from, compared to the rear part, where the red-shifted emission originates. 

The radio light curve (Fig. \ref{fig:uher-stot-tjd}) shows three rather bright maxima compared to the times before and after. They are the possible burst in the  ’1991/1992 peculiar phase’ (peak emission at TJD = 8682), the maximum in 2000 (peak on TJD = 11640) and the maximum in 2007 (peak on TJD = 14389). We consider them as short-term fluctuations rather than as evidence of 'super-periodicity', because the time intervals between the peaks with a duration of 6.8 and 7.3 stellar cycles do not match. The next maximum would have been expected in April - September 2015. We have observations in this time interval, but no information on the brightness levels before and after. It is therefore not possible to decide if the brightness levels observed in 2015 belong to a local maximum or are part of a general increase of the brightnesses.

Besides our monitoring program, U~Her's \water\ maser has been observed with single-dish telescopes only occasionally by other groups. Excluding the ’1991/1992 peculiar phase’, the strongest maser feature was consistently reported at $\sim -14.5$ \kms\ by \cite{comoretto90} for March 1987, by \cite{kim10} for June 2009, and by \cite{neufeld17} for May 2016. In 1991 the strongest peaks were at $-16$ \citep{takaba94} and $-19$ \kms\ \citep{takaba01} in accordance with our observations.

\subsubsection{\label{sdd_LineProfAna}  Velocity variations of individual \water\ maser features}
Besides the regular periodic and long-term brightness variations of U~Her's \water\ maser emission, also small changes in velocity of the maser features are apparent in the FVt-plot (Fig. \ref{fig:uher-fvt}). For their analysis we decomposed the \water\ maser spectra 1987 -- 2015 into separate features by fitting multiple Gaussian line profiles. The details of the fitting technique are described in Paper I. These maser features can be traced over some period of time in several consecutive spectra, fade away, and may reappear at later times perhaps with a slightly different velocity. As in the semi-regular variable stars (Papers I and II), the full width at half maximum (FWHM) of strong features (visible as distinct peaks in the spectra) is $\sim$1~\kms, and therefore features with FWHM $\ga$ 2~\kms\ are most probably blends. 
We assume that maser features in adjacent (in time) spectra with velocity differences $\la$0.5~\kms\ belong to a unique emission region in the \water\ maser shell, which persisted over this period of time (i.e. the time between the two observations) and varied in intensity.

For accounting purposes all spectral features were grouped according to their velocities into {\it maser spectral components}. The assignment of the features to the spectral components in the four velocity intervals ($< -18$, $-18$ to $-16$, $-16$ to $-14$ and $>-14$ \kms\ as introduced in Sect. \ref{sdd_MaserSpec}) is discussed in Appendix A. Tables \ref{tab:compUHerB-E} and \ref{tab:compUHerG-M} list the spectral features identified by the fitting procedure and their assignments.

The spectral components are labeled with capital letters A, B, ... M in order of increasing velocity $V_{\rm los}$. Their labeling is synchronized with the labels of the spatial components (to be introduced in Sect. \ref{id_spatial_c}), so that corresponding spectral and spatial components share the same label. Due to strong blending in velocity space, in each of the four velocity ranges only few (one to four) spectral components could be defined. In total we identified eleven spectral components. Not all spatial components could be identified in the single-dish spectra, especially not the fainter ones, and therefore there are no spectral components matching the spatial components A1, F1+F2, H1+H2, and J1+J2 (cf. Table \ref{tab:components} in Sect. \ref{id_spatial_c}). 
Two spectral components (D at $\sim-18.5$ \kms\ and G at $\sim-15.0$ \kms) are obvious blends with velocity separations less than the FWHM of the features. In Tables \ref{tab:compUHerB-E} and \ref{tab:compUHerG-M}   the subcomponents making up the spectral components D and G were labeled \Done, \Dtwo\ and \Gone, \Gtwo\ respectively. 

The maser spectral components identified in individual spectra (Tables \ref{tab:compUHerB-E} and \ref{tab:compUHerG-M}) are graphically displayed in Fig. \ref{fig:uher-fvt} (right panel), where it can be compared directly with the FVt-plot. Often, changes of the spectral component peak velocities $V_{los}$ (taken from Tables~\ref{tab:compUHerB-E} -- \ref{tab:compUHerG-M}) in all four velocity ranges occur on timescales of many months by more than $0.5$~\kms, although not in a systematic way. An example are the peak velocities of spectral component \Gone\ in 2007 -- 2011 (TJD $\ga 14000$) with velocities $-16.0 \le V_{los} \le -14.8$. We interpret the meandering of the velocities as a superposition of blended maser features varying in brightness asynchronously and coming perhaps over some time from different locations.  

The interpretation of the short-term velocity variations is less ambiguous in the $>-14$ \kms\ velocity range, where fewer spectral features are apparent and therefore blending is less of a problem.  
Component I shows evidence for blending between 1990 and 1996 ($\sim 7900 <$ TJD $< \sim 10200$) with peak velocities varying back and forth by almost 1 \kms\ ($-13.7$ to $-12.7$ \kms; see Table \ref{tab:compUHerG-M}), 
while the velocity remained almost constant at $-12.9$ \kms\ thereafter until TJD $\sim$ 14500. It then reappeared at the same velocity in 2015. Component K has a peak velocity of $\sim-11$ \kms\ and was detectable only after 1994 (TJD $>$ 9750), also without significant velocity variations. Component L was seen only in two epochs 1990--1992 (TJD $<$ 8900) at $-10.2$ \kms\ and 2007--2010 (14000 $<$ TJD $<$ 15300) at $-9.7$ \kms, and it is unclear if the emissions in these two epochs are related to each other. Finally, component M at $\sim-8$ was only seen at the beginning of the monitoring program, in parallel to component L (1990-1992), while the maser emission in U~Her was strong over the full profile. There were too few appearances to draw conclusions on its velocity variations.   

In the velocity range covered by components I to M we would expect shifts of increasing velocity (components becoming redder),
if the emission regions would persist and move with the expanding CSE. This is not the case here, so that the emission of these components must have come from different emission clouds in the course of the monitoring period. 
The absence of long-term velocity shifts of the spectral components will be discussed further in Sect. \ref{region-lifetime}. 

\begin{figure*}
\centering
\resizebox{17cm}{!}{
\includegraphics{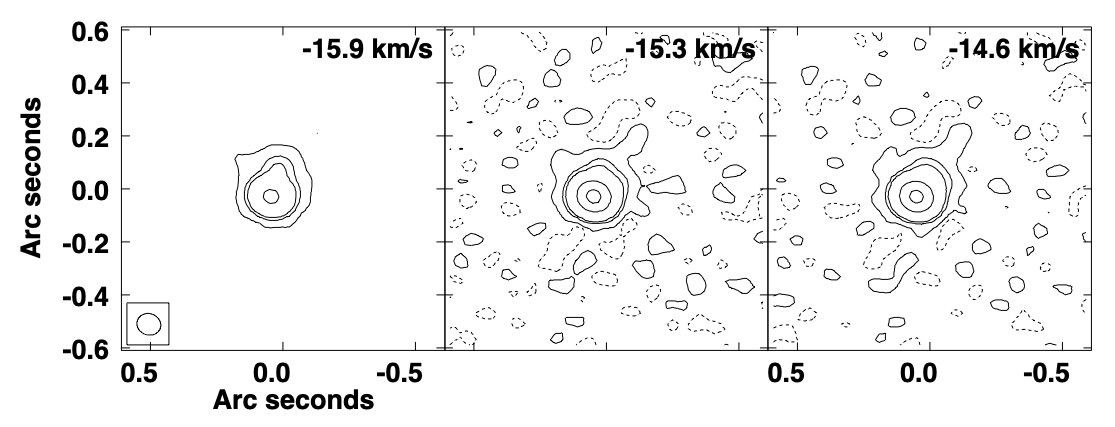}}
\caption{Sample \water\ maser images of U~Her from February 1990 covering the velocity range between $-16$ and $-14$ \kms\ with the strongest emission. The synthesised FWHM beam size (major axis: 0\pas09; minor axis: 0\pas08; position angle of the major axis: 71\degr) is shown in the left panel  (-15.9~\kms). The images are oriented along right ascension and declination and the angular scales are relative to the position \recta{16}{25}{47}{39}, 
\dec{+18}{53}{32.9} (J2000).  Brightness contours are -0.25 (dashed contour), 0.25, 2.5, 5, 50, and 150~Jy per beam area (1.4 \powerten{-13}~sr). }
\label{fig:channel_maps}
\end{figure*}

\subsection{Interferometric data }
The VLA observations of the \water\ masers in U~Her were made with the aim to identify the emission sites in the CSE and breaking the spatial degeneracy in the single dish data. Due to the limited spatial resolution this was only partially successful. The size of the emission sites is not specified a priori, but we refer to the most compact gas clumps hosting maser emission as 'maser clouds'. The data analysis of the images yields maser spatial components, which in general will be superpositions of several maser clouds close to each other in space as well as in velocity.

\begin{table}[!t]
\caption[]{\label{tab:components} Spatial ("Spat") and spectral ("Spec") components of U~Her 1990 -- 1992.}
\begin{tabular}{rlrrrrl}
\hline\noalign{\smallskip}
Spat & Date & $V_{\rm los}$  & S$_p$ & Xoff & Yoff & Spec \\
 &      & [km/s] & [Jy] & [mas] & [mas] &  \\
\noalign{\smallskip}\hline\noalign{\smallskip}
A1 & 91 Oct. & $-$23.8 & 0.4    &  +26  &  +30  &  --  \\[0.1cm]   
B1 & 90 Feb. & $-$21.5 & 2.9    &  +59  &  +32  &  B   \\        
B1 & 90 Jun. & $-$21.2 & 0.1    &  +21  &  +6   &      \\[0.1cm] 
B2 & 91 Oct. & $-$22.0 & 25.0   &  $-$44  &  $-$6   &  B    \\[0.1cm]  
C1 & 90 Feb. & $-$19.9 & 1.5    &  $-$33  &  $-$48  &  C   \\        
C1 & 91 Oct. & $-$19.9 & 41.0   &  $-$38  &  $-$54  &      \\        
C1 & 92 Dec. & $-$20.2 & 0.1    &  $-$56  &  $-$54  &      \\[0.1cm]  
C2 & 90 Feb. & $-$19.9 & 0.6    &  +45  &  +42  &      \\[0.1cm]  
C3 & 91 Oct. & $-$19.2 & 5.0    &  $-$46  &  +58  &      \\[0.1cm] 
D1 & 90 Feb. & $-$18.2 & 52.0   &  $-$9   &  $-$60  &  \Dtwo  \\        
D1 & 90 Jun. & $-$18.1 & 7.3    &  $-$9   &  $-$60  &      \\        
D1 & 92 Dec. & $-$18.2 & 3.0    &  $-$8   &  $-$68  &      \\[0.1cm] 
D2 & 90 Feb. & $-$17.9 & 12.0   &  $-$25  &  +46  &   \Dtwo    \\        
D2 & 90 Jun. & $-$17.9 & 2.5    &  $-$27  &  +46  &      \\
D2 & 91 Oct. & $-$17.9 & 5.0    &  $-$20  &  +52  &      \\[0.1cm] 
E1 & 90 Feb. & $-$17.2 & 0.8    &  +61  &  +62  &      \\[0.1cm] 
E2 & 90 Feb. & $-$16.9 & 2.4    &  $-$63  &  $-$36  &      \\[0.1cm] 
E3 & 90 Feb. & $-$16.9 & 10.0   &  $-$1   &  $-$58  &  E   \\        
E3 & 91 Oct. & $-$17.2 & 12.0   &  $-$32  &  $-$36  &      \\        
E3 & 92 Dec. & $-$17.2 & 2.0    &  $-$8   &  $-$60  &      \\[0.1cm] 
E4 & 90 Jun. & $-$16.6 & 2.7    &  $-$37  &  +10  &      \\[0.1cm] 
E5 & 91 Oct. & $-$16.6 & 1.8    &  $-$44  &  +60  &      \\        
E5 & 92 Dec. & $-$17.2 & 3.5    &  $-$34  &  +66  &      \\[0.1cm] 
F1 & 90 Feb. & $-$15.9 & 5.0    &  $-$25  &  +42  &      \\        
F1 & 91 Oct. & $-$15.9 & 2.2    &  $-$48  &  +30  &      \\[0.1cm]         
F2 & 92 Dec. & $-$15.9 & 1.5    &  +24  &  +32  &      \\[0.1cm] 
G1 & 90 Feb. & $-$14.9 & 230.0  &  +15  &  $-$54  &  \Gone + \Gtwo   \\        
G1 & 90 Jun. & $-$15.0 & 145.0  &  +15  &  $-$52  &   \\ 
G1 & 92 Dec. & $-$15.3 & 14.0   &  +14  &  $-$46  &      \\[0.1cm] 
G2 & 90 Jun. & $-$14.9 & 5.0    &  $-$21  &  +8   &      \\[0.1cm]         
G3 & 92 Dec. & $-$15.5 & 4.0    &  $-$74  &  $-$10  &      \\[0.1cm] 
G4 & 92 Dec. & $-$15.4 & 3.0    &   +8  &  +42  &      \\[0.1cm] 
H1 & 90 Feb. & $-$14.3 & 5.0    &  $-$43  &  +32  &      \\        
H1 & 90 Jun. & $-$13.9 & 3.8    &  $-$51  &  +32  &      \\        
H1 & 92 Dec. & $-$13.9 & 0.7    &  $-$76  &  +10  &      \\[0.1cm]   
H2 & 91 Oct. & $-$14.5 & 0.3    &  $-$44  &  +42  &      \\        
H2 & 92 Dec. & $-$13.9 & 1.0    &   $-$4  &  +54  &      \\[0.1cm] 
I1 & 90 Feb. & $-$12.9 & 1.0    &  +49  &  $-$14  &  I   \\        
I1 & 90 Jun. & $-$13.3 & 0.8    &  +31  &  $-$4   &      \\[0.1cm]  
I2 & 90 Feb. & $-$12.6 & 1.0    &  $-$35  &  +44  &      \\[0.1cm] 
J1 & 90 Jun. & $-$12.0 & 0.3    &  $-$29  &  +34  &      \\         
J1 & 92 Dec. & $-$12.0 & 0.1    &  $-$28  &  +74  &      \\[0.1cm] 
J2 & 92 Dec. & $-$11.6 & 0.1    &  $-$34  &  $-$28  &      \\[0.1cm] 
L1 & 90 Feb. & $-$10.0 & 1.4    &  $-$67  &  +4   &  L    \\       
L1 & 90 Jun. & $-$10.0 & 0.8    &  $-$67  &  +6   &      \\[0.1cm] 
M1 & 90 Feb. & $-$7.7  & 0.3    &  $-$43  &  +36  &  M   \\        
M1 & 90 Jun. & $-$8.0  & 0.2    &  $-$49  &  +32  &      \\[0.1cm] 
M2 & 92 Dec. & $-$8.3  & 0.1    &  $-$20  &  $-$10  &      \\
\noalign{\smallskip}\hline
\end{tabular}
\end{table}

\subsubsection{\label{id_spatial_c} Spatial component identification}
The VLA interferometric data consist of one data cube for each of the four epochs, containing 63 channel maps each.  The maps are separated in velocity by 0.658~\kms. An example is given in Fig. \ref{fig:channel_maps} where maps of the 3 channels with the strongest emission seen in the first VLA epoch (February 1990) are shown. The sensitivities measured in a line-free channel were 12--21 mJy/beam. In order to single out maser components the data cubes were analysed within AIPS\footnote{Astronomical Image Processing System,\\ www.aips.nrao.edu/index.shtml.} in a three step process. First, the individual channel maps were analysed one by one by fitting multiple 2D Gaussians, then spatial components were identified by comparing the fit results in neighbouring channel maps, and finally these components were verified in velocity space. 
The details of this analysis are described in Paper I. 

About twelve spatial maser components were identified in each VLA observing epoch. These components are listed in Table \ref{tab:components}, which gives the spatial component (Spat), the VLA observing epoch, the line-of-sight velocity $V_{\rm los}$ and peak flux density $S_{\rm p}$ of the spatial components identified, the spatial offsets (Xoff and Yoff) from the adopted map centre (defined in Sect. \ref{map-alignment}) and the associated (single dish) spectral component (Spec) from Tables \ref{tab:compUHerB-E} and \ref{tab:compUHerG-M} in the Appendix. Spectral component B had different spatial counterparts in 1990 and 1991. Spectral component \Done\ is not listed as it appeared only after 1990--1992. Spatial component G1 is likely of composite nature, as the corresponding spectral component G is according to our analysis of the spectral profiles made up by two components (\Gone\ and \Gtwo\ in Table~\ref{tab:compUHerG-M}). 

For the epoch June 1990 the components can be compared to those found by \cite{colomer00}, who used the VLA to observe U~Her one day apart from our observation. As in the case of RX~Boo (see Paper I) they found about the same number of components (13 vs. 12). However, their components are spread over a smaller velocity range of $-18.3 \le  V_{\rm los} -10.2$ \kms, because they did not detect the faint components B1 ($V_{\rm los} = -21.2$ \kms, $S_\nu = 0.1$ Jy) 
and M1 ($V_{\rm los} = -8.0$ \kms, $S_\nu = 0.2$ Jy) (cf. Table~\ref{tab:components}).
Our strongest spatial component in June 1990, G1 ($V_{\rm los} = -15.0$ \kms) is split by the 3-dimensional Gaussian fitting program of Colomer et al. into three spatial components in the velocity range $-15.3 < V_{\rm los} < -14.6$ \kms. This corroborates our conclusion that G1 is of composite nature. Common components in both June 1990 maps are present outside the very crowded main velocity range $-16 < V_{\rm los} < -14$ \kms, if flux densities surpassed 1 Jy, whereas weaker spatial components 
were not recognized by the fitting program of Colomer et al.. As discussed in Paper I, the two fitting methods lead to different results for weaker components and regions of high spatial blending. The overall spatial distributions of both maps is however similar, so that the projected angular shell sizes are similar. 

\subsubsection{Alignment of the maps \label{map-alignment}}
The maps taken between 1990 and 1992 were aligned to a common origin using spatial components present over two or more observing epochs and assuming that the components are located in a ring-like structure around the star. Matched components are given a common designation in Table \ref{tab:components}. For example at $V_{\rm los} \approx -20.0$ \kms\  the strong C1 spatial component seen in October 1991, is identified in February 1990 and December 1992 as a weak component, while other spatial components (C2, C3) identified at (or close to) this velocity are clearly coming from different parts of the shell.  
As in the case of RX~Boo (Paper I), the 1990 maps had many components in common, while components in 1991 and 1992 were difficult to identify with components seen in the other years. The identification was further complicated by the poor east-west resolution in October 1991, and the ’1991/1992 peculiar phase', in which the masers were at that time. As discussed in Sect. \ref{sdd_MaserSpec}, the maser emission at $V_{\rm los} \leq -18$ \kms\ was prominent around the turn of the year 1991/1992, while in other epochs this emission was relatively weak and emission from the  $-16 < V_{\rm los} < -14$ \kms\ velocity range prevailed. In October 1991 spatial components B2 and C1 were strongest (Table \ref{tab:components}), while the strongest component during the other three epochs (G1) could not be identified. 
The components used to align the December 1992 map with the maps from 1990 were D1 and G1, which were strong in both years. C1 and D2 were used to align the October 1991 map. After alignment these components scattered in position by $\le12$ mas.

\begin{figure}
\includegraphics[width=\columnwidth]
{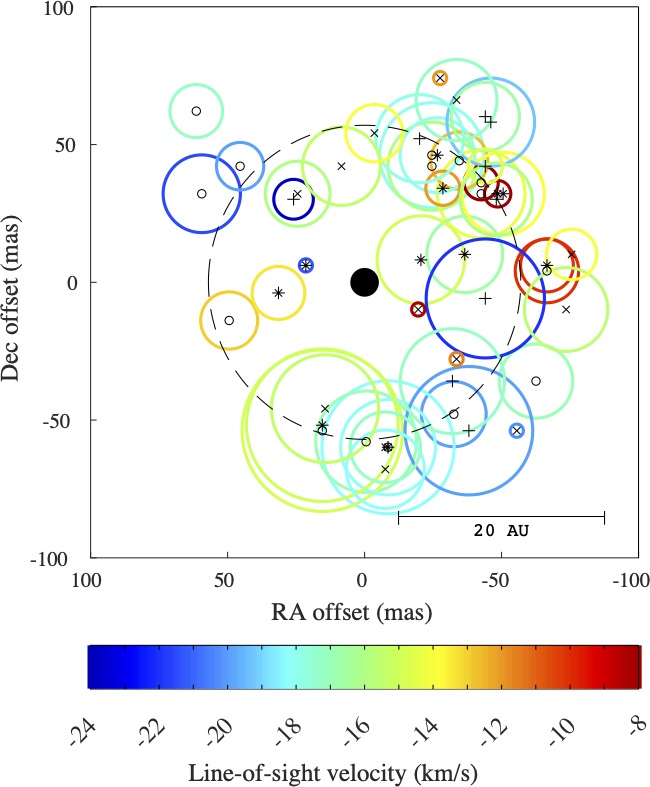}
\caption{All the spatial components of U Her listed in Table 3 plotted on the sky. Each component is represented by a symbol surrounded by a circle with a diameter $d$ depending on flux density $S_\nu$: $d = 8\,({\rm log}\,S_\nu  + 1.3)$. 
The dates are represented by different symbols: Feb. 90 by small circles; Jun. 90 by asterisks; Oct. 91  by plus signs; Dec. 92 by crosses. The circles around the components are colour coded according to the line-of-sight velocity of the component (see the scale below the map). The dashed circle with a radius of 57~mas has been obtained from a fit to the components with line-of-sight velocities $-18 < V_{\rm los} < -14$ \kms\ (see text) and the origin of the plot has been moved to the center of this circle. The filled black circle at the center symbolizes the central star with a diameter of 10.65~mas \citep{vanbelle96}.} 
\label{fig:uher-rainbow}
\end{figure}

A plot of all spatial components on the sky relative to a common origin as given in Table \ref{tab:components} is shown in Fig. \ref{fig:uher-rainbow}\footnote{Note that Fig.~2 in \cite{winnberg11} erroneously shows the mirror image of this distribution.}.  The distribution of the components suggests a ring-like structure. To find the most likely position of the star, a circle was fitted according to a least-squares method to all components having radial velocities between $-18$ and $-14$ \kms. They were considered as being likely 'tangential components', able to outline the ring-like structure of the projected shell. The fit was carried out without weights and the center of the circle was used as our best guess for the stellar position, and as common origin of the plot and of the component offsets in Table \ref{tab:components}. The best fit gave a radius for the circle of 57 mas ($\sim15$ AU).

Spatial coincidences among other components were searched for in the aligned maps.  Coincident spatial components detected in different epochs were given a common label. After subtraction of coincident components we ended up with 28 different spatial components (hereafter 'merged spatial components') of which half were present in at least two maps. Positional deviations between maps were $\la15$~mas, although in a few ambiguous cases we accepted as coincidences also components with deviations up to 45~mas (cf. B1, H2, J1 in Table \ref{tab:components}). 
The number of merged spatial components found and the accuracies in velocities and positions are very similar to the results obtained for the SRV RX~Boo in Paper I. 

\subsubsection{Cross correlation of single-dish and interferometric data 1990–1992}
 The assignment in Table \ref{tab:components} of spectral components identified in Sect. \ref{sdd_LineProfAna} to the spatial components identified in Sect. \ref{id_spatial_c} was made using the velocities in common. Spatial component A1 detected in October 1991 at $-23.8$ \kms\ with a peak flux density of 0.4 Jy was not present in any of our spectra. Its velocity is lower by 0.5 \kms\ than the blue border of the \water\ maser velocity range that we determined in Sect. \ref{sdd_MaserVel} from the single-dish spectra. Due to blending in velocity space also other weaker spatial components ($\le 10$ Jy: spatial components F, H, J) could not be assigned to individual spectral components.
An exception are the spatial components M1 and M2 at $\sim-8$ \kms\ in the extreme red part of the velocity range. Their emission could be detected in the spectra due to absence of  stronger maser emission at neighboring velocities. 
The brighter spectral components \Done\ and K at $-18.9\pm0.2$ and $-10.8\pm0.3$ \kms\ respectively (see Table \ref{tab:compUHerB-E} and \ref{tab:compUHerG-M}) were not seen in our spectra before 1993. Accordingly, spectral component \Done\ was not assigned to any spatial component, and spectral component K is absent from Table \ref{tab:components} because no emission was seen at the corresponding velocities in the VLA maps 1990 -- 1992. Spatial component G1 is a blend of two emission sites, which could be identified as sub-components \Gone\ and \Gtwo\ of spectral component G in the single-dish spectra due to their superior velocity resolution.

For brighter spatial/spectral components the cross-correlation was not unambiguous at several velocities, due to blending in velocity and position. One case is spectral component B with peak velocities $-21.1\pm0.5$ \kms, which
was the strongest in the maser profile probably only for a couple of days 
during the '1991/1992 peculiar phase'.
The VLA map of 20 October 1991 was made close to the maximum of this phase and the emission was detected in the east part of the shell (spatial component B2
at $-22.0$ \kms). The peak velocity of B2 is 1.4 \kms\ lower than the peak velocity $-$20.6 \kms\ of the spectral component B measured 6 and 13 days later (cf. Table~\ref{tab:compUHerB-E}). Such a large velocity difference between spatial and spectral components was not seen by any other component, and may indicate the presence of brief emission bursts on the timescales of many days.

The peculiarity of the '1991/1992 peculiar phase' is evident also from the result that in 1990 the spectral component B maser emission came from a different part of the \water\ maser shell (spatial component B1 in the north-east part of the shell). There was no emission at corresponding velocities in December 1992. In parallel to spectral component B also component C  reached a maximum between October 1991 and April 1992 (see Table \ref{tab:compUHerB-E}). This emission was located in the south-west of the shell, and in this case emission from this part at that velocity was seen also in the maps from 1990 and 1992 (spatial component C1).

 The cross-correlation is also complex for velocities $V_{\rm los} \ge -18$ \kms. Spatial components D1 and D2 ($V_{\rm los} \approx -18$ \kms, velocity difference $\approx 0.2$ \kms) cannot be separated in the single-dish spectra, and coincide in velocity with spectral component \Dtwo. Spectral component \Done\ was not prominent in 1990--1992. Spatial component E3 is the strongest component among five in the velocity range $-17.2  < V_{\rm los} < -16.6$ \kms. It was not identified in the spectra of 1990, but was detected as spectral component E in 1991 and 1992. As E3 comes in velocity space from close to the brighter spectral component \Dtwo, it was not distinguished in the 1990 spectra because of blending. 

In the $-16 < V_{\rm los} < -14$ \kms\ range the dominating spectral feature G, composed of \Gone\ and \Gtwo\ (velocity separation in 1990: 0.7 \kms), was identified spatially as one component G1, 
caused by the insufficient spectral resolution of the VLA data ($\sim$0.7 \kms). The maps show however, that both spectral features came from the same region in the southern part of the shell. Spatial component G1 was detected in 1990 as well as in Dec. 1992 and had been therefore a dominant emission region over a time range of at least 3 years, except for a few months in 1991/1992. 

At velocities $\ge-14.5$ \kms\ of the nine different spatial components listed in Table \ref{tab:components} only three (I1, L1, and M1 from the 1990 maps) could be distinguished also in the spectra (Table \ref{tab:compUHerG-M}). The velocities corresponding to spatial components H1/H2 ($V_{\rm los} \approx -14.1$ \kms) are strongly blended in the spectra by the dominating spectral component \Gtwo . In the 1991 and 1992 maps spatial components at larger velocities ($V_{\rm los} >-14$ \kms) were either absent (October 1991) or extremely weak (0.1 Jy).

\subsection{An \water\ maser shell model for U~Her \label{sec:shell-model}}

\subsubsection{The projected structure}
The distribution of all spatial components observed in U\,Her 1990-1992 (Fig. \ref{fig:uher-rainbow}) is best described as an incomplete ring with most components located in the half of the ring between position angles 170 and 350$^\circ$ counting from North over East.

Following the analysis and discussion of the location of the \water\ masers in the circumstellar envelope of the semi-regular variable RX~Boo (Paper I), we assume that the \water\ masers are embedded in an isotropically expanding envelope. In this case, the relationship between the line-of-sight velocities $V_{\rm los}$ of the maser spatial components relative to that of the star and their projected distances $r_{\rm p}$ from the star is given by
\begin{equation} \label{velocity-law}
\left( \frac{r_{\rm p}}{r}\right) ^2 + \left( \frac{V_{\rm los} - V_*}{V_{\rm out}}\right) ^2 = 1
\end{equation}
where $r$ is the radial distance, and $V_{\rm out}$ is the outflow velocity of the components,
and $V_*$ is the radial velocity of the star. 

\begin{figure}
\includegraphics[width=\columnwidth]
{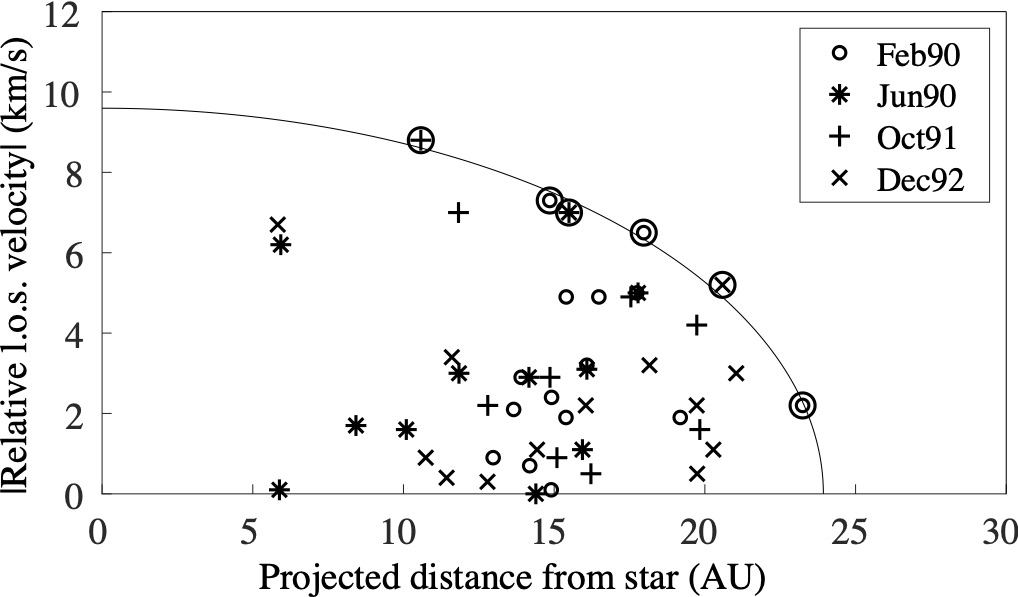}
\caption{Relative line-of-sight velocity $\mid V_{\rm los} - V_*\mid$ versus projected distance $r_{\rm p}$ for all spatial components of U\,Her from Table~\ref{tab:components} identified in the four epochs. The six outer points surrounded by circles were used for the fit with a quart ellipse.}
\label{fig:uher-absvexp}
\end{figure}

Using $V_* = -15.0$ \kms\ (Table \ref{centralcoords}), we plotted in Fig.~\ref{fig:uher-absvexp} the relative line-of-sight velocity $\mid V_{\rm los}-V_*\mid$ in absolute values of all 48 spatial components against their projected distance $r_{\rm p}$. $\mid V_{\rm los}-V_*\mid$ and $r_{\rm p}$ were calculated using $V_{\rm los}$-, Xoff-, Yoff-values from Table\,\ref{tab:components}. For a shell-like distribution of the masers we expect, according to Eq.~(\ref{velocity-law}), elliptical inner and outer boundaries for their locations in Fig.~\ref{fig:uher-absvexp}. Unlike in the corresponding diagram of RX~Boo (Paper\;I), there is no sharp inner boundary, but an outer boundary can be defined, by fitting a quarter of an ellipse to the six outer masers components marked by surrounding circles in Fig.~\ref{fig:uher-absvexp}, using the 'least-squares method'. From this fit we conclude that the outflow velocity in the envelope of U~Her at $\sim$24~AU from the star is about 10~\kms. 

\begin{figure}
\includegraphics[width=\columnwidth]
{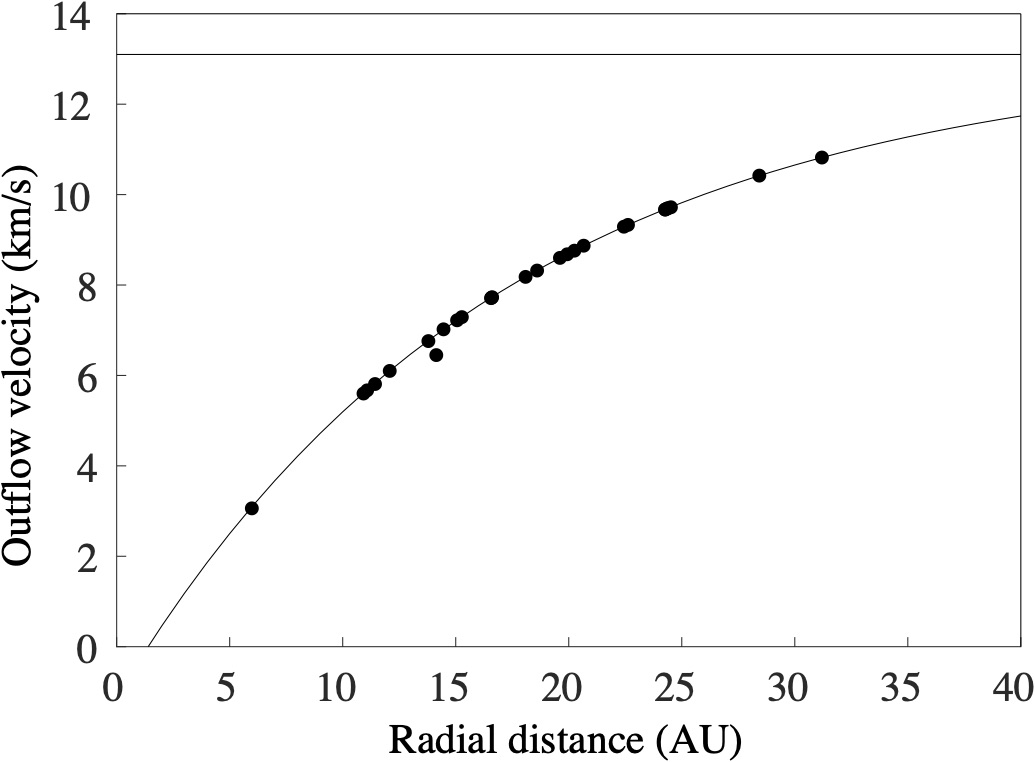}
\caption{The model outflow velocity for U Her as a function of the distance from the star. It approaches the final expansion velocity (horizontal straight line) asymptotically. The dots along the curve give the radial distances and outflow velocities of the 28 merged spatial components (see text).}
\label{fig:uher-expansion}
\end{figure}

\subsubsection{The 3-dimensional shell structure} \label{sec:3D-struct}
As in Paper I we assume that the outflow velocity law is exponential in nature and is leading asymptotically to the final expansion velocity, $V_{\rm exp}$ 
\begin{equation}
V_{\rm out}(r) = V_{\rm exp} \left\lbrace  1 - {\rm exp}\,\left[ -k (r - r_0)\right]
\right\rbrace 
\label{eq:exp_model}
\end{equation} 
where $k$ is a scaling factor and $r_0$ is a radial offset. 

Equation (\ref{eq:exp_model}) contains three constant parameters ($V_{\rm exp}$, $k$ and $r_0$) and in order to find plausible values for them we need to estimate the coordinate values for three points on the outflow law.  For the value of $V_{\rm exp}$ (the outflow velocity at $r = \infty$) we adopted $V_{\rm exp} = 13.1$ \kms\ (Table~\ref{centralcoords}). 
Based on the fit of the ellipse shown in Fig.~\ref{fig:uher-absvexp} to determine an outer boundary of the shell, we adopted an outflow velocity of 9.6~\kms\ at a distance of 23.9~AU from  the star. We also assumed that the outflow velocity at the photosphere $r_0=1.4$ AU is
$V_{\rm out}=0$ \kms.
 
Thus the value of the parameter $k$ can be determined from the condition that the stellar wind passes through the position \mbox{($r_1$, $V_1$) = (23.9, 9.6).}  Using \begin{equation}
k = \frac{{\rm ln}\,V_{\rm exp} - {\rm ln}\,(V_{\rm exp} - V_1)}{r_1 - r_0}
\label{eq:k_value}
\end{equation}
we obtain $k = 0.059\,{\rm AU}^{-1}$. 

For the case of non-tangential movements of maser components, the model allows one to calculate the observed line-of-sight velocities $V_{\rm los}$ and projected distances $r_{\rm p}$ 
from the equations
\begin{equation}
V_{\rm los} - V_* = V_{\rm out}(r)\,{\rm sin}\,\theta
\label{eq:v-los}
\end{equation}
and
\begin{equation}
r_{\rm p} = r\,{\rm cos}\,\theta
\label{eq:r-p}
\end{equation}
where $\theta$ is the aspect angle between the normal to the line of sight and the radius from the star to the maser. The aspect angle is $-90^\circ \le \theta < 0^\circ$ for $V_{\rm los} - V_{*} < 0$ \kms\  and viceversa. 

Combining Eqs. (\ref{eq:exp_model}), (\ref{eq:v-los}), and (\ref{eq:r-p}) gives a relation, which uniquely determines the distance $r$ of the maser cloud from the star for the time of the distance measurement, using the adopted outflow law and the observed quantities ($V_{\rm los} - V_{*}$) and $r_{\rm p}$
\begin{equation}
V_{\rm los} = V_{\rm exp} \left\lbrace  1 - {\rm exp}\,\left[ -k (r - r_0)\right]
\right\rbrace \cdot \sqrt{1 - (r_{\rm p}/r)^2} + V_*
    \label{eq:vlos-r-rp}
\end{equation}

\begin{figure}
\centering
\includegraphics[width=6.25cm]
{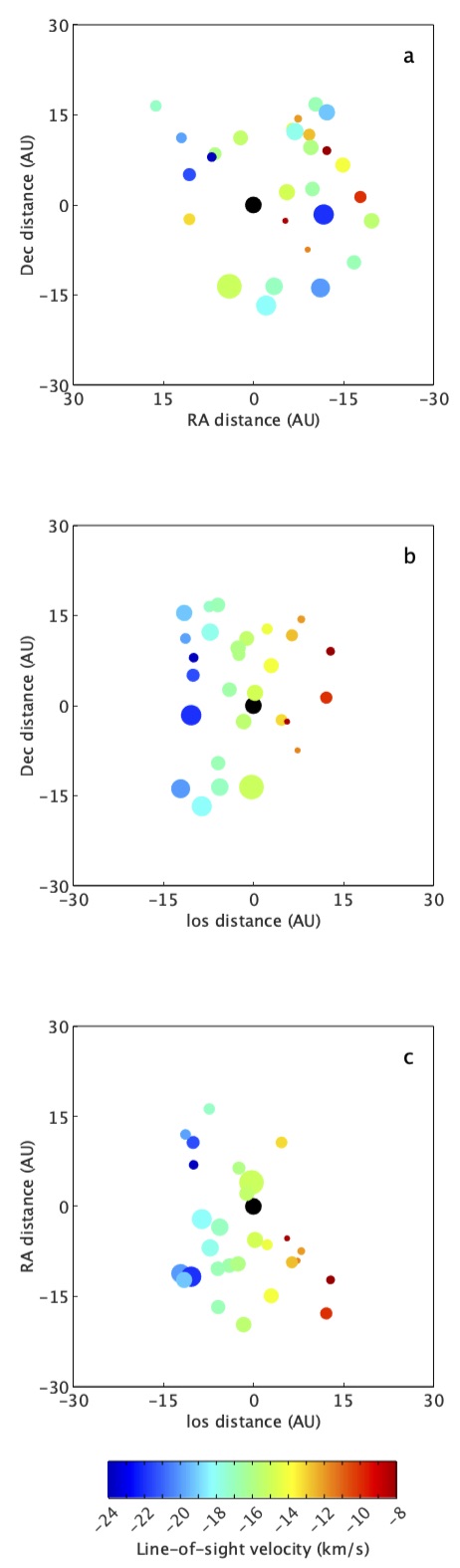}
\caption{The three-dimensional distribution of the U Her merged maser spatial components and their associated velocities according to the outflow velocity law given by Eq. \ref{eq:exp_model}. The distribution is seen from three cardinal directions: on the sky (a), ``from the side'' (b) and ``from above'' (c). In (b) and (c) the observer is to the left. The maser components are represented by filled circles with colours according to their line-of-sight velocities (see scale at bottom). Their diameters are proportional to the logarithm of their flux density.  The central star is symbolised by a black circle at (0,0).
}
\label{fig:uher-3Dvert}
\end{figure}

In Fig.~\ref{fig:uher-expansion}, the adopted outflow velocity law with the parameters $k = 0.059$~AU$^{-1}$, $r_0 = 1.4$ AU and $V_{\rm exp} = 13.1$ \kms\ is shown graphically, together with the radial distances $r$ of the maser spatial components calculated with Eq.\,(\ref{eq:vlos-r-rp}). The projected distances $r_p = \sqrt{{\rm Xoff}^2 + {\rm Yoff}^2}$ used for these calculations were averages over the epochs in which the components were detected (Table\,\ref{tab:components}). The uncertainties in positions and velocities lead to errors in the radial distances $r$ of a few AU, so that their locations on the velocity curve mainly delineate the typical distance range of the maser components. The main conclusion is that the maser shell is primarily located between $\sim$11 and $\sim$25 AU. The outflow velocity within these boundaries increases from 5.6 to 9.8 \kms. The maser components outside this shell ($>25$ AU) are the spatial components A1 and M1 which have line-of-sight velocities $V_{\rm los}$ at the extreme ends of the observed velocity range, and have the largest outflow velocities $V_{\rm out}$. The component inside this shell is the tangential component G2 with  $V_{\rm los}-V_* = -0.1$ \kms\ close to the line-of-sight toward the star itself. The two outliers at $r>25$ AU with flux densities $<1$\,Jy indicate that weak and short-lived maser activity can occur outside the shell delineated by the strong maser components. The same is true for the regions inside the inner boundary of the shell, in which the masers are usually suppressed because there the acceleration of the wind is relatively high. Due to projection effects the shell radius ($r=15$ AU), derived from the spatial distribution of the maser components on the sky (Fig. \ref{fig:uher-rainbow}), is $\sim$80\% of the mean radius of the shell ($r\sim18$ AU) given by the midpoint between inner and outer boundary of the 3D model. 

The 3D-distribution of the merged maser spatial components and their associated velocities is shown in Fig.~\ref{fig:uher-3Dvert} as seen from three different directions: on the sky (a), ‘from the side’ (b) and ‘from above’ (c). The dominance of tangential masers is seen clearly in all three diagrams. 
In panel (c) there is a scarcity of (red-shifted) components seen on the backside of the shell. This is also evident in the FVt-plot (Fig.\,\ref{fig:uher-fvt}, left), where there is less emission at red-shifted velocities, and in Fig. \ref{fig:uher-sratio-tjd} where the blue-shifted integrated emission prevailed between 1991 and 2007. However, one should keep in mind that an observer seeing U Her from directions deviating considerably from the geocentric one would see a different set of maser components, because a maser beams in a preferred direction, which is governed by the direction of greatest elongation of the maser cloud. 
Thus, Fig.~\ref{fig:uher-3Dvert} is merely showing the three-dimensional positions of the maser components that we see from Earth. 
Of course a similar situation occurs for our present point of observation: there may be maser spots that we do not see because their emission may happen to be beamed in the wrong direction from our vantage point.

Having adopted an outflow law it is of interest to investigate the associated time scale for the gas to travel through the maser shell. Thus, we have integrated the function $V_{\rm out}^{-1}$, where $V_{\rm out}$ is described by Eq.\,(\ref{eq:exp_model}), along the radius $r$ between the shell boundaries. We find that it takes $\sim$8.5 years for gas to travel through the shell (11 -- 25 AU), where most of the \water\ masers reside.

\subsection{Other \water\ maser shell observations of U~Her \label{shell-size}}
\subsubsection{Maser shell sizes}
Our shell model can be directly compared to the results of the contemporaneous 22 GHz VLA observations of \cite{colomer00} from June 1990 discussed in Sect. \ref{id_spatial_c}. Using a 3D fitting program they advocate a thick maser shell with an inner radius of 45 mas ($\sim$12 AU) and an outer radius of 70 mas ($\sim$19 AU), in which the molecules flow outwards with a velocity of $\sim$6 \kms. A comparison with our model expansion velocity law (Fig. \ref{fig:uher-expansion}) shows that their shell boundaries delineate a narrow 7~AU-wide shell in the central part of our shell model covering the strongest spatial maser components. The average radius of this shell is $\sim$15.5 AU compared to $\sim$18 AU, the average radius of our shell model. The two independent analyses highlight the uncertainties in  the determination of the boundaries, with the shell boundaries for 1990 of Colomer et al. being a conservative estimate.

After our VLA observations 1990 -- 1992, the \water\ maser emission of U~Her was mapped several times and these observations can be used to search for variations in the spatial distribution of the masers over a longer time range. Interferometric observations with better spatial resolution than achievable with the VLA were made with MERLIN 1994, 2000 and 2001 \citep{bains03, richards12}, and with the VLBA in 1995 \citep{marvel96}. They confirm the overall ring-like geometry of the \water\ maser shell. However, the boundaries of this shell determined from the different observations are not well defined. The most compelling determination is by \cite{richards12} who, for the epochs 2000 and 2001, found  projected inner and outer radii of 10 and 40 AU, respectively. While the inner radius is compatible with the inner boundary determined here and by \cite{colomer00}, the outer radius is significantly larger. 
However, the outer shell at radii $>30$ AU was only sparsely populated by maser components in 2000/2001 \citep{richards12}, and these were not detected  1990 -- 1992.
The decrease in brightness of the maser emission with radial distance makes the determination of the outer shell boundary much more dependent on instrumental sensitivity compared to the inner boundary.
The conclusion from all \water\ maser shell observations available is that the \water\ masers are located in a shell within the expanding spherical wind of U~Her with most of the sites with stronger emission located at a radial distance of about 15--20~AU. Nevertheless, VLBA observations by \cite{vlemmings02b,vlemmings05} challenged the characterization of U~Her's \water\ maser shell as a persistent ring-like distribution of distinct emission regions.  

\subsubsection{Constraints on shell geometry due to spatial resolution effects}
To study magnetic field strengths in the CSE of U~Her, 22~GHz observations with the VLBA were made on 13 December 1998 by \cite{vlemmings02b} and  on 20 April 2003 by \cite{vlemmings05} with an average beam width of 0.5 mas. They noted a significant change in spectrum and spatial distribution between the two observations. The spatial features detected in 1998 (optical phase $\phi_{\rm s} = 0.44$) were at velocities $-19.3$ to $-17.6$ (likely part of our spectral components \Done\ and \Dtwo), while in 2003 ($\phi_{\rm s} = 0.97$) they detected features at  $-15.9$ to $-14.5$ (\Gone\ and \Gtwo). Our single-dish spectra taken close to their observations (12 December  1998 and 2 April 2003, see Fig. \ref{fig:uher_all}) show a stronger profile change only in the red wing ($>-14$ \kms), but not at the velocities with VLBA detections by Vlemmings et al.. During both epochs spectral components \Gone\ and \Gtwo\ were strongest, while \Done+\Dtwo\ was 2--3 times weaker. It is therefore conceivable that the spatial distribution as such did not change significantly between the two epochs, but the sizes of the spatial components did, leading to different components  being resolved out by the VLBA in the two epochs. This explanation is corroborated by the failure of \cite{imai97b} to detect U~Her's \water\ maser emission in 1994/1995 during a VLBI experiment (resolution 2.1 mas), indicating that the dominating spectral components \Gone\ and \Gtwo\ had significantly larger sizes. 
\cite{richards11} give typical sizes 2--5 AU (8--19 mas) for \water\ maser clouds of U~Her, so that losses of spatial features due to the largest recoverable scale for imaging with the VLBA are plausible. 

\subsubsection{Maser amplification in 2001 by stellar emission?}
\cite{vlemmings02a} mapped the maser on 20 May 2001 with MERLIN (beam size $\le30$ mas), when the strongest spectral maser feature was present at $-15.6$ \kms\ (spectral component \Gone, see Table \ref{tab:compUHerG-M} and Fig. \ref{fig:uher_all} in the Appendix). They found the position of the feature to match with the location of the star at this epoch, determined using own and HIPPARCOS proper motion measurements of U\,Her, and argued that the maser feature is amplified by the stellar radiation in the background. They note that this result would place the star not in the center but on the ring-like distribution of the maser spots.

In the monitoring data we find no evidence for extraordinary amplification of the \Gone\ component. If the \Gone\ component would have moved as part of the stellar wind in radial direction within the line-of-sight to the star, a systematic blue-shift of its velocity $V_{\rm los}$ over time is expected, which is not observed. Alternatively, if \Gone\ would have moved in tangential direction the stellar amplification would be only a temporary effect during the time of eclipse. Adopting a U~Her maser cloud size of 2--5 AU \citep{richards11}, a stellar diameter of 2.8 AU \citep{vanbelle96, ragland06} and a velocity of $\sim$7 \kms\ perpendicular to the line-of-sight (see Fig. \ref{fig:uher-expansion})  leads to a duration of the eclipse of several years. Therefore, a monitoring program of such an eclipsing event is expected to observe a temporary year-long flare of the emission on top on the regular maser variations. The occurrence of such a flare of the \Gone\ emission can be  definitely ruled out in the years around 2001. 

Using the stellar position and proper motion given by GAIA DR3 we recalculated the position of the star for the MERLIN observation of \cite{vlemmings02a} and found an offset of the position of the maser spot from the star of $\Delta \alpha = -37.4$ mas and $\Delta \delta = 0.3$ mas, which places the maser spot $\sim$10 AU west from the star and at the inner boundary of the \water\ maser shell as determined earlier.

Based on the lack of brightness and velocity variations of spectral component \Gone, which could be related to an eclipse event, and based on the stellar position in 2001 according to recent GAIA astrometry, we conclude that the assumption of the stellar position close to the center of the ring-like distribution of spatial components, as made in Sect. \ref{map-alignment}, is still a valid approach. The location of the \water\ masers in a shell within the spherically expanding stellar wind, remains therefore the most plausible geometric configuration.

\begin{table*}
\begin{center}
\caption[]{\label{tab:emission-regions} Regions in U\,Her's \water\ maser shell contributing emission to the prominent spectral components G (composed of \Gone\ and \Gtwo\ at $V_{\rm los}\sim-15\pm0.5$ \kms) and \Dtwo\ ($V_{\rm los}\sim-18$ \kms).}
\begin{tabular}{lrrrrrrl}
\hline\noalign{\smallskip}
Map Date &  Xoff & Yoff & $V_{\rm los}$  & $S_\nu$ & Region & Instr. & Reference \\
         & [mas] & [mas]&[\kms]& [Jy]  &        &           \\
\noalign{\smallskip} \hline\noalign{\smallskip}
1988 Dec. &  5   & -46  & -14.7 & $\sim$210 & SE & VLA  & \cite{bowers94}\\
1990 Feb. & 15   & -54  & -14.9 & 230.0   & SE & VLA  & this paper, Table \ref{tab:components} \\
1990 June & 15   & -52  & -15.0 & 145.0   & SE & VLA  & this paper, Table \ref{tab:components} \\
1990 June & 22   & -42  & -14.6 & 117.2   & SE & VLA  & \cite{colomer00} \\
1992 Dec. & 14   & -46  & -15.3 &  14.0   & SE & VLA & this paper, Table \ref{tab:components} \\
1994 Apr. & 21   & -63  & -14.8 &   8.4   & SE & MERLIN  & \cite{richards12} \\
1994 Apr. & 21   & -46  & -14.8 &   2.2   & SE & MERLIN  & \cite{richards12} \\
1995 June & 23   & -55  & -14.8 &   4.1   & SE & VLBA  & \cite{marvel96} \\
2000 May  & -    & -    &  -    &  n.d.   & SE & MERLIN  & \cite{richards12}$^\dagger$\\
2001 Apr. & -    & -    &  -    &  n.d.   & SE & MERLIN  & \cite{richards12}$^\dagger$ \\
\noalign{\smallskip} \hline\noalign{\smallskip}
1994 Apr. & -65   & -8  & -15.2 &  10.7   & W & MERLIN  & \cite{richards12} \\
1995 June & -36   & -21 & -15.5 &  10.6   & W & VLBA  & \cite{marvel96} \\
2000 May  & -50   & -2  & -15.0 & 141.0   & W & MERLIN  & \cite{richards12} \\
2001 Apr. & -59   & -8  & -15.9 &  38.0   & W & MERLIN  & \cite{richards12} \\
\noalign{\smallskip} \hline\noalign{\smallskip}
1988 Dec. & -8   & -51  & -17.6 & $\sim$110 & SW & VLA  & \cite{bowers94}\\
1990 Feb. & -9   & -60  & -18.2 & 52.0    & SW & VLA  & this paper, Table \ref{tab:components} \\
1990 June & -9   & -60  & -18.1 &  7.3    & SW & VLA  & this paper, Table \ref{tab:components} \\
1990 June & -2   & -52  & -18.3 &  11.0   & SW & VLA  & \cite{colomer00} \\
1992 Dec. & -8   & -68  & -18.2 &   3.0 & SW & VLA & this paper, Table \ref{tab:components} \\
1994 Apr. & -30  & -62  & -17.9 &   2.4 & SW & MERLIN  & \cite{richards12} $^{\dagger\dagger}$\\
1995 June & -26  & -44  & -17.9 &   2.0 & SW & VLBA  & \cite{marvel96}$^{\dagger\dagger}$ \\
2000 May  & -69  & -13  & -17.8 &  12.3 & SW & MERLIN  & \cite{richards12}$^{\dagger\dagger}$ \\
2001 Apr. & -17  & -15  & -17.9 &   4.7 & SW & MERLIN  & \cite{richards12}$^{\dagger\dagger}$ \\
\noalign{\smallskip} \hline \noalign{\smallskip}
\end{tabular}\\
\end{center}
{\it Note:} The position of the star in the maps presented by \cite{colomer00} and \cite{marvel96} was set by us to ($-$20,+43) and (+35,+20) mas, respectively,   relative to their map origin.\\ $^\dagger$ n.d. = not detected. The non-detection of component G in 2000/2001 means that it is not seen in the south-east quadrant anymore.\\ $^{\dagger\dagger}$ The spatial components in the south-west quadrant might have originated in different regions before and after 1993.
\end{table*}

\subsection{Lifetime of emission regions \label{region-lifetime}}
The analysis of the line profile variations in Sect. \ref{sdd_LineProfAna} found the velocities of the spectral components remarkably constant over the monitoring period with only small shifts back and forth on timescales of months. The small shifts were assumed to be caused by blending caused by different maser clouds with similar velocities and unrelated brightness fluctuations. They may reflect the asynchronous formation and dissolution of contributing maser clouds or a variation of excitation conditions on these timescales.

We also found that some less blended spectral components could be followed without significant variations in velocity over longer time ranges. An example is component K, which could be detected continuously for 5 years (March 1995 -- April 2000; TJD = 9790 -- 11640; (see Fig. \ref{fig:uher-fvt}, and Table \ref{tab:compUHerG-M} in the Appendix) at $-10.9\pm0.2$ \kms, i.e. with a rather small velocity dispersion. Later it reappeared frequently close to the maxima of the stellar variability cycle. If this emission component comes from an individual maser cloud, one has to conclude  that U\,Her's \water\ maser shell can host individual clouds with a range of lifetimes of 0.5 to many years. Alternatively, short-living clouds would have to appear regularly with always similar line-of-sight velocity $V_{\rm los}$. 

In the first case of a long-living maser cloud, which moves within the expanding CSE, systematic shifts of the projected velocity $V_{\rm los}$ are expected, while the cloud is accelerated outward. This is not seen in any of the spectral components, and so the second case must apply and their emission must come from different short-living spatial components in the course of the monitoring period.

\subsubsection{Maser cloud lifetime constraints \label{region-lifetime-1}}
We will now use the absence of systematic velocity shifts to derive upper limits for the lifetime of individual maser clouds. 

For the outflow velocity curve shown in Fig.~\ref{fig:uher-expansion}, we found in Sect. \ref{sec:3D-struct} a crossing time of $\sim${8.5} years for a mass element moving radially through the \water\ maser shell with inner and outer boundaries of 11 and 25 AU, and with an increase of its outflow velocity from 5.6 to 9.8~\kms. Moving in the line-of-sight, such a hypothetical mass element showing maser emission all the time would show an acceleration of $a = 0.5$~\kms yr$^{-1}$.  

A maser cloud observed in two epochs with a time difference $\Delta t$ and moving along a straight trajectory in the shell with an angle $\theta$ with respect to the plane of the sky will experience a shift in line-of-sight velocity $\delta V_{\rm los} = \delta V_{\rm out}(r) \cdot \sin{\theta}$ (cf. Eq. \ref{eq:v-los}), with $\delta V_{\rm out}(r)$ the increase of the outflow velocity. For simplicity, we approximate the acceleration within the \water\ maser shell with the average velocity increase $a = 0.5$~\kms yr$^{-1}$, i.e.  $\delta V_{\rm out}(r) = a \cdot \Delta t$ leading to the relation
\begin{equation}
    \delta V_{\rm los} =  a \cdot \Delta t \cdot \sin{\theta}.
    \label{eq:vlos-shift}
\end{equation}     
In the following we adopt a conservative value $\mid$$\delta V_{\rm los}$$\mid\ < 0.5$ \kms\ for recognizable velocity shifts due to participation in the stellar outflow of any of U\,Her's spectral components. For a given time interval $\Delta t$, only maser components with an aspect angle obeying 
\begin{equation}
    \mid\sin{\theta}\mid = \delta V_{\rm los} / a / \Delta t < 1.0 / \Delta t
    \label {aspect-angle-limit}
\end{equation}
would not have been detected as having a drifting line-of-sight-velocity. 
A special case of Eq.\,(\ref{aspect-angle-limit}) is a tangential movement of the maser clouds ($\theta \approx 0$), where a constant velocity $V_{\rm los}$ (i.e. $\delta V_{\rm los} \approx 0$) can be expected for maser clouds traceable over long time intervals. Equation (\ref{eq:v-los}) demands that in this case $V_{\rm los} \approx V_*$. Allowing an uncertainty of 0.5 \kms\ for the stellar radial velocity $V_* = -15.0$ \kms, such a tangential movement would be able to explain the absence of velocity shifts for spectral components  \Gone\ ($\sim -15.5$ \kms) and \Gtwo\ ($\sim -14.5$ \kms). However, as our monitoring period is longer than the crossing time in U~Her's \water\ maser shell, more than one cloud must have contributed even for its \Gone+\Gtwo\ spectral components.

For the case of non-tangential movements of clouds and $\mid V_{\rm los} - V_*\mid\ > 0.5$ \kms, the line-of-sight velocity $V_{\rm los}$ is determined by  $(V_{\rm los} - V_*) = V_{\rm out}(r) \cdot \sin{\theta}$ (Eq. \ref{eq:v-los}). A measurement of the projected distance $r_p$ of the cloud from the star and use of Eq.\,(\ref{eq:vlos-r-rp})
uniquely determines the distance $r$ of the maser cloud from the star for the time of the projected distance measurement 
and fixes the aspect angle $\theta$ according to Eq. (\ref{eq:r-p}). After a time interval $\Delta t$ a new measurement of $V_{\rm los}$ should show a shift $\delta V_{\rm los} = a \cdot \Delta t \cdot \sin{\theta}$ (Eq. \ref{eq:vlos-shift}). A limit $\delta V_{\rm los}$(max) on $\delta V_{\rm los}$ constrains the time difference $\Delta t$, for  which Eq. (\ref{eq:vlos-shift}) would not be violated. The maser emission seen in observations separated by more than the corresponding $\Delta t$(max) and having $\delta V_{\rm los} \le \delta V_{\rm los}$(max) can therefore not originate from the same cloud. We interpret therefore $\Delta t(max)$ as the lifetime of the cloud.

As an example, we will discuss the lifetime of spectral component E of U~Her seen almost permanently during the monitoring period. We detected the component in 1990 at $V_{\rm los} = -16.9$~\kms\ (Table \ref{tab:compUHerB-E} in the Appendix) and associated to it the spatial component E3 located $\approx$60 mas ($r_{\rm p} \approx 16$ AU) south from the star, as judged from the February 1990 and December 1992 positions (Table \ref{tab:components}). The projected outflow velocity is $V_{\rm los} - V_* = -1.9$ \kms , 
and using Eq. (\ref{eq:vlos-r-rp}) we can derive a radial distance $r\sim17$~AU, and from this
$V_{\rm out}(r) = 7.9$~\kms\ and $\theta = -14^\circ$, using Eqs. (\ref{eq:exp_model}) and (\ref{eq:v-los}). Adopting $\delta V_{\rm los} < 0.5$ \kms\ and $a = 0.5$~\kms yr$^{-1}$, we find $\Delta t(max) =  \delta V_{\rm los} / a /  \sin{\theta} \approx 4$ yr.

\begin{figure*}
\resizebox{18cm}{!}{
\includegraphics{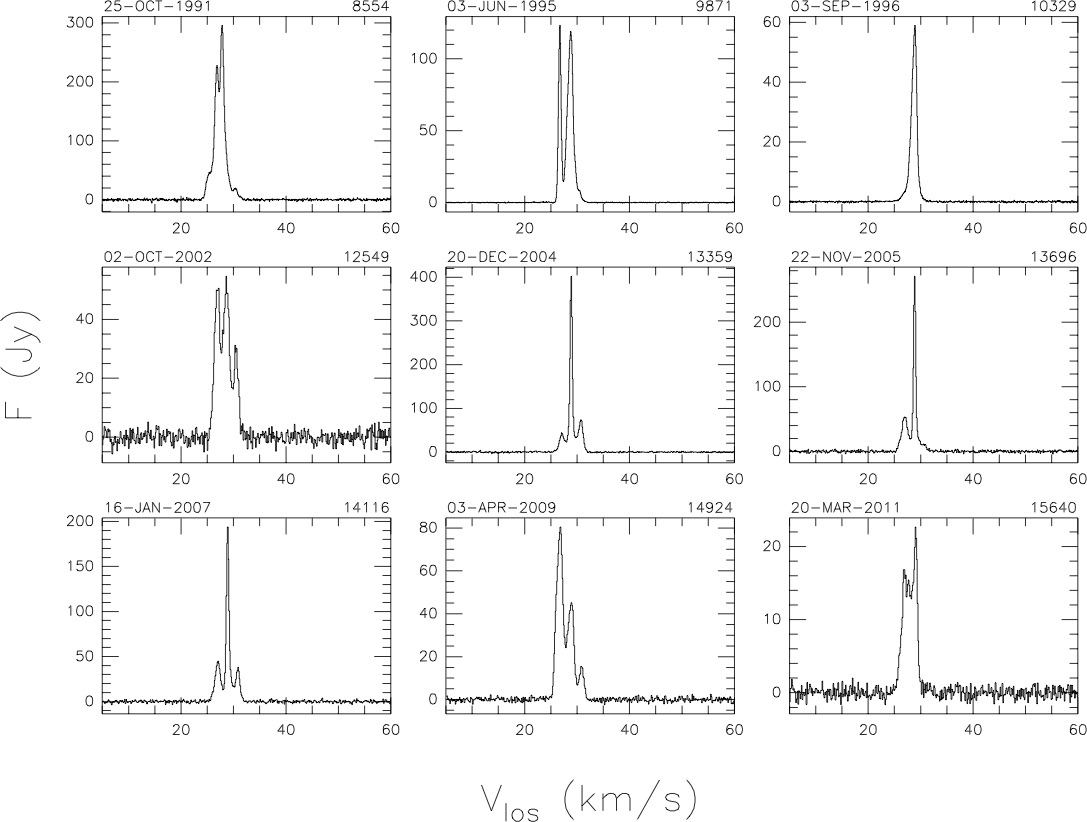}}
\caption{Selected H$_2$O maser spectra of RR~Aql. The calendar date of the observation is indicated on the top left above each panel, the TJD (JD-2440000.5), on the top right.}
\label{fig:rraql_sel}
\end{figure*}

Lifetimes of at most several years are also found for other spectral components and varying the outflow velocity model derived in Sect. \ref{sec:shell-model}. They are compatible with our series of VLA observations 1990 -- 1992, where the identification of common spatial maser components proved to be difficult over even shorter time ranges (see Sect. \ref{map-alignment}). They corroborate the findings of \cite{bains03}, who give lifetimes $\la2$ years for maser clouds of U~Her with typical sizes 2--4 AU. 

\subsubsection{Long-living regions favourable to \water\ maser emission \label{region-lifetime-2}}
The limited lifetime of the individual maser clouds are in apparent contrast to the persistent presence of several spectral components over the full monitoring period (Fig. \ref{fig:uher-fvt}).  The most notable components are \Gone\ and \Gtwo\ in the $-16$ to $-14$ \kms\ velocity region, which maintained their dominant role in the spectral profile after 1991 over more than 20 years. As we have argued in the previous Section, this is due to many consecutive components superposed over time at the same velocity. This suggests that the dominant emission region in U\,Her's \water\ maser shell may not have moved away from its 1990-1992 location.   

In the standard model having a continuous homogeneous outflowing stellar wind there is no reason for the existence of preferred locations in the \water\ maser shell for particular emission sites. If the \Gone\ and \Gtwo\ components would come from short-lived single maser clouds, one would expect that their emission comes from random positions within the shell having the right projected line-of-sight velocity. In Table \ref{tab:emission-regions} we compiled the strongest spatial maser components with velocities compatible with \Gone\ and \Gtwo\ identified in a number of interferometric maps covering the years 1988 - 2002. The table gives the offsets Xoff, Yoff of the spatial components from the position of the star, their velocity $V_{\rm los}$, flux density $S_\nu$, and a label for the region (see below). Furthermore, the date of the map, the interferometer used and the reference are given. The position of the star is either taken as given in the references or determined (see Note to Table \ref{tab:emission-regions}) assuming a ring-like distribution of the maser emission as described in Sect. \ref{map-alignment}.

We found that the strong ($S_\nu > 200$ Jy in 1988--1990) spectral component has spatial counterparts between 1988 and 1995 in the south-east (SE in Tab. \ref{tab:emission-regions}) quadrant of the maps. The flux density decreased to $<10$ Jy levels in 1994/1995 and the component was not detected afterwards. In 2000/2001 emission at \Gone + \Gtwo\ velocities is seen in the west (W) of the MERLIN maps. Spectral component \Dtwo, which decreased in brightness after 1988, can be identified at least until end of 1992 with spatial components found in the south-west (SW) quadrant, while afterwards the relation with spatial components in the same quadrant having velocities compatible with \Dtwo\ is inconclusive due to the large deviations of the offsets from those in the 1988--1992 period.

The frequent interferometic observations of U~Her's \water\ maser emission show that the regions in the CSE with conditions favourable to water maser excitation can survive longer than a few years, in contrast to what was found for single maser clouds. Spectral components with only small variations in velocity are therefore originating from multiple maser clouds, which are preferentially formed in particular regions of the shell, which for at least 6.5 years (in the case of  spectral component G) can maintain their favourable \water\ maser excitation condition. The origin of such regions will be discussed in Sect. \ref{sdd-cse-asym}.

\begin{figure*}
\begin{minipage}[t]{17cm}
 \begin{minipage}[t]{8.5cm}
  \begin{flushleft}
   {\includegraphics*[width=8cm,angle=0]{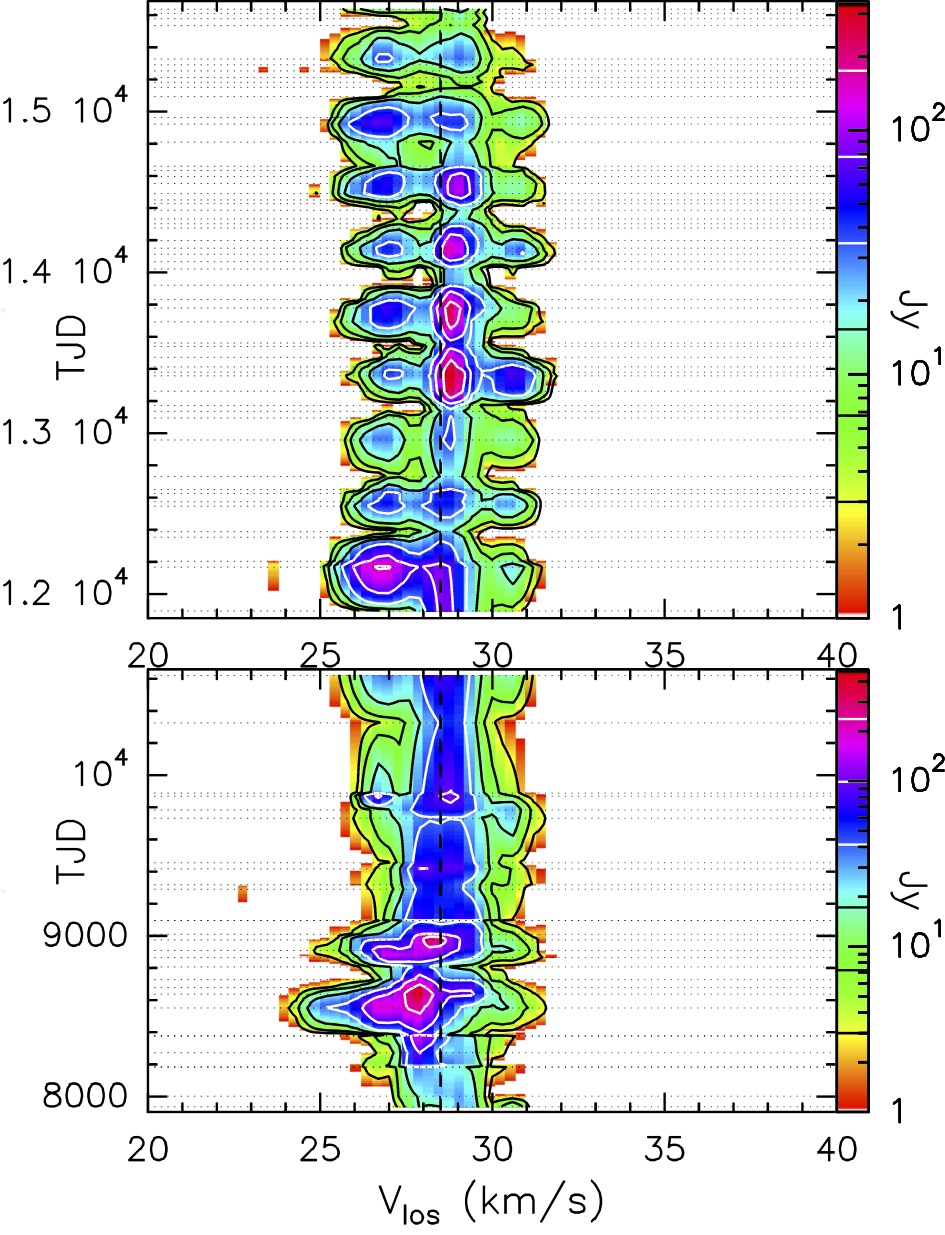}}
     \end{flushleft}
 \end{minipage}
 \begin{minipage}[t]{8.5cm}
  \begin{flushright}
      {\includegraphics*[width=8cm,angle=0,scale=0.9]{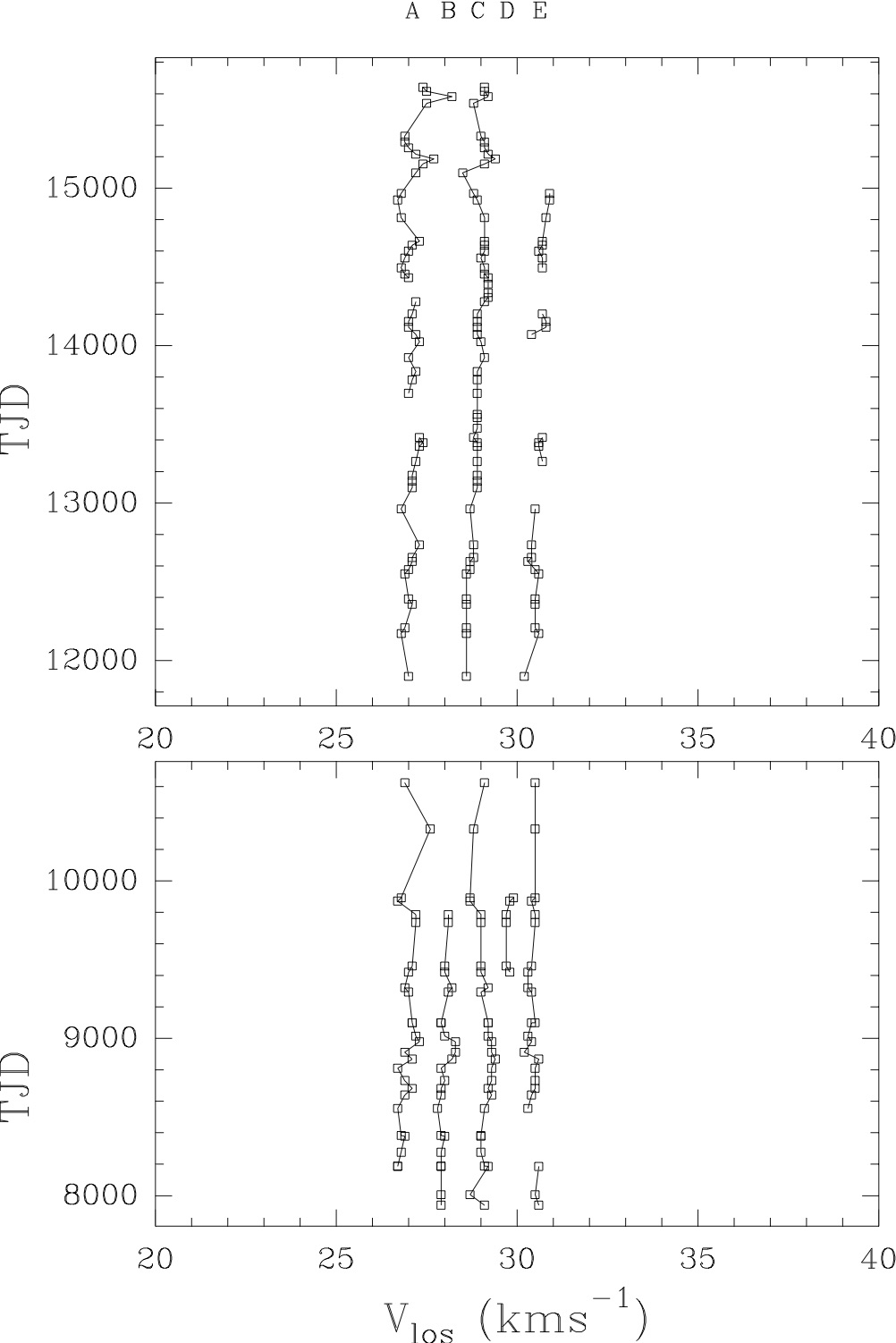}}      
  \end{flushright}
 \end{minipage}
\end{minipage}
\caption{As Fig.~\ref{fig:uher-fvt}, but for RR~Aql. {\it Left:} First spectrum: 17 February 1990; JD = 2447939.5, TJD = 7939. and last spectrum 25 June 1997 (lower panel). First spectrum 20 December 2000 and last spectrum: 20 March 2011 (upper panel). {\it Right:} Spectral components identified by the component fit of the single-dish spectra as listed Table \ref{tab:compRRAql}  in the Appendix.}
\label{fig:rraql-fvt}
\end{figure*}

\section{RR~Aql \label{sec:rraql}}
RR\,Aql is a long-period variable AGB star having a distance of $410^{+12}_{-11}$ pc (Table \ref{centralcoords}), based on the \water\ and SiO maser parallaxes measured by \cite{sun22}. We adopt as radial velocity of the star $V_{\ast} = 28.5\pm0.5$ \kms\ and as final expansion velocity $V_{\rm exp} = 9$ \kms, as determined from circumstellar CO and SiO thermal emission (Table \ref{centralcoords}).

The \water\ maser of RR~Aql was first detected in 1971 by \cite{dickinson73} as a single strongly variable feature at $\sim$27 \kms. A VLBI observation in 1976 by \cite{spencer79} showing a maser feature at 25.9 \kms\ with a flux density 353\,Jy indicated that the maser could reach levels of several hundred Jy. 
Regular observations of the maser were first made within the Pushchino monitoring program starting in 1980 \citep[hereafter B98]{berulis98}. The characteristics of the maser variations until 1997 are discussed in B98, who included also the maser observations published in the literature during this time interval. The brightest \water\ maser features in RR\,Aql usually peaked at velocities $26-29$ \kms\ close to the stellar velocity. The brightness variations were indeed strong with phases where peak flux densities of several hundred Jy were reached, and occasions in the 1970s where the maser was not detected (i.e. $F_\nu \la 10$ Jy). 

The \water\ maser region of RR~Aql was mapped on three occasions between 1981 and 1988 with the VLA.  The diameter of the maser region was confined to about 100 mas  (radius $\sim$21 AU), 
with clustering of the maser components mostly in two locations with a north-south orientation \citep{johnston85, lane87, bowers94}. A VLBI observation of \cite{imai97a} in 1995 was probably resolving out the emission, although the maser seemed to have been in a bright phase. Based on a total-power spectrum, they reported for 17-20 January 1995 a flux density of 296 Jy of the \water\ maser feature at 28.52 \kms, which is a factor of $\sim$4 higher than the brightness level observed before and after this date by B98, who concluded that the VLBI observation must have taken place during a transient flare of the maser. However, also our spectrum from 18 January 1995 showed a peak flux density of only $\sim$40\,Jy at 29.0 \kms, which leaves the high flux densities during the VLBI observations unexplained. VLBA observations in 2017/2018 \citep{sun22}, after the end of the monitoring program reported here, detected distinct maser emission regions with velocities in the range  $26-29$ \kms, which had dominated the maser profile since the discovery of the maser. The latter regions were found at a distance of $\sim$28 AU from the star, which is compatible with the shell dimensions inferred from the 1980s VLA observations.

\subsection{Variations of the \water\ maser profile}
\begin{figure}
\resizebox{9cm}{!}{\rotatebox{270}{
\includegraphics
{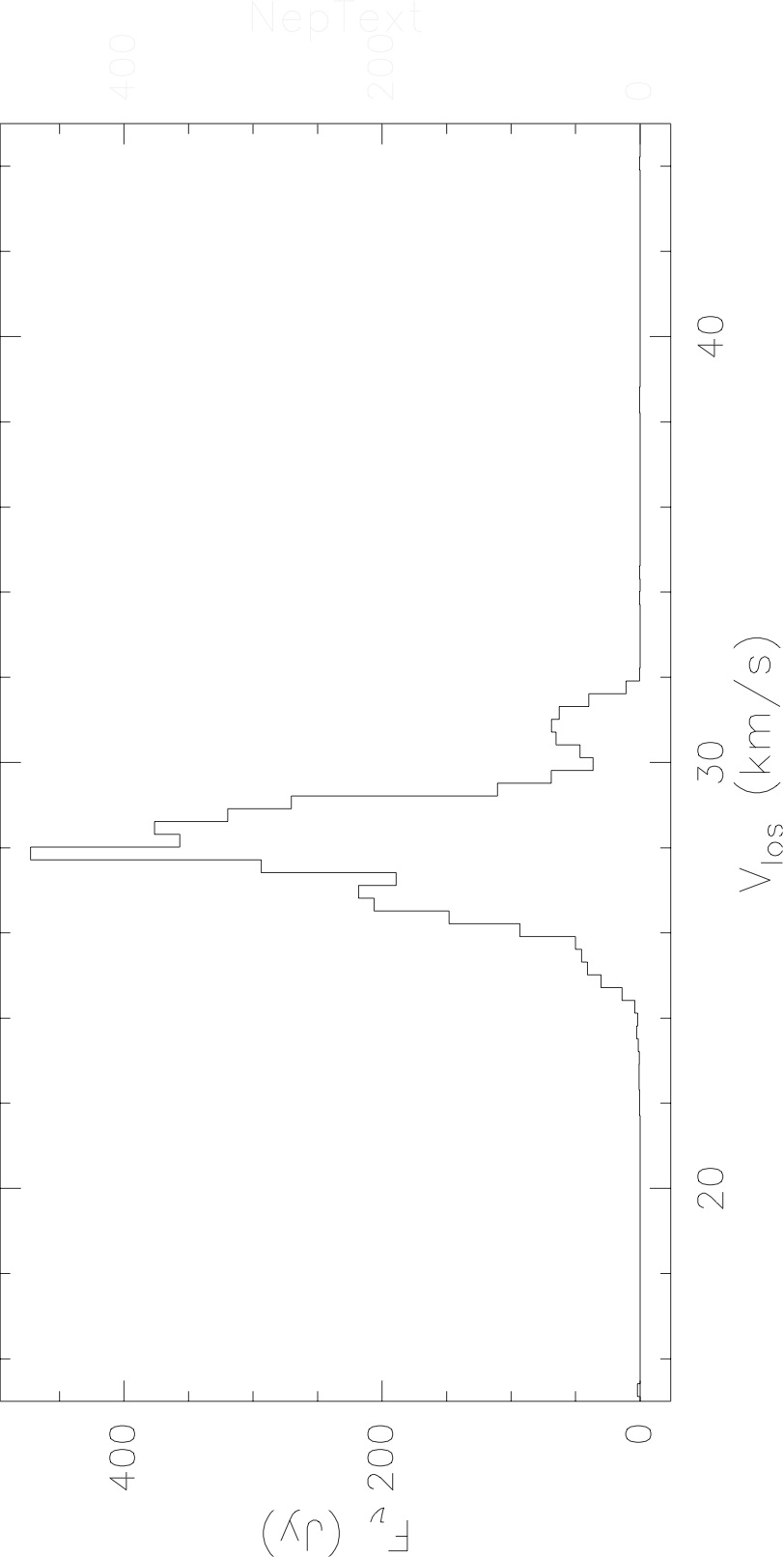}}}
\caption{Upper envelope spectrum for RR~Aql; 1987-2015.}
\label{fig:rraql-upenv}
\end{figure}

\begin{figure}
\resizebox{9cm}{!}{\rotatebox{270}{
\includegraphics
{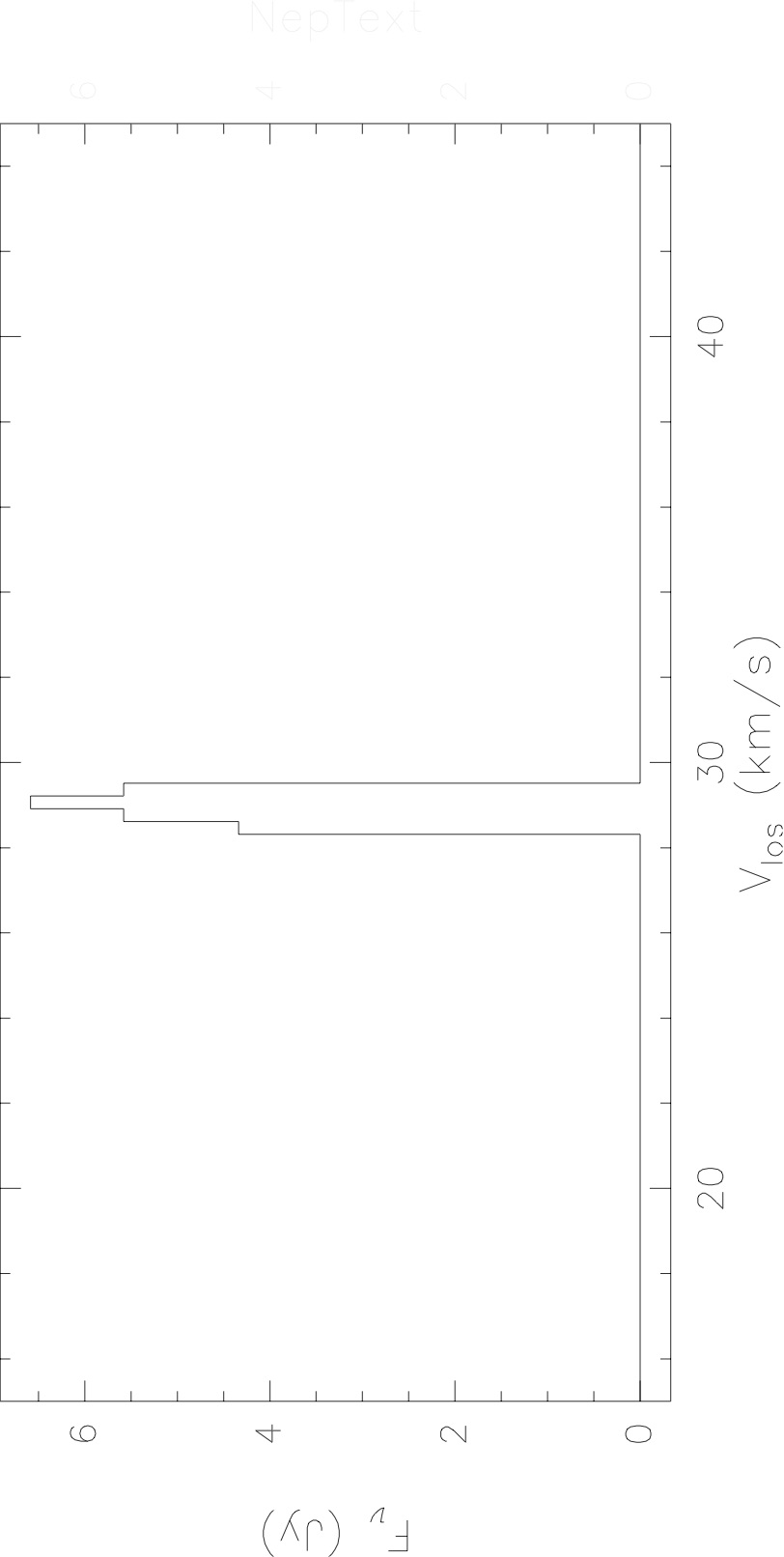}}}
\caption{Lower envelope spectrum for RR~Aql; 1987-2015.}
\label{fig:rraql-loenv}
\end{figure}

RR\,Aql was first observed at Medicina in 1987 \citep{comoretto90}, and we monitored the star regularly between 1990 and 2011. Additional observations were made in 2015. As before 1987, the strongest maser features were always detected close to the stellar velocity ($\sim 26-31$ \kms). In Fig.~\ref{fig:rraql_sel} representative maser spectra from our observations taken between 1990 and 2011 are shown, while the complete set of 89 spectra is displayed in Fig.~\ref{fig:rraql_all} (Appendix). An overview on the general behaviour of the maser variations is shown in the FVt plot (Fig. \ref{fig:rraql-fvt}, left panel). Because of only occasional observations between June 1995 and December 2001, the plot is split into two panels omitting 3.5 years between July 1997 and November 2000. As in U~Her, the profile in general is varying with stellar pulsation in brightness and velocity range  and is dominated by three features (cf. upper envelope spectrum Fig. \ref{fig:rraql-upenv}), which change their relative strengths over time. The central emission at $\sim$29 \kms\ (spectral component C hereafter) is always present (cf. lower envelope spectrum Fig. \ref{fig:rraql-loenv}) and is the dominating emission for most of the time. A second feature at $<28$ \kms\ (spectral component A; Fig. \ref{fig:rraql-fvt}, right panel) is blue-shifted and is the strongest only occasionally: In 1991--1992 (TJD $\sim 8200-8900$) and  September/October 2001 (TJD $\sim 12200$). The third feature at $>$30 \kms\ (spectral component E) is red-shifted and was always weaker than the other two. This feature reached a $\sim$50 Jy level between September 2004 and February 2005 (TJD $\sim 13200-13500$), while it rarely surpassed 10 Jy for the rest of the time. It was often not detected during the faint phase of the optical brightness variations. 

\begin{figure}
\resizebox{9cm}{!}{\rotatebox{270}{
\includegraphics
{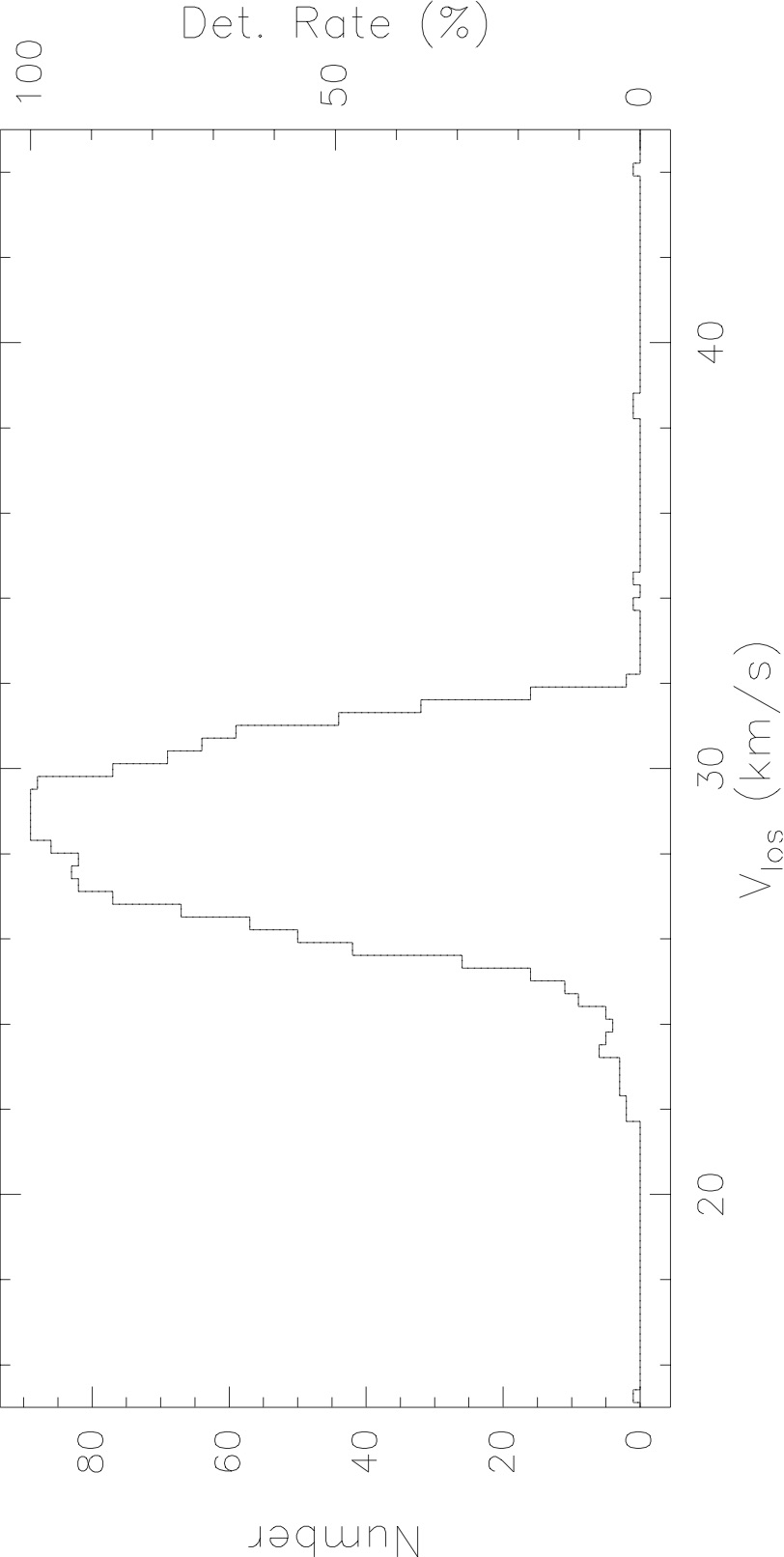}}}
\caption{Detection rate histogram for RR~Aql; 1987-2015.}
\label{fig:rraql-histo}
\end{figure}

\subsection{\water\ maser velocity range  \label{rraql-vel-range}}
As velocity boundaries of the maximum \water\ maser velocity range $\Delta V_{\rm los}$ of RR~Aql we determined $V_{\rm b} = 23.2$ and $V_{\rm r} = 31.9$ \kms\ (Table \ref{centralcoords}) 
from the detection rate histogram (Fig.~\ref{fig:rraql-histo}). The blue boundary is not well defined as the emission between $V_{\rm los} =$ 21.5 and 24 \kms\ is generally weak (seen only marginally in  Fig.~\ref{fig:rraql-upenv}) 
and surpasses flux densities $>$1\,Jy usually only at velocities $V_{\rm los} > 25$ \kms\ as evident from the FVt-plot (Fig. \ref{fig:rraql-fvt}). 

As in U~Her, the FVt-plot shows that the width of the observed velocity range is varying. The width is smaller during the faint part of the stellar variability cycle, when the weaker outer parts of the maser profile fall below the threshold of the FVt-plot ($\sim1$ Jy). Close to the maxima of the stellar light curve, maser emission is seen over the maximum velocity range, except for the emission at $<$25 \kms, which was detectable only in 1991 (TJD$\sim$8600). In 2009, toward the end of the monitoring program, spectral component E at 30.5 \kms\ was becoming weaker and indeed was then only marginally detected in some of our 2010 and 2015 observations close to the optical maxima (cf. Appendix \ref{sdd_LineFitRes_Appendix_RRAql}). 

The maximum \water\ maser velocity range $\Delta V_{\rm los}$ 
between 23.2 and 31.9 \kms\ is asymmetric with respect to $V_{\ast} = 28.5$ with emission from the front side of the shell reaching higher outflow velocities ($V_{\rm out} \sim 4.5-6.5$ \kms) compared to the back side ($V_{\rm out} \sim 3$ \kms). It is quite likely that this velocity range is only representative for the brightest maser features. \cite{johnston85} detected in March 1982 with their sensitive VLA observations (rms $\sim$ 0.35 Jy) emission between 19.7 and 36.9 \kms. Such a large \water\ maser emission range was never confirmed later on, but it may indicate that in RR\,Aql \water\ maser emission can be present over almost the full velocity range $V_{*} -V_{\rm exp} < V_{\rm los} < V_{*} + V_{\rm exp}$ ($19.5-37.5$ \kms; Table \ref{centralcoords}) determined by the stellar wind, but is detectable only under exceptional circumstances.

\begin{figure}
\includegraphics[angle=-90,width=\columnwidth]
{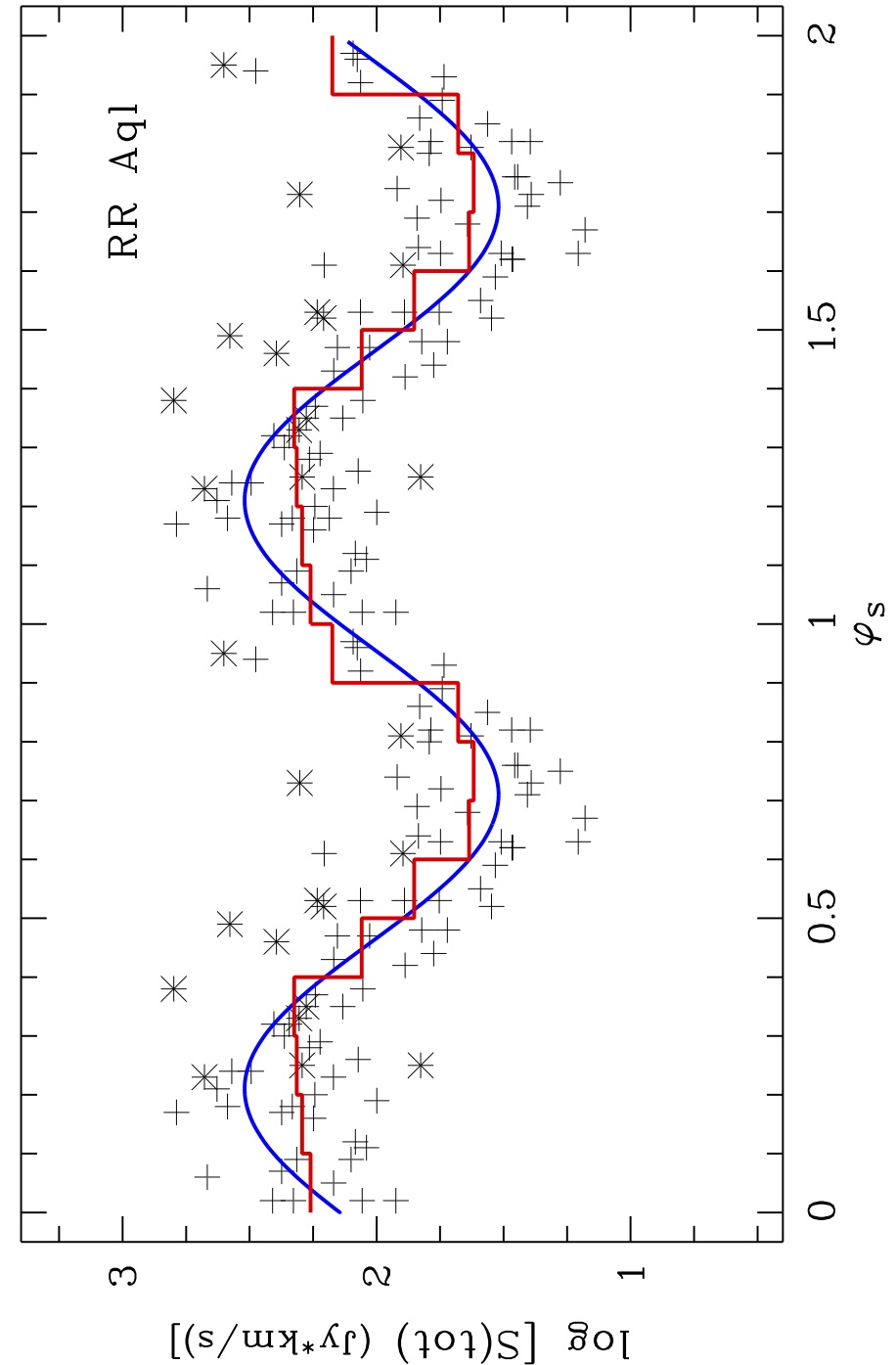}
\caption{ RR~Aql \water\ maser light curve.  See Fig. \ref{fig:uher-lcurve} for details. $S(\rm tot)$ was determined in the velocity range $22 < V_{\rm los} < 34$ \kms. The sine curve (blue) was obtained by a fit to the 1990--2011 radio measurements with a period $P_{\rm opt} = 400$ days and is delayed by $\phi_{\rm lag} = 0.21$, i.e. by 84 days with respect to the optical light curve.}
\label{fig:rraql-lcurve}
\end{figure}

\subsection{\label{sdd_LineProfAna_RRAql} \water\ maser line profile analysis}
To probe velocity variations of the features we made a similar profile analysis as for U\,Her (Sect. \ref{sdd_LineProfAna}) by fitting Gaussian profiles to the spectra. The
assignment of the features to the spectral components is discussed in Appendix
\ref{sdd_LineFitRes_Appendix_RRAql}. We found five features labelled as maser spectral components A--E, which could be isolated by the fitting procedure. These components identified in individual spectra are graphically displayed in Fig. \ref{fig:rraql-fvt} (right panel), where a direct comparison with the FVt-plot is possible. The corresponding flux densities and velocities are listed in Table \ref{tab:compRRAql} 
(Appendix). Components A, C and E are the three dominant features in the maser profile, while components B and D could be isolated by the Gaussian fit only for some time lasting one to six years. 

Three to five spectral components are sufficient to describe the \water\ maser profile in the velocity range $26 < V_{\rm los} <32$.  While in this range weaker maser components could not be identified, their presence is however likely, because such weaker maser emission is seen at velocities $23 - \sim27$ \kms\ (see Appendix \ref{sdd_LineFitRes_Appendix_RRAql}), although we failed to identify individual components. In contrast, there is no evidence in our spectra for emission at velocities larger than $V_{\rm los} > 32$ \kms, which leads to the observed asymmetry of $\Delta V_{\rm los}$ with respect to the stellar velocity.

In RR~Aql's \water\ maser shell, the components A--E represent the regions with the strongest emission only (at least over some time). Due to strong blending in velocity space they overlap in the maser profile and inhibit the identification of a spectral component, if the brightness contrast with respect to stronger neighbouring components becomes too large. We consider therefore the absence of spectral component B as a distinguishable feature in the maser profiles over more than $\sim$15 years (1995 -- 2011, cf. Appendix  \ref{sdd_LineFitRes_Appendix_RRAql}) not as evidence for extinction, but merely as inability to identify the component because of its relative weakness and blending with components A and C. In 2015 the dominant feature had a velocity $V_{\rm los} \approx 28.5$, which is in between the velocities seen for components B and C before, and which we assigned tentatively to component B. Also the identification of component D over only 1.5 years (Fig. \ref{fig:rraql-fvt}, right panel) is mostly due to blending at the other times. 

Blending affects also the determination of the velocities of the spectral components. As is evident from Fig. \ref{fig:rraql-fvt} (right panel) the components showed non-systematic variations in peak velocity within $\sim1$ \kms\ over time. These variations are likely caused by blending of several maser components with velocity differences smaller than the typical line widths. This is corroborated by our experience that the profiles of the spectral components occasionally split into two or even three peaks. It is also expected from interferometric maps of RR Aql's \water\ maser shell, which showed for example in 1988 about two dozen maser spots separated in velocity by 0.3 \kms\ only \citep{bowers94}. 

As in U~Her, a remarkable property of the \water\ maser profile of RR~Aql is the apparent longevity of the spectral components in particular A and C in RR~Aql, which showed also the strongest emission over almost the full monitoring period of $\sim$22 years. No systematic velocity shifts are recognizable, leading to the conclusion that
the emission regions in RR\,Aql have properties as those of U~Her, in particular short lifetimes of at most a few years for individual maser clouds.

\begin{figure}
\includegraphics[angle=-90,width=\columnwidth]
{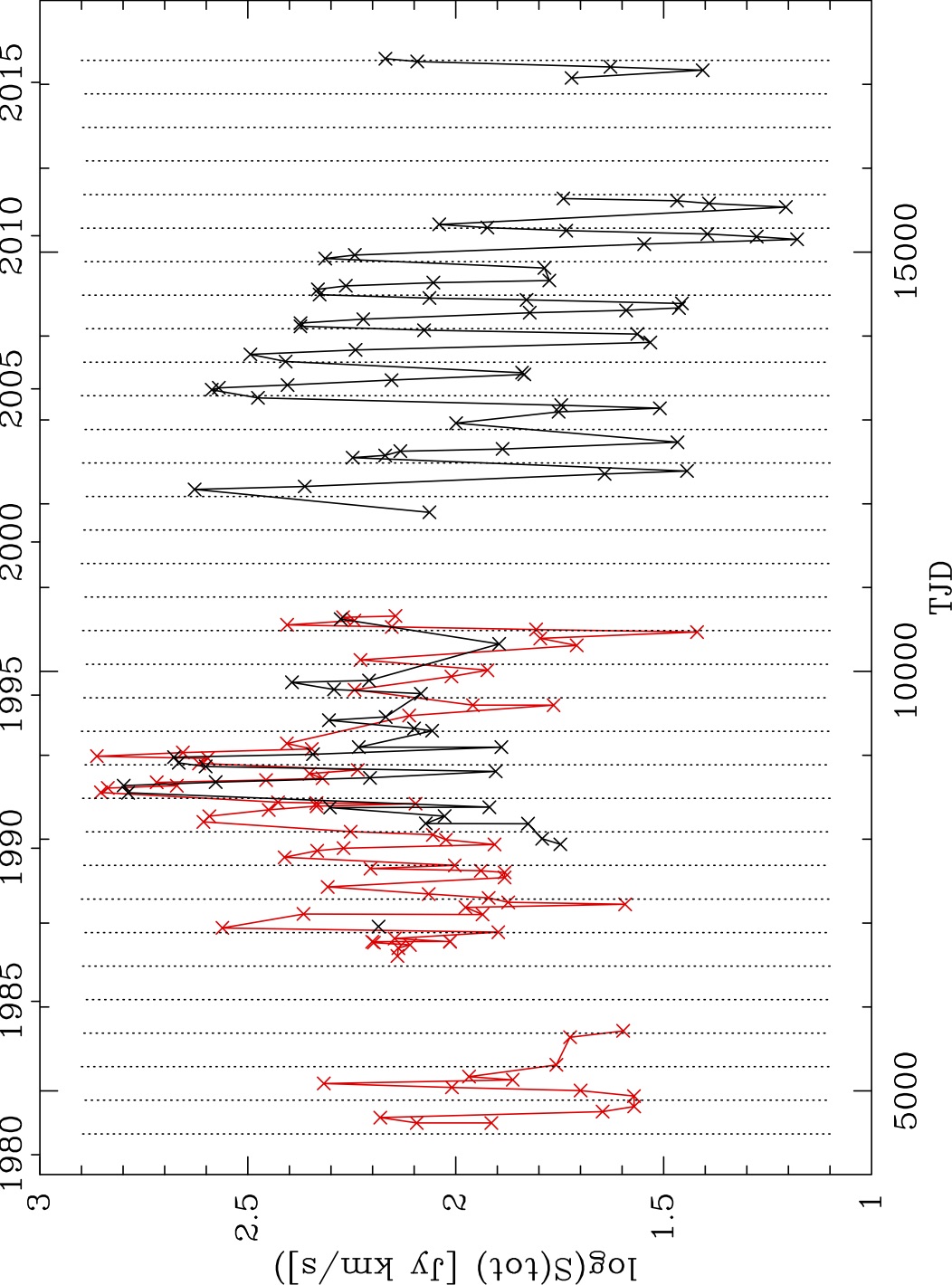}
\caption{RR\,Aql \water\ maser light curves of the integrated flux $S(\rm tot)$ from Pushchino 1980 -- 1997 (red) and Medicina/Effelsberg 1987 -- 2015. Vertical dotted lines are the (modelled) optical maxima with $P = 400$~days.}
\label{fig:rraql-lcurve-TJD}
\end{figure}

\begin{figure}
\includegraphics[angle=-90,width=\columnwidth]
{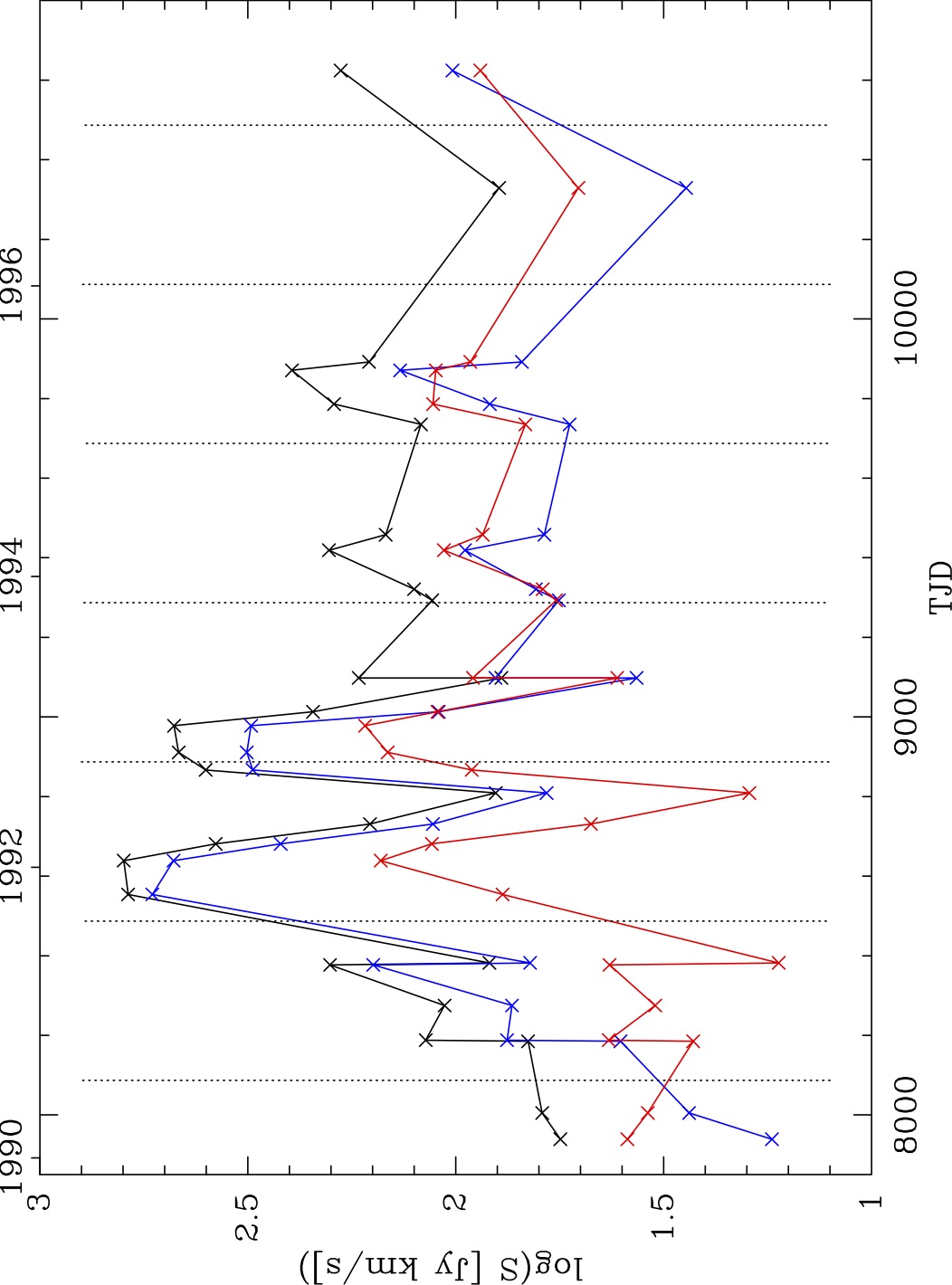}
\caption{RR\,Aql Medicina/Effelsberg \water\ maser light curves 1990 -- 1997 of RR\,Aql showing the emission in the $V_{\rm los} < 28.5$ and $> 28.5$ \kms\ part of the maser velocity range in blue and red color respectively. The sum of both (in black) is the integrated flux $S(\rm tot)$ as shown in Fig. \ref{fig:rraql-lcurve-TJD}. Vertical dotted lines as in Fig.~\ref{fig:rraql-lcurve-TJD}.}
\label{fig:rraql-lcurve-blue-red}
\end{figure}

\begin{figure}
\includegraphics[angle=-90,width=\columnwidth]
{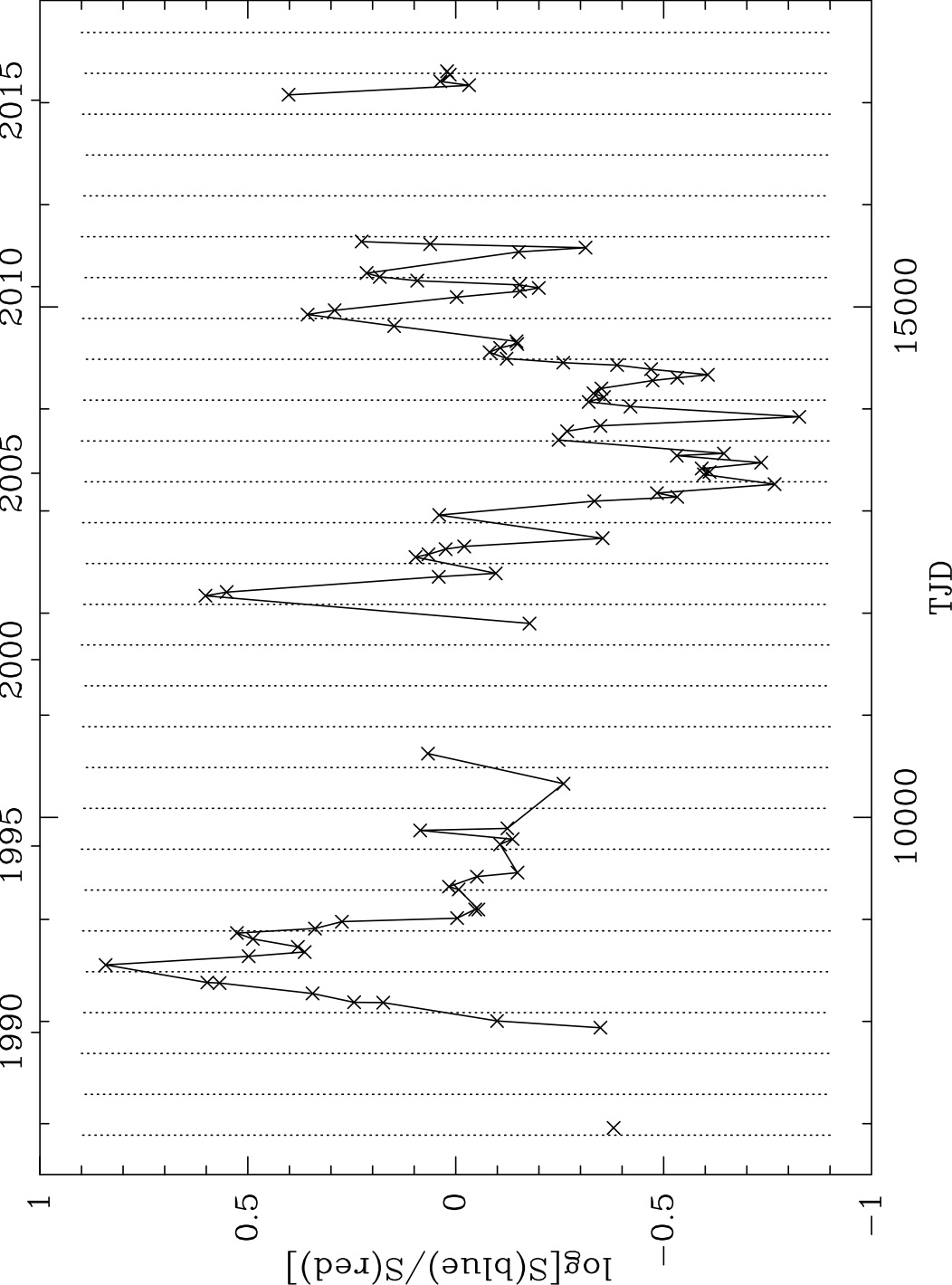}
\caption{Ratio $R = S(\rm blue)$/$S(\rm red)$ of the \water\ maser emission of RR~Aql of the $V_{\rm los} < 28.5$ ($S(\rm blue)$) and $> 28.5$ \kms\ ($S(\rm red)$) part of the maser velocity range in 1987 -- 2015 of RR Aql. The vertical dotted lines are optical maxima as in Fig. \ref{fig:rraql-lcurve-TJD}.}
\label{fig:rraql-lcurve-blue-red-ratio}
\end{figure}

\subsection{\water\ maser light curve}
Figure \ref{fig:rraql-lcurve} shows the maser light curve of RR~Aql using observations between 1990 and 2011, relative to the phase $\varphi_{\rm s}$ of the optical light curve. This radio light curve, based on integrated flux densities $S$(tot), and its relation to the optical light curve was analysed as for U~Her. $S$(tot) was determined over the velocity range  $22 < V_{\rm los} < 34$ \kms. 
The general pattern as seen in the corresponding light curve of U~Her (Fig.~\ref{fig:uher-lcurve}) is present here as well. The selection effect discussed in Section \ref{sdd_phase-lag}, also
in Fig.~\ref{fig:rraql-lcurve} leads to on average higher integrated flux densities $S$(tot) of the Effelsberg- compared to the Medicina observations. While the maser emission varies on average following the optical light curve, the emission for a particular phase shows a large scatter.
A fit of a sine wave to the radio data was made separately for the time ranges 1990 -- 1997 and 2000 -- 2011. We found little evidence for periodic variability in the first time range and a single dominating period of $405\pm10$ days in the second time range. A fit to the complete data set 1990 -- 2011 yielded a period of $400\pm5$ days (Table \ref{centralcoords}), which coincides with the mean optical period in the time range 1987--2015. 
Our optical model light curve for RR\,Aql has a period $P_{\rm opt} = 400\pm2$ days and a reference epoch for maxima TJD$_{max} = 6487 \pm 5$ days (Table \ref{centralcoords}); these numbers were determined as described in Sect. \ref{sdd_OptModelLcurve} for U~Her. 
The lag $\phi_{\rm lag} = 0.21$ (Table \ref{centralcoords}) of the radio light curve is similar to the one of U\,Her ($\phi_{\rm lag} = 0.16$) and other Mira variables of the sample that we monitored (Brand et al., in preparation). 

The significant brightness scatter seen in Fig. \ref{fig:rraql-lcurve} is caused by non-periodic brightness variations, which are caused by relative strength variations of the different spectral components. An overview of the integrated flux density light curve with (truncated) Julian Date is shown in Fig. \ref{fig:rraql-lcurve-TJD}. The curve is dominated by the periodic variations of the maser integrated flux density, but superposed is an apparently bright period (8500 $<$ TJD $<$ 9500; $\sim$1991 -- $\sim$1995) and a relatively faint period after TJD $\approx 13500$ (2005--2011). A zoom-in on the light curve for 1990--1997 is shown in Fig.~ \ref{fig:rraql-lcurve-blue-red}, where we show also the emission contributions $S$(blue) and $S$(red) of the blue-shifted $V_{\rm los} < 28.5$ \kms\ and  red-shifted $>28.5$ velocity range respectively, where $V_{*} = 28.5$ \kms\ is the adopted stellar radial velocity. Almost all the extra emission seen in this period is coming from the blue-shifted velocity range and is caused by the extraordinary strength of spectral component B in this epoch. Except for a few months in 2001 (TJD $\approx 12200$), where spectral component A was strong, the emission in the blue-shifted velocity range remained modest after 2000, and the \water\ maser light curve was reflecting mostly the variations of spectral component C. After 2005 the light curve was following the overall decline of component C. Such long-term trends can be followed on timescales of several years longer than the stellar period (1.1 yr).

Another way to show the role of the secular variations of the individual spectral components is shown in Fig.~\ref{fig:rraql-lcurve-blue-red-ratio}, where the ratio $R = S(\rm blue)$/$S(\rm red)$ is given as function of (truncated) Julian date.  The prevalence of $S(\rm blue)$ is obvious in 8000 $<$ TJD $<$ 9000, but also the phase of dominance of the emission $S(\rm red)$ in the red-shifted velocity range is evident around in 13000 $<$ TJD $<$ 14500. Typical spectra in these phases are shown in Fig.~\ref{fig:rraql_sel}: On 25 October 1991 (TJD = 8554) showing spectral components A and B dominating and in 2005/2007 (TJD = 13359/14116) showing spectral component C dominating.

\subsection{Comparison with the Pushchino and VERA-Iriki monitoring programs \label{sec:push-vera}}
In constrast to U~Her, RR~Aql was monitored also by other groups, allowing verification of our results. The Pushchino \water\ maser monitoring program (B98) of RR Aql between 1980 and 1997 overlapped with ours in 1990 -- 1997. Figure~\ref{fig:rraql-lcurve-TJD} shows the light curve of integrated flux density of the \water\ maser between 1980 and 2011 combining observations of both monitoring programs. Both light curves are consistent with each other, although smaller brightness differences occur. This is caused likely by short-term brightness variations of the maser, leading to small brightness differences for observations made within several days . 

The Pushchino light curve shows  the regular brightness variations connected to the stellar pulsations clearly. In agreement with our average phase lag $\phi_{\rm lag} = 0.21$, B98 could show that these variations are following the optical variations of the star with a delay of 10--30\% of the length of the period. The Pushchino light curve also confirms the long-term change of the average brightness level in $\sim$1991 -- $\sim$1995 (8500 $<$ TJD $<$ 9500), caused by spectral component B at 28~\kms\ reaching flux density levels $>$400~Jy.

Parallel to our monitoring program, observations were made also within the VERA-Iriki monitoring program by \cite{shintani08} between 2003 and 2006. Within the 3.3 years of the VERA-Iriki observations no long-term trends could be studied. In accordance with our observations, spectral components A, C and E were found to be dominant also in their observations. 

The strongest features reported by single epoch observations overlapping with our monitoring program before 2000 were in April/May 1991 at 27.9 \kms\ (our spectral component B) \citep{takaba94} and in October/November 1991 at 26.9~\kms\ (our  spectral component A) \citep{takaba01}.  These velocities are in agreement with our strong components seen at these times. In October 1991 spectral components A and B were competing in brightness, and in our spectrum of 25 October 1991 component B was actually 40\% brighter than component A. After 2000 only one single epoch observation was made. In accordance with our observations that component E became very weak after May 2009 (see Appendix B.3), \cite{kim10} showed from their June 2009 observation that the component was indeed getting fainter.

\section{\water\ maser properties in Mira variable stars \label{sec:miraprop}}

\noindent

\subsection{\water\ maser\ and optical variability -- phase lags}
The prevailing pattern of the  \water\ maser brightness variations in U\,Her and RR\,Aql are the periodic variations, which respond to the brightness variations of the central stars. Their maser variations lag behind the optical variations by $\phi_{\rm lag} = 0.16$ and 0.21, respectively. Similar lags were also found for two other Mira variables o~Cet and R~Cas, monitored by us, and discussed in a separate paper (Brand et al., in preparation). The retarded maser variations are a well known pattern for Mira variables (\citealt{schwartz74}; \citealt{little-marenin91}; \citealt{rudnitskii05} and references therein) and red supergiants \citep{pashchenko99, lekht05}, where the stellar brightness variations have a large amplitude. This differs from the pattern seen in semi-regular variables  of type SRb, such as RX~Boo and SV~Peg (Paper I), R~Crt and RT~Vir (\citealt{lekht99}; Paper II), which do not show large-amplitude periodic variations. The delayed response of maser emission to stellar brightness variations is a general characteristic, and this behaviour is well documented also for SiO masers \citep{pardo04} and OH masers \citep{fillit77, etoka00}.

Often, the lags have been interpreted as the result of the travel time of propagating shock waves through the \water\ maser zone  \citep{lekht01, pashchenko04}. \cite{shintani08} observed periodic velocity variations of maser features in several stars as a general pattern, which they consider as confirmation for the passage of shocks, which first accelerate and later decelerate affected maser clouds. However, lags are also seen in the infrared \citep{lockwood71, smith02, smith06, ita21}. Infrared, SiO and \water\ maser emissions all peak in brightness in the CSE at different distances, so that the lags, having a uniform magnitude, cannot be due to travel time effects.
Guided by the findings of \cite{shintani08} of periodic velocity shifts in a large number of \water\ maser features of several stars, we searched for patterns in the apparent velocity variations of the major spectral components \Gone\ and \Gtwo\ of U~Her and component C of RR~Aql. We found no regularity in the peak velocity changes over time, and conclude that the variations confined to $<0.5$ \kms\ are due to the influence of weaker maser features with similar velocities on the peak velocities derived from the Gaussian fits. 

For the Mira variable BX\,Cam, periodic velocity variations were not found either, but systematic velocity drifts $\le$1.3 \kms\ over timescales of 2--3 years were detected in many maser spectral features, with the blue/red-shifted components decreasing/increasing in line-of-sight velocity $V_{\rm los}$, which is consistent with expansion \citep{xu22}.

If the shock waves propagate radially outwards, their influence on the velocities $V_{\rm los}$ of the stronger spectral components of U\,Her and RR\,Aql furthermore is diminished due the small inclination angles of the outflow directions with respect to the plane of the sky. We were not able to find any evidence for the presence of shock waves in the \water\ maser shells of the two Mira variables. 
We consider therefore the more likely explanation of the lags of the maser brightness variations relative to the optical to be the presence of strong titanium oxide (TiO) absorptions in the visual band at stellar maximum, as was invoked to explain the lags in the infrared by \cite{smith06}. They find that the TiO absorption truncates the rise in the optical light before the maximum in the near-infrared is achieved. Therefore, the phase-lags are not relevant for the understanding of the conditions in the stellar wind, which allow the emergence of \water\ maser emission.
The phases of the \water\ maser light curves are therefore probably better indicators of the phases of the stellar bolometric variations than those of the much more frequently available optical light curves.

\subsection{Short- and long-term maser variability}
We found compelling evidence that the \water\ maser brightness variations in the SRVs R~Crt and RT~Vir 
are a superposition of two types of variations with different timescales. There are short-term fluctuations on timescales $\la$1.5 yr, and long-term variations on timescales of decades (Paper~II). In the case of the Mira variables U~Her and RR~Aql these two types of variations are also present and lead to significant cycle-to-cycle variations of the otherwise periodically varying maser light curves (Figs. \ref{fig:uher-lcurve} and  \ref{fig:rraql-lcurve}). 

 The short term fluctuations seen in integrated flux density are caused by random brightness variations of the individual spectral components, as has been shown for example for the 1991 -- $\sim$1995 absolute maximum of RR~Aql's light curve (Fig. \ref{fig:rraql-lcurve-blue-red} and Sect. \ref{sec:push-vera}). These short-term variations are probably controlled by the coming and going of individual maser clouds, as the limited lifetimes we found for the maser clouds of U~Her indicate.
 
 Very strong short-term fluctuations ("bursts") of individual spectral components at times can dominate the maser profile. Such bursts 
  found in Mira and SR variables by the Pushchino monitoring programs \citep{lekht99, esipov99, pashchenko04} and observed by us for example in RX~Boo (Paper I), in R~Crt and RT~Vir (Paper II), were not seen in the case of U~Her (1990 -- 2015, Fig. \ref{fig:uher-stot-tjd}) or RR~Aql (1980 -- 2015, Fig. \ref{fig:rraql-lcurve-TJD}) (although as mentioned in Sect.~\ref{sdd_longterm_lc} the strong emission during the '1991/1992 peculiar phase' of U\,Her might have been a burst). We assume that bursts are common but that their frequency of occurrence is not large, so that the absence of bursts in the latter two cases could be due to still insufficient time coverage of the maser light curves. 

Long-term maser brightness variations lasting many pulsation cycles are evident for the two Mira variables, if the \water\ maser light curves are displayed as function of time, either integrated over the full profiles or in selected velocity ranges. There is overall dimming and brightening of the different spectral components unrelated to each other, which determine the overall brighteness level but also their relative contributions to it. This basically reflects the change of the excitation conditions in the different parts of the \water\ maser shell, which occur on timescales which are closer to the crossing time (8.5 years in the  case of U\,Her) of material through their \water\ maser shell than to the period of stellar pulsations. 

\subsection{Lifetimes of \water\ maser clouds}
It was not possible to obtain constraints on the lifetimes of individual maser clouds in the winds of the two Mira variables based on the emergence and disappearance of individual maser features in the single-dish spectra. This is due to the wealth of components, which overlap in velocity space. The relative brightness fluctuations lead to small velocity shifts for the spectral components, so that individual maser components may be detectable individually in the spectral profiles only during short times when they are strong. 

The smaller number of spectral components identified in the \water\ maser profile of RR~Aql compared to U~Her is likely also a consequence of blending in velocity space. The final expansion velocity of RR~Aql's wind is about 65\% of that of the wind of U~Her (Table \ref{centralcoords}), so that the acceleration in the \water\ maser shell is smaller. Given the same spectral resolution, separating the spectral components is then more difficult, assuming that the number of maser clouds and the line width distribution of the maser features are comparable in both stars. The spatially resolved maps are the only means to break the degeneracy present in velocity space. We identified nearly 30 different spatial components in the 1990--1992 VLA maps of U~Her compared to fewer than 10 spectral components in the single dish spectra detected during this period. 
Based on the VLA map from \cite{bowers94} a similar discrepancy in the numbers of spatial and spectral components is present also for RR~Aql.

The lifetimes can be constrained by the apparent absence of  acceleration of the spectral components (cf. Sect.\,\ref{region-lifetime-1}). In both stars they showed constant line-of-sight velocities over the monitoring period independently from their location in the maser profile. If the emission clouds participate in the stellar outflow, their outflow velocities would increase with time and hence also their absolute projected velocities, 
modified by the inclination of the outflow direction with respect to the plane of the sky. The expected velocity drifts are not seen (see the FVt-plots Fig. \ref{fig:uher-fvt} and \ref{fig:rraql-fvt}). This observation can be reconciled, if the emission clouds are themselves short-lived and new clouds regularly emerge in a longer-living (stationary?) larger region within the shell, as we have discussed for U\,Her (c.f. Sect. \ref{region-lifetime-2}). The clouds would then be created with similar line-of-sight velocities, and following each other would make up a spectral component with constant velocity. The presence of such regions is also
suggested by the mapping observations of BX\,Cam in 2012--2014 by \cite{matsuno20} and in 2018--2021 by \cite{xu22}, where the emission sites were found in the same parts of the maser shell, despite the stellar wind material in the shell having been exchanged completely between the epochs, given the crossing time of 7.6 years reported by the latter authors.

As the clouds follow the stellar wind with increasing outflow velocities, the spectral components in principle cannot have a strictly constant velocity. Our velocity outflow model (Eq. \ref{eq:exp_model}) for U\,Her predicts an acceleration of 0.5 \kms yr$^{-1}$ in the center of the shell. Adopting a typical lifetime of 2 years \citep{bains03}, a maser cloud therefore will have changed its outflow velocity by 1.0 \kms\ and its line-of-sight  velocity $V_{\rm los}$ by $\le$1 \kms. Velocity drifts of this size are in agreement with the systematic increase/decrease of line-of-sight velocities by up to $\sim$0.4 \kms yr$^{-1}$ observed for the \water\ maser spectral features in BX\,Cam by \cite{xu22}. 
According to our model, the cloud also will have moved radially by $3-4$\, AU during its 2 years lifetime. 
For U\,Her, \cite{richards12} found an increase of the line-of-sight velocities with distance from the star and derived a velocity gradient in the plane of the sky of $0.31\pm0.28$ \kms\ AU$^{-1}$. A gradient of 0.31 \kms\ AU$^{-1}$ translates to a drift in outflow velocity by $0.9-1.2$ \kms\ for a movement over $3-4$\, AU, which is compatible with our estimate of its change in outflow  velocity derived above from the model. We conclude that the line-of-sight velocity drifts of the spectral components, which are expected to be $\le$1 \kms\ over two years are masked in our single-dish data, because the size of the drifts is of the order of the uncertainties in our velocity determinations due to blending with neighbouring features. Components with aspect angles $>30^{\circ}$ would have line-of-sight velocity drifts, which surpass our limit of 0.5~\kms. Of the spatial components plotted in Fig. \ref{fig:uher-3Dvert} that have aspect angles $\theta>30^{\circ}$, only B and C are strong enough to be seen 
contiguously for over more than a year. 
These components are seen between 1990 and 1996 (see Fig. \ref{fig:uher-fvt}). They would be good candidates for the detection of drifts (for B one expects 0.9~\kms\ in two years), but none are seen. They play a prominent role in the '1991/1992 peculiar phase' and their behaviour points to lifetimes $<$2 years, maybe even months (see discussion in Appendix \ref{sdd_LineFitRes_Appendix}). The apparent constant velocity of the spectral components as shown in the U\,Her's FVt-Plot (Fig. \ref{fig:uher-fvt}) is therefore due to the superposition of clouds created with almost equal velocities and replacing each other every few years. The clouds should have systematic velocity drifts over typical maser cloud lifetimes, but these are not detectable in our single-dish spectra.

\subsection{Long-living regions \label{sec:regions}}
The long-term maser brightness variations and the observation  that the clouds responsible for the strongest spectral components in U\,Her are found in the same region of the maser shell over at least 6.5 years (almost 6 stellar pulsation cycles), indicate the presence of inhomogeneities in the otherwise (assumed) spherical stellar wind of Mira variables. One indicator for such inhomogeneities could be parts of the shell, where the conditions are more favourable for exciting the \water\ molecules ('long-living regions') than in others. 
While individual (maser) clouds would be clumps of material with enhanced density moving with the stellar wind, such regions remain stationary in location, at least as long as the wind is not disturbed at these locations. The individual clouds then light up, while they pass through the regions. 
The existence of such regions would then naturally explain, why maser spectral components like component K in U\,Her (see Sect. \ref{region-lifetime}) regularly reappear over about 15 years at almost the same line-of-sight velocity, although different short-living maser clouds must have contributed.

\begin{table*}
\caption{\water\ maser luminosities, stellar luminosities, and mass-loss rates of the semiregular variable stars from Paper~II and the Mira variables U~Her and RR~Aql. Column "D" lists distances as given in Table~\ref{centralcoords}, Paper I and II. Characteristic levels of \water\ maser brightness ($S(\rm tot)$ = integrated fluxes) and maser luminosities (\Lup  [L$_{\odot}$]; $L_{\rm p}$[photons s$^{-1}$]) are listed for the time range 1987-2015 for all sources except RX~Boo and SV~Peg (1987--2005; cf. Paper I). The definition of the levels (High, Mean, Low) is described in the text. Columns $\log$\,$L_{\rm bol}$ and $\log$\,\mdot\ list stellar luminosities and mass-loss rates.} 
\label{table:photon-luminosities}
\begin{center}
\begin{tabular}{lrr|c|rr|rr|rr|c|c}
\hline\noalign{\smallskip}
\multicolumn{1}{c}{Star} & \multicolumn{1}{c}{Type} & \multicolumn{1}{c|}{D} & log\, \Lup &
\multicolumn{6}{c|}{log\,$S(\rm tot)$ , log\,$L_{\rm p}$} & log\,$L_{\rm bol}$ & log\,\mdot \\[0.05cm]
\multicolumn{2}{c}{} & \multicolumn{1}{c|}{[pc]} & [L$_{\odot}$] &
\multicolumn{6}{c|}{[Jy~\kms] , [s$^{-1}$]} & [L$_{\odot}$]    &  [\Myr]   \\
&&&& \multicolumn{2}{c|}{High} & \multicolumn{2}{c|}{Mean}     & \multicolumn{2}{c|}{Low} &&\\
\hline\noalign{\smallskip}
R~Crt    & SRV &  236 & $-$4.88& 3.5    & 44.0& 2.9 & 43.4 &    2.3 &    42.8 &  $4.03\pm0.10$ & $-5.46$ \\
RT~Vir   & SRV &  226 & $-$5.16& 3.3    & 43.8& 2.8 & 43.3 &    2.3 &    42.8 &  $3.70\pm0.09$ & $-6.05$ \\
RX~Boo   & SRV &  136 & $-$6.23& 2.6    & 42.7& 1.9 & 41.9 &    1.5 &    41.5 &  $3.58\pm0.11$ & $-6.12$ \\
SV~Peg   & SRV &  333 & $-$6.42& 1.6    & 42.4& 0.9 & 41.7 & $<$0.8 & $<$41.6 &  $3.93\pm0.20$ & $-6.04$ \\
U~Her    &Mira &  266 & $-$5.77& 2.6    & 43.3    & 2.0    & 42.6    &    1.5 & 42.1   &  $3.71\pm0.17$ & $-6.25$ \\
RR~Aql   &Mira &  410 & $-$5.37& 2.7    & 43.7    & 2.1    & 43.1    &    1.4 & 42.4   &  $3.75\pm0.15$ & $-5.84$ \\
\noalign{\smallskip}\hline
\end{tabular}
\end{center}
\end{table*}

Also in the SRVs RX\,Boo, RT\,Vir and R\,Crt we have found evidence for the existence of long-living regions within the \water\ maser shell, in which preferred conditions for exciting the \water\ molecules exist.  In RX\,Boo they manifested themselves as spatial asymmetries of the \water\ maser emission persistent for at least 11 years (Paper\,I). For RT\,Vir and R\,Crt, the decade long variations of the \water\ maser brightness in particular velocity ranges, was attributed to such regions having possibly higher-than-average densities and being present in the \water\ maser shells for about two decades (Paper\,II). Based on 3D models of \cite{freytag23} (and preceding papers) we argued that such regions could be the remnants of large convective cells, which left the stars as part of the stellar winds. Having reached the \water\ maser shell they may have been inflated to sizes comparable to the width of the maser shells. They may provide the shell sectors with improved maser excitation conditions, in which clouds passing through are preferentially excited. Observationally it would lead to an asymmetric spatial maser distribution within an otherwise spherical symmetric \water\ maser shell. If the presence of (sub)-stellar companions plays a major role in shaping the stellar winds \citep{decin20,gottlieb22}, their influence on the velocity field may be another factor to create long-living regions with improved excitation conditions. These regions could be in parts of the shell relatively far from companions, where the stellar wind is less disturbed. Lifetimes of the long-living regions would probably be of the order of the crossing time of material through the shell or of the orbital periods of the companions. 

While the spherical \water\ maser shell is a persistent feature with some variations of its boundaries, the maser clouds are a transient phenomenon with a lifetime of a few years. The regions, in which favourable excitation conditions for \water\ maser emission occur, are intermediate in the sense that they are part of the \water\ maser shell, but not lasting. They have long but limited lifetimes, because of the inhomogeneity of the stellar wind. Over time, such regions where maser clouds pass through are created at different locations within the shell. 

\subsection{\water\ maser luminosities}
In Table~\ref{table:photon-luminosities}, we give luminosity information on U~Her and RR~Aql, as well as those of the four SRVs treated in Paper I and II. For the details of how the luminosities were derived, we refer to Paper~II. In column 4, we give \Lup, the potential maximum \water\ maser luminosity derived from the upper envelope, which represents the maximum output which the source could produce if all the velocity components we observe were to emit at their maximum level, at the same time and equally in all directions. The table also lists characteristic levels (high, mean, and low) of maser brightnesses, as given by integrated flux densities $S(\rm tot)$ in Jy \kms\ (Cols. 5,7,9) and corresponding maser luminosities $L_{\rm p}$ in photons per second (Cols. 6,8,10). The brightness of the mean level is the median of all integrated flux density measurements 
while the high and low level are represented by the median of the seven highest and lowest integrated flux density measurements, respectively. 

Table~\ref{table:photon-luminosities} shows that the mean maser luminosities of U\,Her and RR\,Aql (as well as \Lup) are in the range of luminosities shown by the SRVs. The ratio between the high and low level of the two Mira variables is 15--20, which is slightly higher than for the four SRVs with ratios $>$6--16 ($>$35--45 between \Lup\ and the low levels of $L_{\rm p}$ compared to $>$25--55 of the SRVs).

As in Paper~II, we compared the \water\ maser photon luminosities with the bolometric luminosities and mass-loss rates of the stars. The bolometric luminosities $L_{\rm bol}$ (Col. 11) were determined from bolometric fluxes, as described in \cite{jimenez15}, and the distances listed in Table~\ref{table:photon-luminosities}. The mass-loss rates \mdot\ (Col. 12) were taken from \cite{loup93} (U~Her) and \cite{danilovich15} (RR~Aql) scaled to the distances used here. All mass-loss rates come from measurements of the CO molecular line and are estimated to have an error of a factor of 3 (= 0.5 dex) prior to the uncertainties introduced by the distances. We note however, that due to a general underestimate of the terminal stellar wind velocities in the past, these mass-loss rates could be systematically overestimated \citep{gottlieb22}. The results are listed together with those of the SRVs (see Paper~II) in the last two columns of Table~5.

We find that the two Mira variables have stellar luminosities and mass-loss rates within the range shown by the SRVs. Not surprisingly also the \water\ maser photon luminosities fit well within the range of those of the SRVs. The stars selected by us for \water\ maser monitoring all showed apparently bright masers since their discovery. Except for SV~Peg, all of them appear among the 10\% brightest galactic \water\ masers in late-type stars known by the year 2000 on the sky north of $\delta > -30^\circ$ \citep{valdettaro01}. Besides a few Red Supergiants this list of late-type stars is composed of optically bright Mira and SR variables.  It is therefore obvious that the stars from our monitoring program represent only the upper end of the photon luminosity distribution of \water\ masers of Mira and SR variables in the solar neighbourhood.

\subsection{\label{sdd-cse-asym} Constraints on the CSE standard model}  
The analysis and discussion of the \water\ maser properties of U\,Her and RR\,Aql are based on the standard model, which assumes the presence of a spherical \water\ maser shell within a circumstellar envelope continuously fed by a homogeneous spherically outflowing stellar wind with smoothly increasing velocities. The stellar pulsations naturally lead to maser brightness variations and connected with this to variations of the velocity range over which emission is detected above the sensitivity limit.  The boundaries of the maser shell are, strictly speaking, boundaries for luminous maser sites, and may change over time due to evolution of the excitation conditions and changes in beaming directions. The time coverage of the maser observations since their discovery is still too short to understand the origins of the long-term variations, but they indicate that there are regions of improved excitation conditions within the maser shell, as discussed in Sect \ref{sec:regions}, and hence inhomogeneities in the  spherical outflow.

The filling factor of detectable maser clouds within the maser shell is rather small \citep{richards12} so that asymmetries seen in maser maps can be due either to accidentally emerging illumination variations, or is another indication that the underlying stellar wind itself is inhomogeneous. 3D models of the formation of dust-driven winds in AGB stars 
by \cite{freytag23}, indicate that the stellar wind starts clumpy above the photosphere. It is not clear how much of the clumpiness is still present at the location of the \water\ maser shell (10--30 AU from the star for SRV and Mira variables), but the related velocity variations of the wind due to shock waves generated close to the photosphere may have become negligible at these distances \citep{bladh19}. Smoothly increasing outflow velocities described by a velocity law with continuous acceleration and approaching a constant outflow at large distances, as for example expressed by Eq. (\ref{eq:exp_model}), may therefore still be applicable. 

However, new observations of several molecular species and transitions in the CSEs of several AGB stars and RSGs challenge the validity of such a law \citep{gottlieb22}. While these authors observe an overall increase of outflow velocities with radial distances, the scatter  is not compatible with a smooth velocity law. The scatter is ascribed to the presence of a (sub)stellar companion which disturbs the velocity field. In addition, they find that the final expansion velocities are up to a factor of two higher than previously adopted. For U\,Her this is $V_{\rm exp} = 19.7$ \kms\ compared to 13.1 \kms\ as adopted by us (Table \ref{centralcoords}). RR\,Aql was not observed. Their outflow velocity law for U\,Her is  $V(r) = V_0 + (V_{\rm exp}-V_0)  \cdot  (1- r_0/r)^\beta$, where $r_0 = 2.6$ R$_*$, $V_0\sim4.4$ \kms\ and $\beta=2.00\pm0.54$. Parameter $r_0$ is the dust condensation radius, where the wind starts, and $V_0$ is the velocity at this radius. This law is considered an approximation and is given for distances in the CSE between $\sim$2.6 and 500 R$_*$ ($\sim4 - 700$ AU). 

Adopting this law in our model of the \water\ maser shell (Sect. \ref{sec:shell-model}), the larger final expansion velocity leads to an increase of the acceleration in particular in the inner shell, where the \water\ masers are located. For example at radial distance r=18 AU the increase is from 0.5 to 1~\kms yr$^{-1}$. Correspondingly the crossing time and the lifetimes derived for individual maser clouds are decreased by a factor of $\sim$1.8.

As long as acceleration is present in the \water\ maser shell, the absence of velocity drifts of the \water\ maser spectral components constrains the lifetimes of the maser emitting clouds. A lifetime larger than the monitoring period of 22 years, would require a rather small acceleration $a < 0.023/ \sin \theta$~\kms yr$^{-1}$ to avoid line-of-sight velocity shifts $|\Delta V_{los}| > 0.5$ \kms\ (see Sect.~\ref{region-lifetime-1}). 
This would need almost a suspension of the acceleration within the \water\ maser shell (i.e. a step-wise increase of the velocity curve as discussed by \cite{Decin15} in the case of the carbon star IRC+10216) or an approach to the final expansion velocity already at the inner maser shell radius. We adopted the latter explanation for the SRV RT\,Vir, where we observed a maser feature showing a constant velocity within $< 0.06$ \kms\ over 7.5 years. We suggested the maser-emitting cloud to move in the outer part of the shell, where the final expansion velocity could have been reached already. The alternative, maser clouds with short lifetimes created again and again, was considered as less likely because of the very small velocity scatter (Paper II). 

So far, a step-wise increase of the velocity curve has not been considered to explain the observed velocities in the \water\ maser shell. 
The model presented in Sect. \ref{sec:shell-model} for U\,Her is strictly constrained only by the outflow velocity $V_{\rm out} = 9.6$ \kms\ at a radial distance of 23.9 AU. Forcing the adopted velocity law to start at the photosphere ($r_0=1.4$ AU, $V_0 = 0$ \kms) is a rather coarse assumption given that in U\,Her's CSE, \water\ line emission from the (0,2,0) $6_{52} - 7_{43}$ transition at 268 GHz line was detected over a velocity range $-24.2$ to $-4.5$ \kms\  by \cite{baudry23} within a projected distance of $\sim$3.2 AU from the center of the star. This implies gas motions with outflow and infall of up to 10 \kms\ in the inner shell, and a smooth velocity law may not be applicable within several stellar radii from the star.

In principle, the outflow velocity could reach $V_{\rm out} = 9.6$ \kms\ at the inner maser shell boundary, stay constant up to the outer maser shell boundary, and increase further beyond. 
In the case of a step-wise increase of the velocity curve, a section with constant velocity of 9.6 \kms\ will have improved velocity coherence, and naturally delineate the shell, where \water\ maser emission preferentially occurs. The boundaries at $\sim5$ and $\sim30$ AU would be determined by the loss of velocity coherence outside the section with constant outflow velocity. However, marginal acceleration would favour radial amplification paths and one would expect double-peaked maser profiles strongest close to $\pm V_{\rm out}$ (in our case $\pm9.6$ \kms) as seen for 1612 MHz OH masers, which are excited at much larger distances from the star, where the outflow velocity also in the standard model is almost constant. This is however not observed, as \water\ maser profiles of SR- and Mira variables generally peak close to the stellar velocity \citep{takaba94}, meaning that tangential maser amplifications paths are preferred to radial ones. \cite{richards12} also argue against an absence of acceleration within the \water\ maser shell, as they found that in the shells of all observed AGB and RSG stars the outflow velocity increases by a factor of two. Following the numerous indications that the velocity field of the stellar winds is likely disturbed over a great part of the CSE, we consider therefore an almost constant velocity over several AU delineating the \water\ maser shell unlikely. 

\section{Conclusions \label{sec:conclusions}}
We analysed the properties of the \water\ maser emission of two Mira variables U~Her and RR~Aql, which both show maser emission spread in velocity over several \kms, with usually bright individual features ($>100$ Jy). Their masers are among the strongest \water\ masers seen in late-type stars on the sky.

The variability of the maser emission is dominated by regular brightness variations synchronised with the stellar pulsation. As observed generally in Mira variables, the maser brightness variations are delayed relative to the optical variations, in the case of U\,Her and RR\,Aql by two and nearly three months, respectively. We attribute the cause of this phase-lag to the influence of absorption by molecular bands on the optical light curve. Superposed on the regular variations are brightness fluctuations on shorter time scales, which are due to secular variations of individual spectral components. Also on longer time scales variations are seen of the average brightness level in both stars. The brightness levels in the blue- and red-shifted velocity ranges show (on these time scales) variations that are independent from each other. This indicates that the stellar wind is inhomogeneous, which leads to varying excitation conditions in the front and rear cap of the \water\ maser shell. This change of excitation conditions can lead to a completely different spectral profile, where in the case of U\,Her the usually dominating profile peak at -15 \kms\ was exceeded for 1.5 years by peaks at $< -18$ \kms\ ('1991/1992 peculiar phase'). Apart from that, the \water\ maser variability patterns of U\,Her and RR\,Aql are remarkably similar.

Based on our VLA observations 1990-1992, we find that the \water\ masers of U\,Her were located in a spherical shell with a size of 11-25 AU. Additional weak maser emission was found inside and outside these boundaries. Comparing this result with interferometric observations from the literature obtained at other epochs shows some variations of the size, but in general the strongest emission comes from radial distances of $\sim15-20$ AU. With a radius of $\sim$21 AU reported for RR\,Aql's \water\ maser shell, also the shell dimensions corroborate the similarity of the \water\ maser properties between U\,Her and RR\,Aql.

The absence of velocity drifts strongly argues for short lifetimes of maser clouds (about $<4$ years). The location of the main emission features of U~Her in the same part of the \water\ maser shell over at least 6.5 years and the regular reappearance of emission at the same velocity (see component K in U~Her), support the idea of there being long-living regions, sections of the maser shell, with favourable excitation conditions which last over time scales longer than the stellar period and the lifetime of individual maser clouds.

The \water\ maser luminosities of the two Mira variables are within the range of luminosities derived in Paper\,I and II for SRV's. They have variation properties very similar to the SRV's, except for their periodic variations as a consequence of the stellar pulsation (via pumping variations).

Variability is present on several timescales, so that the conclusions drawn from our monitoring program lasting more than two decades are strictly valid only for the scales covered. For U\,Her, the length of the monitoring program is about twice the time needed for the stellar wind to cross the \water\ maser shell. Variations in the mass-loss rates on longer timescales (hundreds to thousands of years) may lead to a loss of U~Her's and RR~Aql's prominent status among the strongest circumstellar \water\ masers observed from Earth. They may even join the much larger group of Mira variables which are currently not detected with \water\ masers, while in other Mira variables \water\ maser emission may brighten and/or be beamed into our line of sight. 

With the validity of a smooth acceleration of the stellar wind put into question recently, the advantage of using \water\ masers to trace the velocity field of the inner wind region of AGB stars ($r\la20$  R$_*$) will gain importance.

\begin{acknowledgements}
The Medicina 32-m data presented here are part of a long-term monitoring program, which concerned both late-type stars and star-forming regions. Thanks to those who helped with the observations.
We are grateful to the staff at Medicina observatory for their expert assistence and technical problem-solving. The Medicina radio telescope is funded by the Ministry of University and Research (MUR) and is operated as National Facility by the National Institute for Astrophysics (INAF). This research is partly based on observations with the 100-m telescope of the MPIfR (Max-Planck-Institut für Radioastronomie) at Effelsberg, and the VLA (Very Large Array). 
The VLA is operated by the National Radio Astronomy Observatory, which is a facility of the National Science Foundation operated under cooperative agreement by Associated Universities, Inc. This research has made use of the SIMBAD database and the VizieR catalogue access tool, operated at CDS, Strasbourg, France, and of NASA's Astrophysics Data System. For data reduction and the preparation of figures GILDAS software available at www.iram.fr/IRAMFR/GILDAS was used.
We acknowledge with gratitude the variable star observations from the AAVSO International Database contributed by observers worldwide and used in this research.
The spectra are available in fits-format at CDS; also the VLA data cubes can be downloaded from there.
\end{acknowledgements}

\bibliographystyle{aa}
\bibliography{Winnberg-etal-jpg.bbl}

\Online

\appendix
\input{appendix.tex}

\end{document}

%% file: appendix.tex
\section{\label{sdd_LineFitRes_Appendix} U Her Spectral line fitting results}
The maser features identified between 1987 and 2015 in the individual spectra are listed in Tables \ref{tab:compUHerB-E} and \ref{tab:compUHerG-M}, where they were assigned to different spectral components separated through Gaussian fits. For each spectrum we give the Gregorian and Julian dates, its \textsl{rms} noise level in Jansky, the integrated flux $S$(tot) in Jy\,\kms, the phase of the optical lightcurve $\varphi_s$, and 
the velocity ($V_{\rm p}$) and peak flux density ($S_{\rm p}$) for each component. Flux density values marked by ``:'' and upper limits were measured with the cursor on the computer screen. Velocity values marked by ``:'' are uncertain.
$\varphi_{\rm s}$ was calculated for each observing date using the optical period and a reference epoch for the optical maxima TJD$_{max}$ (i.e. $\varphi_{\rm s} = 0$), as given in Table \ref{centralcoords}. In the following we discuss in detail the variations in flux density and velocity of the spectral components, for the main velocity ranges discussed in the main body of the paper.

\subsection{The $V_{\rm los}<-18$ \kms\ velocity range}
The most blue-shifted emission observed by us in 1987 -- 2015 was detected at $-23.8$ \kms\ in the Oct.\,1991 VLA observation (spatial component A1 in Table \ref{tab:components}). The emission was weak ($\approx$0.4 Jy/beam) and not seen in the single-dish spectra of this epoch. Emission at velocities $<-22$ \kms\ was seen in the single dish spectra only occasionally and was blended with the stronger spectral component B. The most blue-shifted emission detected in Effelsberg and Medicina was seen in 1995 with a peak at $-22$ \kms and extending down to $-23.3$ \kms, while the star was increasing its optical brightness ($0.7 < \varphi_{\rm s} < 1.0$). The emission was on the $\approx$1 Jy level. We conclude therefore that in the velocity range $-24 < V_{\rm los} < -22$ \kms\ (the location of a putative spectral component A) maser emission was probably present, although below our detection limits most of the time.

In addition to this low-level emission we identified three spectral components B--D at velocities $V_{\rm los}<-18$ \kms. The emission in the $-22< V_{\rm los} <-20$ \kms\ region (spectral component B) was seen only until 1996 and was in general about a factor of 10--100 weaker than emission at higher velocities. Only during a very brief period (the '1991/1992 peculiar phase') in Oct./Nov.\,1991 (TJD = 8959) component B (then at $-20.6$ \kms) was the strongest of all spectral components (Fig.~\ref{fig:uher_sel}). The component was not detected in May 1991 and declined by a factor of $\sim15$ until January 1992, constraining the duration of the '1991/1992 peculiar phase' to a few months.

During the '1991/1992 peculiar phase', emission at $-20< V_{\rm los} <-19$ \kms\ (spectral component C) increased in intensity after May 1991 (TJD $\sim 8400$), reached its peak in Jan.\,1992  (TJD $\sim 8640$) and faded to the level seen before and afterwards until March 1992 (TJD $\sim 8700$).  A similar short dominance was seen in Sep.\,1996 (TJD = 10353), where spectral velocity component \Done\ ($V_{\rm los} = -18.8$ \kms) was the strongest feature in the profile. The brightenings seen in spectral components B, C, and \Done\ lasted typically 6 to 12 months.  

The emission between $V_{\rm los} = -19$ and $-18$ \kms\ was dominated by spectral components \Done\ (strong in 3/1995 -- 2/1997, TJD = 9790 -- 10486) and \Dtwo, which often were difficult to separate in velocity space. Therefore Tables~\ref{tab:compUHerB-E}  and \ref{tab:compUHerG-M} usually list only one of both components depending on the peak velocity provided by the Gaussian fit, although most probably both components where present. In February 1992 (TJD = 8682; $\varphi_{\rm s} = 0.97$) component \Dtwo\ reached the strongest flux density ($>300$ Jy) ever observed between 1987 and 2015 for an individual spectral feature. Until October 1992 it decreased to a level of $\sim10$ Jy and remained on that level until it disappeared after 2001. 

\subsection{The $-18 < V_{\rm los} <-16$ \kms\ velocity range}
Often blended with spectral components \Dtwo\ at $-18.3$ \kms\ and \Gone\ at $-15.5$ \kms\ the emission in between was almost always detected. Peak velocities of the Gaussian fits varied between $-17.5$ and $-16.5$ \kms, indicating that more than a single emission feature contributed. As these features in general could not be decomposed, they are listed together in 
Table~\ref{tab:compUHerB-E} as spectral component E.  In the interferometric maps 1990 -- 1992 we identified five spatial components E1 -- E5 (Table \ref{tab:components}).

\subsection{The $-16 < V_{\rm los} <-14$ \kms\ velocity range}
The dominant \water\ maser emission stemmed almost continuously from velocities $-16 < V_{\rm los} <-14$ \kms. Within this velocity range two or more spectral features contributed. Due to the variations of the relative intensities, the central velocities of the Gaussian fits drifted. Usually two features could be identified, which we designated spectral components \Gone\ and \Gtwo\ in 
Table~\ref{tab:compUHerG-M}. In 1990, at the beginning of the systematic monitoring program two features with $>200$ Jy at $-15.3$ (\Gone) and $-14.7$ (\Gtwo) \kms\ dominated the profile, with component \Gtwo\ being the stronger one. This changed after 1991 and lasted until 2006, when again component \Gtwo\ became stronger than \Gone. This behaviour is the cause for the apparent curvature of the ridge of the emission in the FVt-plot (Fig.~\ref{fig:uher-fvt}) within the velocity interval $-16 < V_{\rm los} <-14$ \kms. In 2007 -- 2011 (TJD $\ga 14000$) the evidence that \Gone\ is a blend of maser lines is particularly strong, as the peak velocity varied in the interval $-16.0 \le V_{los} \le -14.8$ and showed shifts between the extremes of this interval within four months (August - December 2007).

\subsection{The $V_{\rm los} >-14$ \kms\ velocity range}
In this fourth velocity range only weaker emission was present which was assigned to spectral components I--M in Table~\ref{tab:compUHerG-M}. Spectral component I was centered most of the time at $-12.9\pm0.1$ \kms. Only between 1993 and 1995 (TJD $\sim8700 - 10200$) the center was at $-13.5\pm0.2$ \kms. After 1995 the component surpassed the detection limit only close to the maxima of the stellar lightcurve. The last four years after 2007 (TJD $>14450$) it was not detectable anymore, but re-appeared in 2015 during an optical maximum (Fig.~\ref{fig:uher_all}). 

In the period 1995 -- 2010 ($9700<$ TJD $<11700$) emission was detected at $\sim-11.0$ \kms\ (component K), often at phases when also component I was detected and similar in strength (Fig.~\ref{fig:uher-fvt}). At velocities $\sim-10$ \kms, in 1990/1991 (TJD $<8400$) emission with flux densities of a few Jansky was present at $-10.2$ \kms. Later in a short phase of about 6 months in 2009/2010 emission was detected at at $-9.7$ \kms\ (Table~\ref{tab:compUHerG-M}). Due to their proximity in velocity we assigned to them a common spectral component L. Components I, K and L are responsible for the distinguished feature in the FVt-plot (Fig.~\ref{fig:uher-fvt}) at $\sim-10$ \kms\ at TJD $\sim15100-15300$.

Also seen for few months only (Oct. 1990 -- May 1991; Table~\ref{tab:compUHerG-M}) was spectral component M, the most red-shifted component in the \water\ maser profile. It appeared at $\sim-8$ \kms\ with flux densities $\sim$1 Jy and is marginally visible at TJD $\sim8300$ in the FVt-plot (Fig.~\ref{fig:uher-fvt}). \newline

\newpage
\onecolumn

\begin{sidewaystable}
\caption{Maser spectral components B -- E of U\,Her}
\begin{tabular}{rrrrrrrrrrrrrrr}
\label{tab:compUHerB-E} 
           &         &         &          &  $    $ &\multicolumn{2}{c}{B}&\multicolumn{2}{c}{C}&\multicolumn{2}{c}{\Done}&\multicolumn{2}{c}{\Dtwo}&\multicolumn{2}{c}{E} \\[0.1cm]
Date & TJD & rms & $S$(tot) & $\varphi_{\rm s}$ &\multicolumn{10}{c}{\hrulefill\ $V_{\rm los}$\gm $S_{\rm p}$ \hrulefill}  \\[0.1cm]
           &         &  [Jy]   &[Jy*\kms] &         & \multicolumn{10}{c}{\hrulefill\ [km\,s$^{-1}$, Jy] \hrulefill} \\[0.1cm]
\noalign{\smallskip}\hline\noalign{\smallskip}
26.03.87  &     6881  &  3.17  &  22   & 0.53   &  $-$ & $-$ & $-$ & $-$ & $-$ & $-$  &  $-$  &  $-$ &  $-$  &  $-$ \\
31.03.87  &     6886  &  2.49  &  27   & 0.54   &  $-$ & $-$ & $-$ & $-$ & $-$ & $-$  &  $-$  &  $-$ &  $-$  &  $-$ \\
17.06.87  &     6964  &  3.13  &  40   & 0.73   &  $-$ & $-$ & $-$ & $-$ & $-$ & $-$  &  $-$  &  $-$ &  $-$  &  $-$ \\
05.09.87  &     7044  &  5.14  & 113   & 0.93   &  $-$ & $-$ & $-$ & $-$ & $-$ & $-$  &  $-$  &  $-$ &  $-17.3$  &  24.3 \\
17.02.90  &     7940  &  0.12  &  433  &  0.14  &  $-21.3$ & 3.8 & $-20$ & $<2$ & $-$ & $-$  &  $-18.0$  &  73.7 &  $-$  &  $-$ \\
31.03.90  &     7982  &  0.17  &  395  &  0.24  &  $-21.2$ & 1.9 & $-20$ & $<2$ & $-$ & $-$  &  $-18.0$  &  57.6 &  $-$  &  $-$ \\
24.04.90  &     8006  &  2.78  &  429  &  0.30  &  $-$ & $-$ & $-$ & $-$ & $-$ & $-$  &  $-18.1$  &  44.8 &  $-$  &  $-$ \\
12.05.90  &     8024  &  0.29  &  294  &  0.35  &  $-21.0$ & $<1$ & $-20.0$ & $<1$ & $-$ & $-$  &  $-18.1$  &  23.9 &  $-$  &  $-$ \\
21.10.90  &     8186  &  0.14  &  122  &  0.75  &  $-21.0$ & 0.5 & $-20.0$ & 0.5 & $-$ & $-$  &  $-18.1$  &  17.8 &  $-16.5$  &  9.4 \\
24.10.90  &     8189  &  1.80  &  97  &  0.76  &  $-$ & $-$ & $-$ & $-$ & $-$ & $-$  &  $-18.3$  &  21.5 &  $-16.5$  &  $<9$ \\
18.01.91  &     8275  &  1.76  &  271  &  0.97  &  $-21.3$ & 4.9 & $-$ & $-$ & $-$ & $-$  &  $-18.1$  &  85.5 &  $-17.1$  &  25.7 \\
31.03.91  &     8347  &  0.11  &  267  &  0.15  &  $-21.2$ & 3.6 & $-20.3$ & 2.6 & $-$ & $-$  &  $-18.2$  &  76.7 &  $-17.2$  &  32.1 \\
01.05.91  &     8378  &  0.23  &  257  &  0.22  &  $-21.2$ & 7.0 & $-20.2$ & 2.9 & $-$ & $-$  &  $-18.1$  &  76.2 &  $-17.2$  &  32.7 \\
17.05.91  &     8394  &  2.14  &  150  &  0.26  &  $-$ & $-$ & $-$ & $-$ & $-$ & $-$  &  $-18.1$  &  50.6 &  $-17.1$  &  21.3 \\
31.05.91  &     8408  &  3.83  &  238  &  0.30  &  $-$ & $-$ & $-$ & $-$ & $-$ & $-$  &  $-18.2$  &  63.1 &  $-16.9$  &  25.3 \\
27.10.91  &     8557  &  0.83  &  170  &  0.66  &  $-20.6$ & 53.7 & $-19.5$ & 31.3 & $-$ & $-$  &  $-18.0$  &  51.8 &  $-$  &  $-$ \\
02.11.91  &     8563  &  1.52  &  175  &  0.68  &  $-20.6$ & 71.0 & $-19.5$ & 29.0 & $-$ & $-$  &  $-18.0$  &  54.5 &  $-$  &  $-$ \\
11.01.92  &     8633  &  0.74  &  275  &  0.85  &  $-21.5$ & 4.2 & $-19.5$ & 51.1 & $-$ & $-$  &  $-18.3$  &  107.0 &  $-17.3$  &  $<55$ \\
18.01.92  &     8640  &  0.18  &  480  &  0.87  &  $-21.3$ & 8.7 & $-19.5$ & 90.6 & $-$ & $-$  &  $-18.2$  &  197.2 &  $-17.3$  &  $<95$ \\
05.02.92  &     8658  &  1.27  &  242  &  0.91  &  $-21.6$ & 5.4 & $-19.5$ & $<22$ & $-$ & $-$  &  $-18.4$  &  116.0 &  $-17.3$  &  $<45$ \\
29.02.92  &     8682  &  0.13  &  632  &  0.97  &  $-21.3$ & 20.5 & $-19.8$ & 20.3 & $-$ & $-$  &  $-18.3$  &  304.0 &  $-17.3$  &  $<120$ \\
18.04.92  &     8731  &  1.49  &  386  &  0.09  &  $-21.5$ & 14.2 & $-19.5$ & $<23$ & $-$ & $-$  &  $-18.3$  &  175.0 &  $-17.3$  &  $<65$ \\
05.07.92  &     8809  &  0.17  &  237  &  0.29  &  $-21.3$ & 4.6 & $-19.7$ & 5.5 & $-$ & $-$  &  $-18.3$  &  66.8 &  $-17.5$  &  41.6 \\
01.09.92  &     8867  &  0.16  &  69  &  0.43  &  $-$ & $-$ & $-$ & $-$ & $-$ & $-$  &  $-18.3$  &  8.0 &  $-17.3$  &  8.5 \\
23.10.92  &     8919  &  1.25  &  88  &  0.56  &  $-$ & $-$ & $-$ & $-$ & $-$ & $-$  &  $-18.2$  &  $<9$ &  $-17.2$  &  $<10$ \\
22.12.92  &     8979  &  0.22  &  41  &  0.71  &  $-$ & $-$ & $-$ & $-$ & $-$ & $-$  &  $-18.3$  &  4.2 &  $-17.4$  &  5.9 \\
26.01.93  &     9014  &  0.66  &  63  &  0.79  &  $-$ & $-$ & $-$ & $-$ & $-$ & $-$  &  $-18.3$  &  4.0 &  $-17.2$  &  9.8 \\
20.04.93  &     9098  &  1.07  &  77  &  0.00  &  $-21.8$ & 3.1 & $-$ & $-$ & $-18.9$  &  7.2  &  $-$ & $-$ &  $-17.6$  &  14.3 \\
21.04.93  &     9099  &  0.15  &  145  &  0.00  &  $-21.5$ & 4.1 & $-19.8$ & 2.0 & $-$ & $-$  &  $-18.4$  &  15.2 &  $-17.4$  &  24.5 \\
13.05.93  &     9121  &  1.17  &  89  &  0.06  &  $-21.6$ & 2.5 & $-$ & $-$ & $-$ & $-$  &  $-18.5$  &  9.3 &  $-17.2$  &  14.2 \\
03.11.93  &     9295  &  0.71  &  82  &  0.49  &  $-$ & $-$ & $-$ & $-$ & $-$ & $-$  &  $-$  &  $-$ &  $-17.0$  &  $<6$ \\
30.11.93  &     9322  &  0.38  &  53  &  0.55  &  $-$ & $-$ & $-$ & $-$ & $-$ & $-$  &  $-$  &  $-$ &  $-17.0$  &  $<5$ \\
08.03.94  &     9420  &  0.21  &  92  &  0.80  &  $-20.9$ & 1.5 & $-19.9$ & 2.9 & $-$ & $-$  &  $-18.5$  &  4.9 &  $-16.9$  &  9.4 \\
16.04.94  &     9459  &  1.18  &  45  &  0.89  &  $-$ & $-$ & $-19.8$ & 2.7 & $-$ & $-$  &  $-18.4$  &  3.3 &  $-17.2$  &  5.3 \\
08.09.94  &     9604  &  1.51  &  120  &  0.25  &  $-$ & $-$ & $-19.4$ & 11.6 & $-$ & $-$  &  $-$  &  $-$ &  $-17.0$  &  12.4 \\
28.10.94  &     9654  &  1.23  &  117  &  0.37  &  $-$ & $-$ & $-19.9$ & 8.1 & $-$ & $-$  &  $-$  &  $-$ &  $-17.4$  &  $<11$ \\
18.01.95  &     9736  &  1.08  &  54  &  0.58  &  $-$ & $-$ & $-19.6$ & 3.6 & $-$ & $-$  &  $-$  &  $-$ &  $-$  &  $-$ \\
13.03.95  &     9790  &  0.05  &  113  &  0.71  &  $-20.7$ & 2.1 & $-19.7$ & 8.8 & $-19.0$ & 5.5  &  $-18.4$  &  5.1 &  $-17.2$  &  12.0 \\
27.03.95  &     9804  &  0.29  &  126  &  0.74  &  $-20.8$ & 1.9 & $-19.7$ & 7.4 & $-19.1$ & 9.4  &  $-18.3$  &  4.5 &  $-17.2$  &  6.1 \\
03.06.95  &     9872  &  0.24  &  235  &  0.91  &  $-20.8$ & 3.0 & $-19.6$ & 14.6 & $-19.0$ & 17.8  &  $-18.3$  &  13.2 &  $-17.2$  &  32.8 \\
23.06.95  &     9892  &  0.24  &  301  &  0.96  &  $-20.8$ & 3.4 & $-19.6$ & 18.2 & $-19.0$ & 21.9  &  $-18.3$  &  15.7 &  $-17.1$  &  45.6 \\
\noalign{\smallskip}\hline
\end{tabular}
\end{sidewaystable}

\addtocounter{table}{-1}

\newpage

\begin{sidewaystable}
\caption{Maser spectral components B -- E of U\,Her (continued)}
\begin{tabular}{rrrrrrrrrrrrrrr}
\label{tab:compUHerB-E} 
           &         &         &          &  $    $ &\multicolumn{2}{c}{B}&\multicolumn{2}{c}{C}&\multicolumn{2}{c}{\Done}&\multicolumn{2}{c}{\Dtwo}&\multicolumn{2}{c}{E} \\[0.1cm]
Date & TJD & rms & $S$(tot) & $\varphi_{\rm s}$ &\multicolumn{10}{c}{\hrulefill\ $V_{\rm los}$\gm $S_{\rm p}$ \hrulefill}  \\[0.1cm]
           &         &  [Jy]   &[Jy*\kms] &         & \multicolumn{10}{c}{\hrulefill\ [km\,s$^{-1}$, Jy] \hrulefill} \\[0.1cm]
\noalign{\smallskip}\hline\noalign{\smallskip}
17.09.95  &     9978  &  0.97  &  328  &  0.17  &  $-$ & $-$ & $-$ & $-$ & $-19.1$ & 21.7  &  $-18.2$  &  $<16$ &  $-17.1$  &  59.2 \\
23.01.96  &    10106  &  1.31  &  122  &  0.49  &  $-$ & $-$ & $-$ & $-$ & $-18.9$  &  8.9  &  $-$  &  $-$ &  $-$  &  $-$ \\
01.03.96  &    10144  &  0.93  &  60  &  0.58  &  $-$ & $-$ & $-$ & $-$ & $-18.6$  &  5.3  &  $-$  &  $-$ &  $-$  &  $-$ \\
02.04.96  &    10176  &  1.75  &  100  &  0.66  &  $-$ & $-$ & $-$ & $-$ & $-18.9$  &  12.1  &  $-$  &  $-$ &  $-17.2$  &  5.1: \\
26.09.96  &    10353  &  0.33  &  273  &  0.10  &  $-20.8$ & 5.2 & $-$ & $-$ & $-18.8$  &  74.3  &  $-$  &  $-$ &  $-16.9$  &  14.1 \\
06.02.97  &    10486  &  0.69  &  136  &  0.43  &  $-$ & $-$ & $-$ & $-$ & $-18.8$  &  5.3  &  $-$  &  $-$ &  $-16.5$  &  5.9 \\
12.03.97  &    10520  &  0.53  &  106  &  0.51  &  $-$ & $-$ & $-$ & $-$ & $-$  &  $-$  &  $-$  &  $<$3 &  $-16.4$  &  7.0 \\
14.03.97  &    10522  &  0.14  &  118  &  0.52  &  $-$ & $-$ & $-$ & $-$ & $-$  &  $-$  &  $-18.2$  &  2.8 &  $-16.5$  &  5.8 \\
03.05.97  &    10572  &  0.74  &  73  &  0.64  &  $-$ & $-$ & $-$ & $-$ & $-$  &  $-$  &  $-$  &  $<$3 &  $-16.9$  &  3.8 \\
23.10.97  &    10745  &  0.57  &  99  &  0.07  &  $-$ & $-$ & $-$ & $-$ & $-$  &  $-$  &  $-18.7$  &  8.2 &  $-17.0$  &  12.1 \\
16.12.97  &    10799  &  0.47  &  171  &  0.20  &  $-$ & $-$ & $-$ & $-$ & $-$  &  $-$  &  $-18.7$  &  9.4 &  $-16.8$  &  22.6 \\
29.01.98  &    10843  &  0.55  &  126  &  0.31  &  $-$ & $-$ & $-$ & $-$ & $-$  &  $-$  &  $-18.7$  &  4.5 &  $-16.8$  &  11.3 \\
20.03.98  &    10893  &  0.47  &  128  &  0.43  &  $-$ & $-$ & $-$ & $-$ & $-$  &  $-$  &  $-18.6$  &  2.7 &  $-16.7$  &  8.0 \\
07.04.98  &    10911  &  0.79  &  105  &  0.48  &  $-$ & $-$ & $-$ & $-$ & $-$  &  $-$  &  $-18.8$  &  3.0: &  $-16.7$  &  5.4 \\
11.05.98  &    10945  &  0.93  &  78  &  0.56  &  $-$ & $-$ & $-$ & $-$ & $-$  &  $-$  &  $-18.2$  &  2.2 &  $-16.6$  &  4.9 \\
12.12.98  &    11160  &  0.45  &  151  &  0.09  &  $-$ & $-$ & $-$ & $-$ & $-$  &  $-$  &  $-18.5$  &  10.8 &  $-17.1$  &  16.0 \\
19.01.99  &    11198  &  0.74  &  132  &  0.19  &  $-$ & $-$ & $-$ & $-$ & $-$  &  $-$  &  $-18.6$  &  7.5 &  $-16.7$  &  13.4 \\
18.03.99  &    11256  &  0.43  &  121  &  0.33  &  $-$ & $-$ & $-$ & $-$ & $-$  &  $-$  &  $-18.4$  &  6.2 &  $-16.9$  &  9.7 \\
12.05.99  &    11311  &  0.75  &  115  &  0.46  &  $-$ & $-$ & $-$ & $-$ & $-$  &  $-$  &  $-$  &  $<$5 &  $-16.5$  &  7.7 \\
31.07.99  &    11391  &  0.20  &  76  &  0.66  &  $-$ & $-$ & $-$ & $-$ & $-$  &  $-$  &  $-18.2$  &  2.7 &  $-16.7$  &  4.0 \\
29.10.99  &    11481  &  1.66  &  143  &  0.88  &  $-$ & $-$ & $-$ & $-$ & $-$  &  $-$  &  $-18.5$  &  7.6 &  $-16.9$  &  15.7 \\
28.12.99  &    11541  &  0.30  &  198  &  0.03  &  $-$ & $-$ & $-$ & $-$ & $-$  &  $-$  &  $-18.3$  &  9.8 &  $-17.0$  &  31.8 \\
15.01.00  &    11559  &  0.80  &  260  &  0.08  &  $-$ & $-$ & $-$ & $-$ & $-$  &  $-$  &  $-18.4$  &  11.4 &  $-17.0$  &  41.6 \\
05.04.00  &    11640  &  0.92  &  275  &  0.28  &  $-$ & $-$ & $-$ & $-$ & $-$  &  $-$  &  $-18.5$  &  7.0 &  $-17.0$  &  34.9 \\
27.10.00  &    11845  &  0.82  &  93  &  0.78  &  $-$ & $-$ & $-$ & $-$ & $-$  &  $-$  &  $-$  &  $<$3 &  $-16.4$  &  7.1: \\
18.12.00  &    11897  &  0.73  &  72  &  0.91  &  $-$ & $-$ & $-$ & $-$ & $-$  &  $-$  &  $-$  &  $-$ &  $-16.8$  &  6.1 \\
27.01.01  &    11937  &  0.77  &  86  &  0.01  &  $-$ & $-$ & $-$ & $-$ & $-$  &  $-$  &  $-18.5$  &  5.8 &  $-17.1$  &  9.9 \\
19.04.01  &    12019  &  1.27  &  118  &  0.21  &  $-$ & $-$ & $-$ & $-$ & $-$  &  $-$  &  $-18.5$  &  5.9 &  $-17.3$  &  14.3 \\
03.05.01  &    12033  &  1.85  &  89  &  0.25  &  $-$ & $-$ & $-$ & $-$ & $-$  &  $-$  &  $-18.6$  &  6.0 &  $-17.2$  &  11.0 \\
19.09.01  &    12172  &  0.41  &  51  &  0.59  &  $-$ & $-$ & $-$ & $-$ & $-$  &  $-$  &  $-$  &  $-$ &  $-$  &  $-$ \\
24.10.01  &    12207  &  1.02  &  52  &  0.68  &  $-$ & $-$ & $-$ & $-$ & $-$  &  $-$  &  $-$  &  $-$ &  $-16.4$: &  4.1 \\
28.01.02  &    12303  &  0.86  &  46  &  0.91  &  $-$ & $-$ & $-$ & $-$ & $-$  &  $-$  &  $-$  &  $-$ &  $-17.1$  &  6.4 \\
20.03.02  &    12354  &  0.89  &  63  &  0.04  &  $-$ & $-$ & $-$ & $-$ & $-$  &  $-$  &  $-$  &  $-$ &  $-17.0$  &  14.4 \\
24.04.02  &    12389  &  1.05  &  96  &  0.13  &  $-$ & $-$ & $-$ & $-$ & $-$  &  $-$  &  $-$  &  $-$ &  $-17.1$  &  21.2 \\
20.06.02  &    12446  &  0.44  &  107  &  0.27  &  $-$ & $-$ & $-$ & $-$ & $-$  &  $-$  &  $-18.2$  &  1.6 &  $-17.0$  &  15.7 \\
26.06.02  &    12452  &  1.21  &  88  &  0.28  &  $-$ & $-$ & $-$ & $-$ & $-$  &  $-$  &  $-$  &  $-$ &  $-17.0$  &  12.8 \\
01.10.02  &    12549  &  0.69  &  66  &  0.52  &  $-$ & $-$ & $-$ & $-$ & $-$  &  $-$  &  $-$  &  $-$ &  $-$  &  $-$ \\
24.10.02  &    12572  &  0.88  &  66  &  0.58  &  $-$ & $-$ & $-$ & $-$ & $-$  &  $-$  &  $-$  &  $-$ &  $-17.0$  &  4.5 \\
19.12.02  &    12628  &  0.77  &  75  &  0.72  &  $-$ & $-$ & $-$ & $-$ & $-$  &  $-$  &  $-$  &  $-$ &  $-16.8$  &  7.6 \\
14.01.03  &    12654  &  1.36  &  80  &  0.78  &  $-$ & $-$ & $-$ & $-$ & $-$  &  $-$  &  $-$  &  $-$ &  $-17.0$  &  7.2 \\
02.04.03  &    12732  &  0.67  &  132  &  0.97  &  $-$ & $-$ & $-$ & $-$ & $-$  &  $-$  &  $-18.0$  &  4.6 &  $-17.1$  &  16.6 \\
\noalign{\smallskip}\hline
\end{tabular}
\end{sidewaystable}

\addtocounter{table}{-1}

\newpage

\begin{sidewaystable}
\caption{Maser spectral components B -- E of U\,Her (continued)}
\begin{tabular}{rrrrrrrrrrrrrrr}
\label{tab:compUHerB-E} 
           &         &         &          &  $    $ &\multicolumn{2}{c}{B}&\multicolumn{2}{c}{C}&\multicolumn{2}{c}{\Done}&\multicolumn{2}{c}{\Dtwo}&\multicolumn{2}{c}{E} \\[0.1cm]
Date & TJD & rms & $S$(tot) & $\varphi_{\rm s}$ &\multicolumn{10}{c}{\hrulefill\ $V_{\rm los}$\gm $S_{\rm p}$ \hrulefill}  \\[0.1cm]
           &         &  [Jy]   &[Jy*\kms] &         & \multicolumn{10}{c}{\hrulefill\ [km\,s$^{-1}$, Jy] \hrulefill} \\[0.1cm]
\noalign{\smallskip}\hline\noalign{\smallskip}
19.11.03  &    12963  &  1.85  &  70  &  0.54  &  $-$ & $-$ & $-$ & $-$ & $-$  &  $-$  &  $-$  &  $-$ &  $-$  &  $-$ \\
24.01.04  &    13029  &  0.66  &  52  &  0.72  &  $-$ & $-$ & $-$ & $-$ & $-$  &  $-$  &  $-$  &  $-$ &  $-16.7$:  &  \\
31.03.04  &    13096  &  0.83  &  60  &  0.87  &  $-$ & $-$ & $-$ & $-$ & $-$  &  $-$  &  $-$  &  $-$ &  $-16.9$  &  5.7 \\
11.05.04  &    13137  &  0.45  &  71  &  0.97  &  $-$ & $-$ & $-$ & $-$ & $-$  &  $-$  &  $-$  &  $-$ &  $-17.0$  &  10.0 \\
18.06.04  &    13175  &  0.57  &  103  &  0.07  &  $-$ & $-$ & $-$ & $-$ & $-$  &  $-$  &  $-$  &  $-$ &  $-17.1$  &  14.2 \\
17.09.04  &    13266  &  0.94  &  81  &  0.29  &  $-$ & $-$ & $-$ & $-$ & $-$  &  $-$  &  $-$  &  $-$ &  $-17.0$  &  9.7 \\
18.12.04  &    13358  &  1.44  &  39  &  0.52  &  $-$ & $-$ & $-$ & $-$ & $-$  &  $-$  &  $-$  &  $-$ &  $-$  &  $-$ \\
12.01.05  &    13383  &  1.17  &  40  &  0.58  &  $-$ & $-$ & $-$ & $-$ & $-$  &  $-$  &  $-$  &  $-$ &  $-$  &  $-$ \\
15.02.05  &    13417  &  0.81  &  37  &  0.66  &  $-$ & $-$ & $-$ & $-$ & $-$  &  $-$  &  $-$  &  $-$ &  $-17.0$  &  2.6 \\
13.04.05  &    13474  &  1.31  &  44  &  0.80  &  $-$ & $-$ & $-$ & $-$ & $-$  &  $-$  &  $-$  &  $-$ &  $-17.0$  &  3.8 \\
21.06.05  &    13543  &  0.71  &  63  &  0.98  &  $-$ & $-$ & $-$ & $-$ & $-$  &  $-$  &  $-$  &  $-$ &  $-17.1$  &  5.9 \\
11.07.05  &    13563  &  1.18  &  58  &  0.02  &  $-$ & $-$ & $-$ & $-$ & $-$  &  $-$  &  $-$  &  $-$ &  $-17.1$  &  5.8 \\
23.11.05  &    13698  &  0.87  &  67  &  0.36  &  $-$ & $-$ & $-$ & $-$ & $-$  &  $-$  &  $-$  &  $-$ &  $-17.1$  &  3.8 \\
14.02.06  &    13781  &  1.36  &  40  &  0.56  &  $-$ & $-$ & $-$ & $-$ & $-$  &  $-$  &  $-$  &  $-$ &  $-17.1$  &  2.2 \\
07.04.06  &    13833  &  1.44  &  45  &  0.69  &  $-$ & $-$ & $-$ & $-$ & $-$  &  $-$  &  $-$  &  $-$ &  $-$  &  $-$ \\
05.07.06  &    13922  &  0.84  &  53  &  0.91  &  $-$ & $-$ & $-$ & $-$ & $-$  &  $-$  &  $-$  &  $-$ &  $-$  &  $-$ \\
01.09.06  &    13980  &  1.24  &  98  &  0.05  &  $-$ & $-$ & $-$ & $-$ & $-$  &  $-$  &  $-$  &  $-$ &  $-17.3$  &  4.3 \\
17.10.06  &    14026  &  0.54  &  136  &  0.17  &  $-$ & $-$ & $-$ & $-$ & $-$  &  $-$  &  $-$  &  $-$ &  $-17.1$  &  6.1 \\
01.12.06  &    14071  &  1.21  &  100  &  0.28  &  $-$ & $-$ & $-$ & $-$ & $-$  &  $-$  &  $-$  &  $-$ &  $-17.4$  &  3.5 \\
17.01.07  &    14118  &  1.44  &  68  &  0.40  &  $-$ & $-$ & $-$ & $-$ & $-$  &  $-$  &  $-$  &  $-$ &  $-$  &  $-$ \\
23.02.07  &    14155  &  0.99  &  47  &  0.49  &  $-$ & $-$ & $-$ & $-$ & $-$  &  $-$  &  $-$  &  $-$ &  $-$  &  $-$ \\
10.04.07  &    14201  &  1.55  &  61  &  0.60  &  $-$ & $-$ & $-$ & $-$ & $-$  &  $-$  &  $-$  &  $-$ &  $-$  &  $-$ \\
28.06.07  &    14280  &  0.86  &  87  &  0.80  &  $-$ & $-$ & $-$ & $-$ & $-$  &  $-$  &  $-$  &  $-$ &  $-$  &  $-$ \\
24.07.07  &    14306  &  1.60  &  89  &  0.86  &  $-$ & $-$ & $-$ & $-$ & $-$  &  $-$  &  $-$  &  $-$ &  $-$  &  $-$ \\
24.08.07  &    14337  &  0.98  &  130  &  0.94  &  $-$ & $-$ & $-$ & $-$ & $-$  &  $-$  &  $-$  &  $-$ &  $-$  &  $-$ \\
15.10.07  &    14389  &  0.87  &  229  &  0.06  &  $-$ & $-$ & $-$ & $-$ & $-$  &  $-$  &  $-$  &  $-$ &  $-$  &  $-$ \\
28.11.07  &    14433  &  0.70  &  121  &  0.17  &  $-$ & $-$ & $-$ & $-$ & $-$  &  $-$  &  $-$  &  $-$ &  $-$  &  $<$5 \\
18.12.07  &    14453  &  0.85  &  107  &  0.22  &  $-$ & $-$ & $-$ & $-$ & $-$  &  $-$  &  $-$  &  $-$ &  $-$  &  $-$ \\
29.01.08  &    14495  &  1.12  &  103  &  0.33  &  $-$ & $-$ & $-$ & $-$ & $-$  &  $-$  &  $-$  &  $-$ &  $-$  &  $-$ \\
31.03.08  &    14557  &  1.43  &  80  &  0.48  &  $-$ & $-$ & $-$ & $-$ & $-$  &  $-$  &  $-$  &  $-$ &  $-$  &  $-$ \\
13.05.08  &    14600  &  1.22  &  74  &  0.59  &  $-$ & $-$ & $-$ & $-$ & $-$  &  $-$  &  $-$  &  $-$ &  $-$  &  $-$ \\
19.06.08  &    14637  &  1.49  &  55  &  0.68  &  $-$ & $-$ & $-$ & $-$ & $-$  &  $-$  &  $-$  &  $-$ &  $-17.6$  &  6.4 \\
15.07.08  &    14663  &  0.59  &  49  &  0.74  &  $-$ & $-$ & $-$ & $-$ & $-$  &  $-$  &  $-$  &  $-$ &  $-17.7$  &  12.1 \\
12.12.08  &    14813  &  1.07  &  28  &  0.11  &  $-$ & $-$ & $-$ & $-$ & $-$  &  $-$  &  $-$  &  $-$ &  $-17.4$  &  8.5 \\
03.04.09  &    14925  &  0.64  &  54  &  0.39  &  $-$ & $-$ & $-$ & $-$ & $-$  &  $-$  &  $-$  &  $-$ &  $-$  &  $-$ \\
13.05.09  &    14965  &  0.70  &  57  &  0.49  &  $-$ & $-$ & $-$ & $-$ & $-$  &  $-$  &  $-$  &  $-$ &  $-$  &  $<$2 \\
23.09.09  &    15098  &  0.55  &  34  &  0.81  &  $-$ & $-$ & $-$ & $-$ & $-$  &  $-$  &  $-$  &  $-$ &  $-$  &  $-$ \\
17.11.09  &    15153  &  0.94  &  63  &  0.95  &  $-$ & $-$ & $-$ & $-$ & $-$  &  $-$  &  $-$  &  $-$ &  $-$  &  $-$ \\
09.12.09  &    15175  &  0.43  &  76  &  0.00  &  $-$ & $-$ & $-$ & $-$ & $-$  &  $-$  &  $-$  &  $-$ &  $-$  &  $-$ \\
19.01.10  &    15216  &  0.21  &  80  &  0.11  &  $-$ & $-$ & $-$ & $-$ & $-$  &  $-$  &  $-$  &  $-$ &  $-17.2$  &  2.1 \\
02.03.10  &    15258  &  0.70  &  61  &  0.21  &  $-$ & $-$ & $-$ & $-$ & $-$  &  $-$  &  $-$  &  $-$ &  $-17.2$  &  2.5 \\
\noalign{\smallskip}\hline
\end{tabular}
\end{sidewaystable}

\addtocounter{table}{-1}

\newpage

\begin{sidewaystable}
\caption{Maser spectral components B -- E of U\,Her (continued)}
\begin{tabular}{rrrrrrrrrrrrrrr}
\label{tab:compUHerB-E} 
           &         &         &          &  $    $ &\multicolumn{2}{c}{B}&\multicolumn{2}{c}{C}&\multicolumn{2}{c}{\Done}&\multicolumn{2}{c}{\Dtwo}&\multicolumn{2}{c}{E} \\[0.1cm]
Date & TJD & rms & $S$(tot) & $\varphi_{\rm s}$ &\multicolumn{10}{c}{\hrulefill\ $V_{\rm los}$\gm $S_{\rm p}$ \hrulefill}  \\[0.1cm]
           &         &  [Jy]   &[Jy*\kms] &         & \multicolumn{10}{c}{\hrulefill\ [km\,s$^{-1}$, Jy] \hrulefill} \\[0.1cm]
\noalign{\smallskip}\hline\noalign{\smallskip}
07.04.10  &    15294  &  0.65  &  60  &  0.30  &  $-$ & $-$ & $-$ & $-$ & $-$  &  $-$  &  $-$  &  $-$ &  $-17.3$  &  2.2 \\
12.05.10  &    15329  &  0.54  &  73  &  0.39  &  $-$ & $-$ & $-$ & $-$ & $-$  &  $-$  &  $-$  &  $-$ &  $-17.2$  &  2.3 \\
08.12.10  &    15539  &  0.73  &  29  &  0.90  &  $-$ & $-$ & $-$ & $-$ & $-$  &  $-$  &  $-$  &  $-$ &  $-$  &  $-$ \\
22.02.11  &    15615  &  0.49  &  34  &  0.09  &  $-$ & $-$ & $-$ & $-$ & $-$  &  $-$  &  $-$  &  $-$ &  $-$  &  $-$ \\
20.03.11  &    15641  &  0.62  &  33  &  0.16  &  $-$ & $-$ & $-$ & $-$ & $-$  &  $-$  &  $-$  &  $-$ &  $-$  &  $-$ \\

24.02.15  &    17078  &  0.61  & 116  &  0.70  &  $-$ & $-$ & $-$ & $-$ & $-$  &  $-$  &  $-$  &  $-$ &  $-17.5$  &  7.0 \\
27.05.15  &    17170  &  1.01  & 143  &  0.93  &  $-$ & $-$ & $-$ & $-$ & $-$  &  $-$  &  $-$  &  $-$ &  $-$  &  $-$ \\
06.07.15  &    17210  &  0.61  &  95  &  0.03  &  $-$ & $-$ & $-$ & $-$ & $-$  &  $-$  &  $-$  &  $-$ &  $-17.1$  &  4.5 \\
08.09.15  &    17274  &  0.33  & 139  &  0.19  &  $-$ & $-$ & $-$ & $-$ & $-$  &  $-$  &  $-$  &  $-$ &  $-16.9$  & 11.7 \\
12.10.15  &    17308  &  0.47  & 144  &  0.27  &  $-$ & $-$ & $-$ & $-$ & $-$  &  $-$  &  $-$  &  $-$ &  $-17.1$  & 10.0 \\
\noalign{\smallskip}\hline\
\end{tabular}
\end{sidewaystable}

\newpage

\begin{sidewaystable}
\caption{Maser spectral components G -- M of U\,Her}
\begin{tabular}{rrrrrrrrrrrrrrrrr}
\label{tab:compUHerG-M} 
           &         &         &          &  $    $ &\multicolumn{2}{c}{\Gone}&\multicolumn{2}{c}{\Gtwo}&\multicolumn{2}{c}{I}&\multicolumn{2}{c}{K}&\multicolumn{2}{c}{L}&\multicolumn{2}{c}{M} \\[0.1cm]
Date & TJD & rms & $S$(tot) & $\varphi_{\rm s}$ &\multicolumn{12}{c}{\hrulefill\ $V_{\rm los}$\gm $S_{\rm p}$ \hrulefill}  \\[0.1cm]
           &         &  [Jy]   &[Jy*\kms] &         & \multicolumn{12}{c}{\hrulefill\ [km\,s$^{-1}$, Jy] \hrulefill} \\[0.1cm]
\noalign{\smallskip}\hline\noalign{\smallskip}
26.03.87  &     6881  &  3.17  &  22   & 0.53   &   $-$ & $-$ &             $-14.6$  &   18.7 & $-$ & $-$  &  $-$  &  $-$ &  $-$  &  $-$ &	$-$	&	$-$	\\
31.03.87  &     6886  &  2.49  &  27   & 0.54   &   $-$ & $-$ &             $-14.5$  &   27.3 & $-$ & $-$  &  $-$  &  $-$ &  $-$  &  $-$ &	$-$	&	$-$	\\
17.06.87  &     6964  &  3.13  &  40   & 0.73   &   $-$ & $-$ &             $-14.4$  &   33.6 & $-$ & $-$  &  $-$  &  $-$ &  $-$  &  $-$ &	$-$	&	$-$	\\
05.09.87  &     7044  &  5.14  & 113   & 0.93   &   $-16$ & $<20$ &         $-14.3$  &   62.9 & $-$ & $-$  &  $-$  &  $-$ &  $-$  &  $-$ &	$-$	&	$-$	\\
17.02.90  &     7940  &  0.12  &  433  &  0.14  & 	$-15.3$	 & 	184.6	 & 	$-14.6$	 & 	277.1	 & 	$-12.9$	 & 	3.0	 & 	$-$	 & 	$-$	& 	$-10.2$	&	2.0	&	$-$	&	$-$	 \\
31.03.90  &     7982  &  0.17  &  395  &  0.24  & 	$-15.2$	 & 	$<200$	 & 	$-14.6$	 & 	266.8	 & 	$-12.7$	 & 	3.1	 & 	$-$	 & 	$-$	& 	$-10.2$	&	2.1	&	$-$	&	$-$	 \\
24.04.90  &     8006  &  2.78  &  429  &  0.30  & 	$-15.0$	 & 	284.5	 & 	$-$	     & $-$	 & 	$-$	 & 	$-$	 & 	$-$	 & 	$-$	& 	$-$	&	$-$	&	$-$	&	$-$	 \\
12.05.90  &     8024  &  0.29  &  294  &  0.35  & 	$-14.9$	 & 	201.9	 & 	$-$	     & $-$	 & 	$-12.9$	     & 	$<3$	 & 	$-$	 & 	$-$	& 	$-10.2$	&	1.9	&	$-$	&	$-$	 \\
21.10.90  &     8186  &  0.14  &  122  &  0.75  & 	$-15.3$	 & 	$<45$	 & 	$-14.8$	 & 	68.5	 & 	$-13.1$	 & 	5.5	 & 	$-$	 & 	$-$	& 	$-10.2$	&	1.5	&	$-7.5$	&	1.1	 \\
24.10.90  &     8189  &  1.80  &   97  &  0.76  & 	$-15.0$	 & 	$<45$	 & 	$-14.8$	 & 	56.2	 & 	$-13.1$	 & 	$<6$	 & 	$-$	 & 	$-$	& 	$-$	&	$-$	&	$-$	&	$-$	 \\
18.01.91  &     8275  &  1.76  &  271  &  0.97  & 	$-$	 & 	$-$	 & 	$-14.6$	 & 	94.6	 & 	$-13.0$	 & 	$<17$	 & 	$-$	 & 	$-$	& 	$-10.2$	&	$8.0$	&	$-8.1$	&	$<4$	 \\
31.03.91  &     8347  &  0.11  &  267  &  0.15  & 	$-$	 & 	$-$	 & 	$-14.8$	 & 	80.5	 & 	$-13.0$	 & 	$<13$	 & 	$-$	 & 	$-$	& 	$-10.2$	&	4.1	&	$-8.0$	&	1.2	 \\
01.05.91  &     8378  &  0.23  &  257  &  0.22  & 	$-15.3$	 & 	$<50$	 & 	$-14.8$	 & 	73.2	 & 	$-13.0$	 & 	$<13$	 & 	$-$	 & 	$-$	& 	$-10.2$	&	2.9	&	$-8.0$	&	0.7	 \\
17.05.91  &     8394  &  2.14  &  150  &  0.26  & 	$-$	 & 	$-$	 & 	$-14.9$	 & 	43.7	 & 	$-$	 & 	$-$	 & 	$-$	 & 	$-$	& 	$-$	&	$-$	&	$-$	&	$-$	 \\
31.05.91  &     8408  &  3.83  &  238  &  0.30  & 	$-$	 & 	$-$	 & 	$-15.0$	 & 	66.5	 & 	$-$	 & 	$-$	 & 	$-$	 & 	$-$	& 	$-$	&	$-$	&	$-$	&	$-$	 \\
27.10.91  &     8557  &  0.83  &  170  &  0.66  & 	$-15.6$	 & 	16.9	 & 	$-15.0$	 & 	17.2	 & 	$-$	 & 	$-$	 & 	$-$	 & 	$-$	& 	$-$	&	$-$	&	$-$	&	$-$	 \\
02.11.91  &     8563  &  1.52  &  175  &  0.68  & 	$-15.4$	 & 	18.8	 & 	$-$	 & 	$-$	 & 	$-$	 & 	$-$	 & 	$-$	 & 	$-$	& 	$-$	&	$-$	& 	$-$	& 	$-$	 \\
11.01.92  &     8633  &  0.74  &  275  &  0.85  & 	$-15.8$	 & 	36.7	 & 	$-15.2$	 & 	34.5	 & 	$-$	 & 	$-$	 & 	$-$	 & 	$-$	& 	$-$	&	$-$	&	$-$	&	$-$	 \\
18.01.92  &     8640  &  0.18  &  480  &  0.87  & 	$-15.8$	 & 	64.1	 & 	$-15.2$	 & 	60.5	 & 	$-$	 & 	$-$	 & 	$-$	 & 	$-$	& 	$-$	&	$-$	&	$-$	&	$-$	 \\
05.02.92  &     8658  &  1.27  &  242  &  0.91  & 	$-15.7$	 & 	35.8	 & 	$-15.1$	 & 	35.8	 & 	$-$	 & 	$-$	 & 	$-$	 & 	$-$	& 	$-$	&	$-$	&	$-$	&	$-$	 \\
29.02.92  &     8682  &  0.13  &  632  &  0.97  & 	$-15.8$	 & 	99.9	 & 	$-15.2$	 & 	101.0	 & 	$-$	 & 	$-$	 & 	$-$	 & 	$-$	& 	$-$	&	$-$	&	$-$	&	$-$	 \\
18.04.92  &     8731  &  1.49  &  386  &  0.09  & 	$-15.6$	 & 	$<72$	 & 	$-15.1$	 & 	82.6	 & 	$-13.4$	 & 	5.1	 & 	$-$	 & 	$-$	& 	$-$	&	$-$	&	$-$	&	$-$	 \\
05.07.92  &     8809  &  0.17  &  237  &  0.29  & 	$-15.2$	 & 	91.3	 & 	$-$	 & 	$-$	 & 	$-$	 & 	$-$	 & 	$-$	 & 	$-$	& 	$-$	&	$-$	&	$-$	&	$-$	 \\
01.09.92  &     8867  &  0.16  &   69  &  0.43  & 	$-15.3$	 & 	47.0	 & 	$-$	 & 	$-$	 & 	$-$	 & 	$-$	 & 	$-$	 & 	$-$	& 	$-10.2$	&	0.4:	&	$-$	&	$-$	 \\
23.10.92  &     8919  &  1.25  &   88  &  0.56  & 	$-15.4$	 & 	58.0	 & 	$-$	 & 	$-$	 & 	$-$	 & 	$-$	 & 	$-$	 & 	$-$	& 	$-$	&	$-$	&	$-$	&	$-$	 \\
22.12.92  &     8979  &  0.22  &   41  &  0.71  & 	$-15.2$	 & 	24.0	 & 	$-$	 & 	$-$	 & 	$-13.5$	 & 	$<1.5$	 & 	$-$	 & 	$-$	& 	$-$	&	$-$	&	$-7.5$	&	0.6:	 \\
26.01.93  &     9014  &  0.66  &   63  &  0.79  & 	$-15.4$	 & 	39.0	 & 	$-$	 & 	$-$	 & 	$-13.5$	 & 	2.7	 & 	$-$	 & 	$-$	& 	$-$	&	$-$	&	$-$	&	$-$	 \\
20.04.93  &     9098  &  1.07  &   77  &  0.00  & 	$-15.6$	 & 	31.7	 & 	$-$	 & 	$-$	 & 	$-13.6$	 & 	7.6	 & 	$-$	 & 	$-$	& 	$-$	&	$-$	&	$-$	&	$-$	 \\
21.04.93  &     9099  &  0.15  &  145  &  0.00  & 	$-15.3$	 & 	49.8	 & 	$-$	 & 	$-$	 & 	$-13.4$	 & 	13.1	 & 	$-$	 & 	$-$	& 	$-$	&	$-$	&	$-$	&	$-$	 \\
13.05.93  &     9121  &  1.17  &   89  &  0.06  & 	$-15.4$	 & 	34.0	 & 	$-$	 & 	$-$	 & 	$-13.3$	 & 	8.4	 & 	$-$	 & 	$-$	&	$-$	&	$-$	&	$-$	&	$-$	 \\
03.11.93  &     9295  &  0.71  &   82  &  0.49  & 	$-15.4$	 & 	66.4	 & 	$-$	 & 	$-$	 & 	$-$	 & 	$-$	 & 	$-$	 & 	$-$	&	$-$	&	$-$	&	$-$	&	$-$	 \\
30.11.93  &     9322  &  0.38  &   53  &  0.55  & 	$-15.4$	 & 	38.3	 & 	$-$	 & 	$-$	 & 	$-$	 & 	$-$	 & 	$-$	 & 	$-$	&	$-$	&	$-$	&	$-$	&	$-$	 \\
08.03.94  &     9420  &  0.21  &   92  &  0.80  & 	$-15.4$	 & 	53.0	 & 	$-$	 & 	$-$	 & 	$-13.7$	 & 	1.9	 & 	$-$	 & 	$-$	&	$-$	&	$-$	&	$-$	&	$-$	 \\
16.04.94  &     9459  &  1.18  &   45  &  0.89  & 	$-15.5$	 & 	19.1	 & 	$-$	 & 	$-$	 & 	$-$	 & 	$-$	 & 	$-$	 & 	$-$	&	$-$	&	$-$	&	$-$	&	$-$	 \\
08.09.94  &     9604  &  1.51  &  120  &  0.25  & 	$-15.4$	 & 	67.0	 & 	$-$	 & 	$-$	 & 	$-$	 & 	$-$	 & 	$-$	 & 	$-$	&	$-$	&	$-$	&	$-$	&	$-$	 \\
28.10.94  &     9654  &  1.23  &  117  &  0.37  & 	$-15.5$	 & 	68.7	 & 	$-$	 & 	$-$	 & 	$-$	 & 	$-$	 & 	$-$	 & 	$-$	&	$-$	&	$-$	&	$-$	&	$-$	 \\
18.01.95  &     9736  &  1.08  &   54  &  0.58  & 	$-15.5$	 & 	40.0	 & 	$-$	 & 	$-$	 & 	$-$	 & 	$-$	 & 	$-$	 & 	$-$	&	$-$	&	$-$	&	$-$	&	$-$	 \\
13.03.95  &     9790  &  0.05  &  113  &  0.71  & 	$-15.5$	 & 	56.3	 & 	$-$	 & 	$-$	 & 	$-13.3$	 & 	5.0	 & 	$-10.8$	 & 	1.2	&	$-$	&	$-$	&	$-$	&	$-$	 \\
27.03.95  &     9804  &  0.29  &  126  &  0.74  & 	$-15.6$	 & 	63.2	 & 	$-$	 & 	$-$	 & 	$-13.4$	 & 	5.5	 & 	$-10.8$	 & 	1.4	&	$-$	&	$-$	&	$-$	&	$-$	 \\
03.06.95  &     9872  &  0.24  &  235  &  0.91  & 	$-15.6$	 & 	96.8	 & 	$-$	 & 	$-$	 & 	$-13.5$	 & 	9.5	 & 	$-10.9$	 & 	2.4	&	$-$	&	$-$	&	$-$	&	$-$	 \\
23.06.95  &     9892  &  0.24  &  301  &  0.96  & 	$-15.6$	 & 	117.6	 & 	$-$	 & 	$-$	 & 	$-13.5$	 & 	14.9	 & 	$-10.8$	 & 	2.7	&	$-$	&	$-$	&	$-$	&	$-$	 \\
\noalign{\smallskip}\hline
\end{tabular}
\end{sidewaystable}

\addtocounter{table}{-1}

\newpage

\begin{sidewaystable}
\caption{Maser spectral components G -- M of U\,Her (continued)}
\begin{tabular}{rrrrrrrrrrrrrrrrr}
\label{tab:compUHerG-M} 
           &         &         &          &  $    $ &\multicolumn{2}{c}{\Gone}&\multicolumn{2}{c}{\Gtwo}&\multicolumn{2}{c}{I}&\multicolumn{2}{c}{K}&\multicolumn{2}{c}{L}&\multicolumn{2}{c}{M} \\[0.1cm]
Date & TJD & rms & $S$(tot) & $\varphi_{\rm s}$ &\multicolumn{12}{c}{\hrulefill\ $V_{\rm los}$\gm $S_{\rm p}$ \hrulefill}  \\[0.1cm]
           &         &  [Jy]   &[Jy*\kms] &         & \multicolumn{12}{c}{\hrulefill\ [km\,s$^{-1}$, Jy] \hrulefill} \\[0.1cm]
\noalign{\smallskip}\hline\noalign{\smallskip}
17.09.95  &     9978  &  0.97  &  328  &  0.17  & 	$-15.6$	 & 	126.7	 & 	$-$	 & 	$-$	 & 	$-13.5$	 & 	$<22$	 & 	$-10.8$: & 	7.9	&	$-$	&	$-$	&	$-$	&	$-$	 \\
23.01.96  &    10106  &  1.31  &  122  &  0.49  & 	$-15.5$	 & 	70.5	 & 	$-14.6$	 & 	14.5	 & 	$-13.4$	 & 	4.4	 & 	$-$	 & 	$-$	&	$-$	&	$-$	&	$-$	&	$-$	 \\
01.03.96  &    10144  &  0.93  &   60  &  0.58  & 	$-15.5$	 & 	34.2	 & 	$-$	 & 	$-$	 & 	$-$	 & 	$-$	 & 	$-$	 & 	$-$	&	$-$	&	$-$	&	$-$	&	$-$	 \\
02.04.96  &    10176  &  1.75  &  100  &  0.66  & 	$-15.3$	 & 	49.9	 & 	$-$	 & 	$-$	 & 	$-$	 & 	$-$	 & 	$-$	 & 	$-$	&	$-$	&	$-$	&	$-$	&	$-$	 \\
26.09.96  &    10353  &  0.33  &  273  &  0.10  & 	$-15.3$	 & 	74.0	 & 	$-$	 & 	$-$	 & 	$-12.8$	 & 	13.7	 & 	$-10.9$	 & 	10.1	&	$-$	&	$-$	&	$-$	&	$-$	 \\
06.02.97  &    10486  &  0.69  &  136  &  0.43  & 	$-15.5$	 & 	86.6	 & 	$-14.4$	 & 	37.0	 & 	$-12.8$	 & 	2.8	 & 	$-10.7$	 & 	1.7	&	$-$	&	$-$	&	$-$	&	$-$	 \\
12.03.97  &    10520  &  0.53  &  106  &  0.51  & 	$-15.5$	 & 	69.1	 & 	$-14.4$	 & 	35.9	 & 	$-$	 & 	$-$	 & 	$-10.8$	 & 	2.9	&	$-$	&	$-$	&	$-$	&	$-$	 \\
14.03.97  &    10522  &  0.14  &  118  &  0.52  & 	$-15.5$	 & 	77.0	 & 	$-14.4$	 & 	39.2	 & 	$-13.0$	 & 	1.8	 & 	$-11.0$	 & 	1.5	&	$-$	&	$-$	&	$-$	&	$-$	 \\
03.05.97  &    10572  &  0.74  &   73  &  0.64  & 	$-15.5$	 & 	41.4	 & 	$-14.3$	 & 	28.9	 & 	$-12.8$	 & 	2.4	 & 	$-$	 & 	$<$2	&	$-$	&	$-$	&	$-$	&	$-$	 \\
23.10.97  &    10745  &  0.57  &   99  &  0.07  & 	$-15.7$	 & 	19.4	 & 	$-14.7$	 & 	14.1	 & 	$-12.8$	 & 	13.1	 & 	$-10.9$	 & 	10.5	&	$-$	&	$-$	&	$-$	&	$-$	 \\
16.12.97  &    10799  &  0.47  &  171  &  0.20  & 	$-15.6$	 & 	44.2	 & 	$-14.8$	 & 	32.3	 & 	$-12.8$	 & 	17.5	 & 	$-10.9$	 & 	13.7	&	$-$	&	$-$	&	$-$	&	$-$	 \\
29.01.98  &    10843  &  0.55  &  126  &  0.31  & 	$-15.5$	 & 	62.2	 & 	$-14.5$	 & 	25.9	 & 	$-12.8$	 & 	8.8	 & 	$-11.1$	 & 	6.2	&	$-$	&	$-$	&	$-$	&	$-$	 \\
20.03.98  &    10893  &  0.47  &  128  &  0.43  & 	$-15.5$	 & 	90.4	 & 	$-14.5$	 & 	33.4	 & 	$-13.1$	 & 	2.8	 & 	$-10.8$	 & 	2.1	&	$-$	&	$-$	&	$-$	&	$-$	 \\
07.04.98  &    10911  &  0.79  &  105  &  0.48  & 	$-15.5$	 & 	79.0	 & 	$-14.5$	 & 	27.7	 & 	$-$	 & 	$-$	 & 	$-$	 & 	$-$	&	$-$	&	$-$	&	$-$	&	$-$	 \\
11.05.98  &    10945  &  0.93  &   78  &  0.56  & 	$-15.5$	 & 	62.9	 & 	$-14.5$	 & 	21.5	 & 	$-$	 & 	$-$	 & 	$-$	 & 	$-$	&	$-$	&	$-$	&	$-$	&	$-$	 \\
12.12.98  &    11160  &  0.45  &  151  &  0.09  & 	$-15.7$	 & 	54.2	 & 	$-14.6$	 & 	23.5	 & 	$-13.2$	 & 	30.6	 & 	$-10.9$	 & 	5.2	&	$-$	&	$-$	&	$-$	&	$-$	 \\
19.01.99  &    11198  &  0.74  &  132  &  0.19  & 	$-15.7$	 & 	65.0	 & 	$-14.6$	 & 	23.7	 & 	$-13.1$	 & 	13.3	 & 	$-10.8$	 & 	5.6	&	$-$	&	$-$	&	$-$	&	$-$	 \\
18.03.99  &    11256  &  0.43  &  121  &  0.33  & 	$-15.6$	 & 	68.0	 & 	$-14.5$	 & 	27.2	 & 	$-13.0$	 & 	7.4	 & 	$-$	 & 	$<$3	&	$-$	&	$-$	&	$-$	&	$-$	 \\
12.05.99  &    11311  &  0.75  &  115  &  0.46  & 	$-15.5$	 & 	87.9	 & 	$-14.4$	 & 	25.3	 & 	$-$	 & 	$<$5	 & 	$-$	 & 	$-$	&	$-$	&	$-$	&	$-$	&	$-$	 \\
31.07.99  &    11391  &  0.20  &   76  &  0.66  & 	$-15.6$	 & 	59.6	 & 	$-14.5$	 & 	16.1	 & 	$-13.0$	 & 	2.2	 & 	$-11.0$	 & 	0.9	&	$-$	&	$-$	&	$-$	&	$-$	 \\
29.10.99  &    11481  &  1.66  &  143  &  0.88  & 	$-15.6$	 & 	93.4	 & 	$-14.5$	 & 	25.7	 & 	$-12.9$	 & 	8.6	 & 	$-$	 & 	$<$4	&	$-$	&	$-$	&	$-$	&	$-$	 \\
28.12.99  &    11541  &  0.30  &  198  &  0.03  & 	$-15.7$	 & 	107.3	 & 	$-14.5$	 & 	26.0	 & 	$-12.9$	 & 	14.7	 & 	$-11.0$	 & 	7.9	&	$-$	&	$-$	&	$-$	&	$-$	 \\
15.01.00  &    11559  &  0.80  &  260  &  0.08  & 	$-15.7$	 & 	142.3	 & 	$-14.7$	 & 	32.3	 & 	$-12.8$	 & 	21.1	 & 	$-10.9$	 & 	10.9	&	$-$	&	$-$	&	$-$	&	$-$	 \\
05.04.00  &    11640  &  0.92  &  275  &  0.28  & 	$-15.6$	 & 	172.7	 & 	$-14.6$	 & 	41.1	 & 	$-12.9$	 & 	19.4	 & 	$-10.8$	 & 	8.3	&	$-$	&	$-$	&	$-$	&	$-$	 \\
27.10.00  &    11845  &  0.82  &   93  &  0.78  & 	$-15.5$	 & 	81.9	 & 	$-14.5$	 & 	15.0	 & 	$-$	 & 	$-$	 & 	$-$	 & 	$-$	&	$-$	&	$-$	&	$-$	&	$-$	 \\
18.12.00  &    11897  &  0.73  &   72  &  0.91  & 	$-15.6$	 & 	53.1	 & 	$-14.5$	 & 	11.2	 & 	$-$	 & 	$-$	 & 	$-$	 & 	$-$	&	$-$	&	$-$	&	$-$	&	$-$	 \\
27.01.01  &    11937  &  0.77  &   86  &  0.01  & 	$-15.6$	 & 	50.7	 & 	$-14.5$	 & 	11.3	 & 	$-$	 & 	$-$	 & 	$-$	 & 	$-$	&	$-$	&	$-$	&	$-$	&	$-$	 \\
19.04.01  &    12019  &  1.27  &  118  &  0.21  & 	$-15.6$	 & 	76.0	 & 	$-14.4$	 & 	12.6	 & 	$-$	 & 	$-$	 & 	$-$	 & 	$-$	&	$-$	&	$-$	&	$-$	&	$-$	 \\
03.05.01  &    12033  &  1.85  &   89  &  0.25  & 	$-15.6$	 & 	64.4	 & 	$-14.5$	 & 	9.7	 & 	$-$	 & 	$-$	 & 	$-$	 & 	$-$	&	$-$	&	$-$	&	$-$	&	$-$	 \\
19.09.01  &    12172  &  0.41  &   51  &  0.59  & 	$-15.4$	 & 	54.7	 & 	$-14.6$	 & 	7.2	 & 	$-$	 & 	$-$	 & 	$-$	 & 	$-$	&	$-$	&	$-$	&	$-$	&	$-$	 \\
24.10.01  &    12207  &  1.02  &   52  &  0.68  & 	$-15.5$	 & 	48.5	 & 	$-14.5$	 & 	7.2	 & 	$-$	 & 	$-$	 & 	$-$	 & 	$-$	&	$-$	&	$-$	&	$-$	&	$-$	 \\
28.01.02  &    12303  &  0.86  &   46  &  0.91  & 	$-15.5$	 & 	27.3	 & 	$-14.3$	 & 	8.2	 & 	$-$	 & 	$-$	 & 	$-$	 & 	$-$	&	$-$	&	$-$	&	$-$	&	$-$	 \\
20.03.02  &    12354  &  0.89  &   63  &  0.04  & 	$-15.7$	 & 	19.2	 & 	$-14.6$	 & 	16.3	 & 	$-$	 & 	$-$	 & 	$-$	 & 	$-$	&	$-$	&	$-$	&	$-$	&	$-$	 \\
24.04.02  &    12389  &  1.05  &   96  &  0.13  & 	$-15.6$	 & 	31.6	 & 	$-14.5$	 & 	23.1	 & 	$-$	 & 	$-$	 & 	$-$	 & 	$-$	&	$-$	&	$-$	&	$-$	&	$-$	 \\
20.06.02  &    12446  &  0.44  &  107  &  0.27  & 	$-15.5$	 & 	57.6	 & 	$-14.5$	 & 	23.6	 & 	$-12.7$	 & 	3.4	 & 	$-$	 & 	$-$	&	$-$	&	$-$	&	$-$	&	$-$	 \\
26.06.02  &    12452  &  1.21  &   88  &  0.28  & 	$-15.5$	 & 	52.8	 & 	$-14.5$	 & 	17.8	 & 	$-$	 & 	$-$	 & 	$-$	 & 	$-$	&	$-$	&	$-$	&	$-$	&	$-$	 \\
01.10.02  &    12549  &  0.69  &   66  &  0.52  & 	$-15.4$	 & 	60.2	 & 	$-14.7$	 & 	$<$10	 & 	$-$	 & 	$-$	 & 	$-$	 & 	$-$	&	$-$	&	$-$	&	$-$	&	$-$	 \\
24.10.02  &    12572  &  0.88  &   66  &  0.58  & 	$-15.5$	 & 	50.4	 & 	$-14.5$	 & 	10.9	 & 	$-$	 & 	$-$	 & 	$-$	 & 	$-$	&	$-$	&	$-$	&	$-$	&	$-$	 \\
19.12.02  &    12628  &  0.77  &   75  &  0.72  & 	$-15.6$	 & 	44.5	 & 	$-14.6$	 & 	23.9	 & 	$-$	 & 	$-$	 & 	$-$	 & 	$-$	&	$-$	&	$-$	&	$-$	&	$-$	 \\
14.01.03  &    12654  &  1.36  &   80  &  0.78  & 	$-15.6$	 & 	41.1	 & 	$-14.5$	 & 	22.1	 & 	$-$	 & 	$-$	 & 	$-$	 & 	$-$	&	$-$	&	$-$	&	$-$	&	$-$	 \\
02.04.03  &    12732  &  0.67  &  132  &  0.97  & 	$-15.7$	 & 	39.0	 & 	$-14.5$	 & 	44.6	 & 	$-12.8$	 & 	3.6	 & 	$-11.2$	 & 	2.0:	&	$-$	&	$-$	&	$-$	&	$-$	 \\
\noalign{\smallskip}\hline
\end{tabular}
\end{sidewaystable}

\addtocounter{table}{-1}

\newpage

\begin{sidewaystable}
\caption{Maser spectral components G -- M of U\,Her (continued)}
\begin{tabular}{rrrrrrrrrrrrrrrrr}
\label{tab:compUHerG-M} 
           &         &         &          &  $    $ &\multicolumn{2}{c}{\Gone}&\multicolumn{2}{c}{\Gtwo}&\multicolumn{2}{c}{I}&\multicolumn{2}{c}{K}&\multicolumn{2}{c}{L}&\multicolumn{2}{c}{M} \\[0.1cm]
Date & TJD & rms & $S$(tot) & $\varphi_{\rm s}$ &\multicolumn{12}{c}{\hrulefill\ $V_{\rm los}$\gm $S_{\rm p}$ \hrulefill}  \\[0.1cm]
           &         &  [Jy]   &[Jy*\kms] &         & \multicolumn{12}{c}{\hrulefill\ [km\,s$^{-1}$, Jy] \hrulefill} \\[0.1cm]
\noalign{\smallskip}\hline\noalign{\smallskip}
19.11.03  &    12963  &  1.85  &   70  &  0.54  & 	$-15.5$	 & 	74.0	 & 	$-$	 & 	$-$	 & 	$-$	 & 	$-$	 & 	$-$	 & 	$-$	&	$-$	&	$-$	&	$-$	&	$-$	 \\
24.01.04  &    13029  &  0.66  &   52  &  0.72  &       $-15.5$	 & 	50.9	 & 	$-14.5$	 & 	5.9	 & 	$-$	 & 	$-$	 & 	$-$	 & 	$-$	&	$-$	&	$-$	&	$-$	&	$-$	 \\
31.03.04  &    13096  &  0.83  &   60  &  0.87  & 	$-15.5$	 & 	43.4	 & 	$-14.6$	 & 	11.6	 & 	$-12.8$	 & 	4.7	 & 	$-10.9$	 & 	3.5:	&	$-$	&	$-$	&	$-$	&	$-$	 \\
11.05.04  &    13137  &  0.45  &   71  &  0.97  & 	$-15.5$	 & 	33.7	 & 	$-14.8$	 & 	21.8	 & 	$-12.9$	 & 	7.9	 & 	$-11.0$	 & 	4.7	&	$-$	&	$-$	&	$-$	&	$-$	 \\
18.06.04  &    13175  &  0.57  &  103  &  0.07  & 	$-15.5$	 & 	29.4	 & 	$-14.6$	 & 	32.3	 & 	$-12.9$	 & 	12.6	 & 	$-11.0$	 & 	8.5	&	$-$	&	$-$	&	$-$	&	$-$	 \\
17.09.04  &    13266  &  0.94  &   81  &  0.29  & 	$-15.5$	 & 	32.5	 & 	$-14.7$	 & 	20.8	 & 	$-12.9$	 & 	5.3	 & 	$-10.9$	 & 	5.3	&	$-$	&	$-$	&	$-$	&	$-$	 \\
18.12.04  &    13358  &  1.44  &   39  &  0.52  & 	$-15.4$	 & 	31.0	 & 	$-$	 & 	$<$10	 & 	$-$	 & 	$-$	 & 	$-$	 & 	$-$	&	$-$	&	$-$	&	$-$	&	$-$	 \\
12.01.05  &    13383  &  1.17  &   40  &  0.58  & 	$-15.4$	 & 	34.0	 & 	$-$	 & 	$<$10	 & 	$-$	 & 	$-$	 & 	$-$	 & 	$-$	&	$-$	&	$-$	&	$-$	&	$-$	 \\
15.02.05  &    13417  &  0.81  &   37  &  0.66  & 	$-15.5$	 & 	30.4	 & 	$-14.7$	 & 	7.3	 & 	$-$	 & 	$-$	 & 	$-$	 & 	$-$	&	$-$	&	$-$	&	$-$	&	$-$	 \\
13.04.05  &    13474  &  1.31  &   44  &  0.80  & 	$-15.5$	 & 	34.5	 & 	$-14.5$	 & 	9.4	 & 	$-$	 & 	$-$	 & 	$-10.9$	 & 	2.9	&	$-$	&	$-$	&	$-$	&	$-$	 \\
21.06.05  &    13543  &  0.71  &   63  &  0.98  & 	$-15.5$	 & 	23.3	 & 	$-14.4$	 & 	21.0	 & 	$-13.0$	 & 	4.4	 & 	$-$	 & 	$-$	&	$-$	&	$-$	&	$-$	&	$-$	 \\
11.07.05  &    13563  &  1.18  &   58  &  0.02  & 	$-15.5$	 & 	16.3	 & 	$-14.7$	 & 	24.6	 & 	$-13.1$	 & 	3.0	 & 	$-$	 & 	$-$	&	$-$	&	$-$	&	$-$	&	$-$	 \\
23.11.05  &    13698  &  0.87  &   67  &  0.36  & 	$-15.4$	 & 	38.6	 & 	$-14.6$	 & 	29.0	 & 	$-12.8$	 & 	3.4	 & 	$-$	 & 	$-$	&	$-$	&	$-$	&	$-$	&	$-$	 \\
14.02.06  &    13781  &  1.36  &   40  &  0.56  & 	$-15.4$	 & 	31.5	 & 	$-14.4$	 & 	6.7	 & 	$-$	 & 	$-$	 & 	$-$	 & 	$-$	&	$-$	&	$-$	&	$-$	&	$-$	 \\
07.04.06  &    13833  &  1.44  &   45  &  0.69  & 	$-15.4$	 & 	26.8	 & 	$-$	 & 	$<$15	 & 	$-$	 & 	$-$	 & 	$-$	 & 	$-$	&	$-$	&	$-$	&	$-$	&	$-$	 \\
05.07.06  &    13922  &  0.84  &   53  &  0.91  & 	$-15.4$	 & 	17.8	 & 	$-14.4$	 & 	20.9	 & 	$-$	 & 	$-$	 & 	$-$	 & 	$-$	&	$-$	&	$-$	&	$-$	&	$-$	 \\
01.09.06  &    13980  &  1.24  &   98  &  0.05  & 	$-$	 & 	$-$	 & 	$-14.4$	 & 	50.5	 & 	$-$	 & 	$-$	 & 	$-11.1$	 & 	3.9	&	$-$	&	$-$	&	$-$	&	$-$	 \\
17.10.06  &    14026  &  0.54  &  136  &  0.17  & 	$-$	 & 	$-$	 & 	$-14.4$	 & 	63.4	 & 	$-$	 & 	$-$	 & 	$-11.1$	 & 	4.3	&	$-$	&	$-$	&	$-$	&	$-$	 \\
01.12.06  &    14071  &  1.21  &  100  &  0.28  & 	$-$	 & 	$-$	 & 	$-14.6$	 & 	41.2	 & 	$-$	 & 	$-$	 & 	$-$	 & 	$-$	&	$-$	&	$-$	&	$-$	&	$-$	 \\
17.01.07  &    14118  &  1.44  &   68  &  0.40  & 	$-15.5$	 & 	16.6	 & 	$-14.4$	 & 	28.4	 & 	$-$	 & 	$-$	 & 	$-$	 & 	$-$	&	$-$	&	$-$	&	$-$	&	$-$	 \\
23.02.07  &    14155  &  0.99  &   47  &  0.49  & 	$-$	 & 	$-$	 & 	$-14.8$	 & 	22.3	 & 	$-$	 & 	$-$	 & 	$-$	 & 	$-$	&	$-$	&	$-$	&	$-$	&	$-$	 \\
10.04.07  &    14201  &  1.55  &   61  &  0.60  & 	$-15.4$	 & 	20.2	 & 	$-14.3$	 & 	31.9	 & 	$-$	 & 	$-$	 & 	$-$	 & 	$-$	&	$-$	&	$-$	&	$-$	&	$-$	 \\
28.06.07  &    14280  &  0.86  &   87  &  0.80  & 	$-15.6$	 & 	21.0	 & 	$-14.2$	 & 	46.5	 & 	$-$	 & 	$-$	 & 	$-$	 & 	$<$3	&	$-$	&	$-$	&	$-$	&	$-$	 \\
24.07.07  &    14306  &  1.60  &   89  &  0.86  & 	$-15.9$	 & 	25.3	 & 	$-14.2$	 & 	46.6	 & 	$-$	 & 	$-$	 & 	$-$	 & 	$-$	&	$-$	&	$-$	&       $-$	&	$-$	 \\
24.08.07  &    14337  &  0.98  &  130  &  0.94  & 	$-16.0$	 & 	52.6	 & 	$-14.2$	 & 	57.4	 & 	$-12.9$: & 	5.7	 & 	$-10.9$	 & 	6.6	&	$-$	&	$-$	&	$-$	&	$-$	 \\
15.10.07  &    14389  &  0.87  &  229  &  0.06  & 	$-15.9$	 & 	84.1	 & 	$-14.3$	 & 	77.1	 & 	$-12.9$	 & 	11.2	 & 	$-10.9$	 & 	20.5	&	$-$	&	$-$	&	$-$	&	$-$	 \\
28.11.07  &    14433  &  0.70  &  121  &  0.17  & 	$-15.4$	 & 	25.2	 & 	$-14.3$	 & 	53.1	 & 	$-12.9$: & 	6.5	 & 	$-10.8$	 & 	12.0	&	$-9.5$:	&	2.9	&	$-$	&	$-$	 \\
18.12.07  &    14453  &  0.85  &  107  &  0.22  & 	$-14.8$	 & 	28.5	 & 	$-14.3$	 & 	36.1	 & 	$-$	 & 	$-$	 & 	$-11.0$	 & 	7.4	&	$-$	&	$-$	&	$-$	&	$-$	 \\
29.01.08  &    14495  &  1.12  &  103  &  0.33  & 	$-14.9$	 & 	29.0	 & 	$-14.3$	 & 	46.0	 & 	$-$	 & 	$-$	 & 	$-$	 & 	$-$	&	$-$	&	$-$	&	$-$	&	$-$	 \\
31.03.08  &    14557  &  1.43  &   80  &  0.48  & 	$-14.9$	 & 	23.5	 & 	$-14.4$	 & 	51.5	 & 	$-$	 & 	$-$	 & 	$-$	 & 	$-$	&	$-$	&	$-$	&	$-$	&	$-$	 \\
13.05.08  &    14600  &  1.22  &   74  &  0.59  & 	$-14.9$	 & 	24.0	 & 	$-14.3$	 & 	32.0	 & 	$-$	 & 	$-$	 & 	$-$	 & 	$-$	&	$-$	&	$-$	&	$-$	&	$-$	 \\
19.06.08  &    14637  &  1.49  &   55  &  0.68  & 	$-15.4$	 & 	14.5	 & 	$-14.4$	 & 	35.4	 & 	$-$	 & 	$-$	 & 	$-$	 & 	$-$	&	$-$	&	$-$	&	$-$	&	$-$	 \\
15.07.08  &    14663  &  0.59  &   49  &  0.74  & 	$-14.9$	 & 	14.0	 & 	$-14.3$	 & 	13.4	 & 	$-$	 & 	$-$	 & 	$-$	 & 	$-$	&	$-$	&	$-$	&	$-$	&	$-$	 \\
12.12.08  &    14813  &  1.07  &   28  &  0.11  & 	$-15.0$	 & 	6.5	 & 	$-$	 & 	$-$	 & 	$-$	 & 	$-$	 & 	$-$	 & 	$-$	&	$-$	&	$-$	&	$-$	&	$-$	 \\
03.04.09  &    14925  &  0.64  &   54  &  0.39  & 	$-15.6$	 & 	13.7	 & 	$-14.5$	 & 	32.4	 & 	$-$	 & 	$-$	 & 	$-$	 & 	$<3$	&	$-$	&	$<2$	&	$-$	&	$-$	 \\
13.05.09  &    14965  &  0.70  &   57  &  0.49  & 	$-15.7$	 & 	15.0	 & 	$-14.6$	 & 	34.5	 & 	$-$	 & 	$-$	 & 	$-$	 & 	$<3$	&	$-$	&	$<2$	&	$-$	&	$-$	 \\
23.09.09  &    15098  &  0.55  &   34  &  0.81  & 	$-15.7$	 & 	13.5	 & 	$-14.5$	 & 	9.7	 & 	$-$	 & 	$-$	 & 	$-11.1$	 & 	2.8	&	$-9.8$	&	2.1	&	$-$	&	$-$	 \\
17.11.09  &    15153  &  0.94  &   63  &  0.95  & 	$-15.8$	 & 	22.8	 & 	$-14.6$	 & 	8.2	 & 	$-$	 & 	$-$	 & 	$-10.5$	 & 	21.8	&	$-9.7$	&	13.9	&	$-$	&	$-$	 \\
09.12.09  &    15175  &  0.43  &   76  &  0.00  & 	$-15.8$	 & 	21.5	 & 	$-15.3$	 & 	13.0	 & 	$-$	 & 	$-$	 & 	$-10.7$	 & 	10.5	&	$-9.7$	&	22.2	&	$-$	&	$-$	 \\
19.01.10  &    15216  &  0.21  &   80  &  0.11  & 	$-15.8$	 & 	37.4	 & 	$-14.5$	 & 	13.1	 & 	$-$	 & 	$-$	 & 	$-10.8$	 & 	6.3	&	$-9.7$	&	16.4	&	$-$	&	$-$	 \\
02.03.10  &    15258  &  0.70  &   61  &  0.21  & 	$-15.8$	 & 	28.1	 & 	$-14.5$	 & 	18.7	 & 	$-$	 & 	$-$	 & 	$-10.9$	 & 	3.1	&	$-9.5$:	&	3.6	&	$-$	&	$-$	 \\
\noalign{\smallskip}\hline
\end{tabular}
\end{sidewaystable}

\addtocounter{table}{-1}

\newpage

\begin{sidewaystable}
\caption{Maser spectral components G -- M of U\,Her (continued)}
\begin{tabular}{rrrrrrrrrrrrrrrrr}
\label{tab:compUHerG-M} 
           &         &         &          &  $    $ &\multicolumn{2}{c}{\Gone}&\multicolumn{2}{c}{\Gtwo}&\multicolumn{2}{c}{I}&\multicolumn{2}{c}{K}&\multicolumn{2}{c}{L}&\multicolumn{2}{c}{M} \\[0.1cm]
Date & TJD & rms & $S$(tot) & $\varphi_{\rm s}$ &\multicolumn{12}{c}{\hrulefill\ $V_{\rm los}$\gm $S_{\rm p}$ \hrulefill}  \\[0.1cm]
           &         &  [Jy]   &[Jy*\kms] &         & \multicolumn{12}{c}{\hrulefill\ [km\,s$^{-1}$, Jy] \hrulefill} \\[0.1cm]
\noalign{\smallskip}\hline\noalign{\smallskip}
07.04.10  &    15294  &  0.65  &   60  &  0.30  & 	$-15.7$	 & 	22.3	 & 	$-14.5$	 & 	30.9	 & 	$-$	 & 	$-$	 & 	$-$	 & 	$-$	&	$-$	&	$-$	&	$-$	&	$-$	 \\
12.05.10  &    15329  &  0.54  &   73  &  0.39  & 	$-15.3$	 & 	24.5	 & 	$-14.4$	 & 	39.8	 & 	$-$	 & 	$-$	 & 	$-$	 & 	$-$	&	$-$	&	$-$	&	$-$	&	$-$	 \\
08.12.10  &    15539  &  0.73  &   29  &  0.90  & 	$-15.4$	 & 	12.7	 & 	$-14.3$	 & 	15.6	 & 	$-$	 & 	$-$	 & 	$-$	 & 	$-$	&	$-$	&	$-$	&	$-$	&	$-$	 \\
22.02.11  &    15615  &  0.49  &   34  &  0.09  & 	$-15.1$	 & 	12.6	 & 	$-14.4$	 & 	12.5	 & 	$-$	 & 	$-$	 & 	$-$	 & 	$-$	&	$-$	&	$-$	&	$-$	&	$-$	 \\
20.03.11  &    15641  &  0.62  &   33  &  0.16  & 	$-15.4$	 & 	12.4	 & 	$-14.4$	 & 	19.8	 & 	$-$	    &  $-$  &  $-$  &  $-$ &  $-$  &  $-$ &  $-$  &	$-$	 \\
24.02.15  &    17078  &  0.61  & 116  &  0.70   &   $-16.1$  &  29.3     & $-14.9$   &  83.7     & $-$      &  $-$  &  $-$  &  $-$ &  $-$  &  $-$ &  $-$  &  $-$ \\
27.05.15  &    17170  &  1.01  & 143  &  0.93   &   $-16.0$  &  19.3     & $-14.8$   & 139.1     & $-12.9$  &  4.7  &  $-$  &  $-$ &  $-$  &  $-$ &  $-$  &  $-$ \\
06.07.15  &    17210  &  0.61  &  95  &  0.03   &   $-15.8$  &  18.5     & $-14.8$   &  74.8     & $-13.0$  &  4.4  &  $-$  &  $-$ &  $-$  &  $-$ &  $-$  &  $-$ \\
08.09.15  &    17274  &  0.33  & 139  &  0.19   &   $-16.0$  &  55.1     & $-14.9$   &  61.1     & $-12.8$  &  5.5  &  $-$  &  $-$ &  $-$  &  $-$ &  $-$  &  $-$ \\
12.10.15  &    17308  &  0.47  & 144  &  0.27   &   $-15.9$  &  91.6     & $-14.8$   &  50.9     & $-13.0$  &  3.6  &  $-$  &  $-$ &  $-$  &  $-$ &  $-$  &  $-$ \\
\noalign{\smallskip}\hline\
\end{tabular}
\end{sidewaystable}

\twocolumn

\newpage

\section{\label{sdd_LineFitRes_Appendix_RRAql} RR Aql Spectral line fitting results}
The maser features identified between 1987 and 2015 in individual spectra are listed in Table~ \ref{tab:compRRAql}, where they were assigned to five different spectral components labeled A--E. The corresponding Gaussian fits were made using the original spectral resolution of the individual spectra.  The meaning of the columns in Table \ref{tab:compRRAql} are as in Tables \ref{tab:compUHerB-E} and \ref{tab:compUHerG-M}.  In the following we discuss the variations in flux density and velocity of the spectral components.

\subsection{The $V_{\rm los} < 28$ \kms\ velocity range}
The spectral component, which was almost always detectable in this velocity range is component A centered on $V_{\rm los} = 27$ \kms. It could be identified during the entire monitoring period 1990 -- 2011, except for a few months in 2005 and 2007. During these months emission was present at this velocity, but was too weak to be identified by the Gaussian fitting procedure as a separate maser feature. The same applies for 2015, when the velocity of component A was within the blue wing of the maser profile. Component A occasionally became the strongest feature in the spectral profile. If it reached high brightness levels ($>$150 Jy), this occurred ususally around the maximum of the visual light curve $0.0 \le \varphi_{\rm s} \le 0.3$. 

Between 1990 and the beginning of 1994 spectral component B, centered on $V_{\rm los} = 28$ \kms, was the strongest component in the profiles (see the 1991 spectrum in Fig.~\ref{fig:rraql_sel}). With brightness levels surpassing 400 Jy (0.2 < $\varphi_{\rm s} <0.4$), it reached also the highest brightness among all spectral components between 1987 and 2015. In 1994, emission at this velocity became weaker. After March 1995, blending by the neighbouring spectral components A and C became so severe that component B could not be identified unambiguously with the Gaussian fitting procedure anymore. After the observing gap 2011 -- 2015 component B was the dominating spectral component again, but with a slightly higher velocity of $V_{\rm los} = 28.5$ \kms. 

At velocities $V_{\rm los} < 27$ \kms\ the emission was always a factor of $\ge10$ weaker than spectral components A or B. No convincing spectral components covering more than a few spectra could be identified with the Gaussian fitting procedure. 

\subsection{The $28 < V_{\rm los} \le 30$ \kms\ velocity range}
This velocity range includes the stellar radial velocity $V_{\ast} = 28.5$ \kms\ (see Table \ref{centralcoords}). The prominent feature in this velocity range is spectral component C at $V_{\rm los} = 29$ \kms. 
It was always present, even in 2015, when the emission at
29 \kms\ was blended with the single feature seen at $V_{\rm los} = 28.5$ \kms\, and which we assigned to spectral component B. Component C was the strongest feature in the years 2000--2009. In 2010 and 2011 it had a similar brightness as component A, resulting sometimes in characteristic double-peaked profiles (cf. in Fig.~\ref{fig:rraql_all} in the Appendix).

For about a year in 1994/1995 we could identify a component with the Gaussian fits with a velocity of $V_{\rm los} = 29.8$ \kms. This component D is blended at other times with the stronger components C and E. Although emission is always present at this velocity, a separate feature could not be identified unambiguously with the Gaussian fit procedure. Component D is therefore similar to component B. When they are significantly weaker than their neighbouring components (or maybe even absent) they could not be identified due to blending in velocity space. 

\subsection{The $V_{\rm los} > 30$ \kms\ velocity range}
Velocities $V_{\rm los} > 30$ \kms\ are represented by component E at $V_{\rm los} = 30.5$ \kms. In general the emission is significantly weaker than those of spectral components A, B, and C. In only one occasion component E was found to be the strongest: In our first (isolated) spectrum taken 12.6.1987 and reported also by \cite{comoretto90}, this feature reached a peak flux density $\sim60$ Jy.
The maximum brightness observed during the whole monitoring time was in December 2004 / January 2005 with a flux density $\sim$65 Jy. This indicates that in 1987 the other components were unusually weak, rather than that component E was particularly bright. The component E emission was prominent in the FVt-plot and could be identified with the Gaussian fit procedure  mostly at phases after the visual maximum, when the star was bright. After May 2009 and including 2015, the spectral component disappeared from the spectra, except between March and May 2010 and in September/October 2015 (at phase $\varphi_{\rm s} = 0.0\pm0.1$ in both cases), when emission $<5$ Jy was observed at its velocity.

\newpage
\onecolumn

\begin{sidewaystable}
\caption{Maser spectral components of RR\,Aql}
\label{tab:compRRAql} 
\begin{tabular}{rrrrrrrrrrrrrrr}
           &         &         &          &  $    $ &\multicolumn{2}{c}{A}&\multicolumn{2}{c}{B}&\multicolumn{2}{c}{C}&\multicolumn{2}{c}{D}&\multicolumn{2}{c}{E} \\[0.1cm]
Date & TJD & rms & $S$ & $\varphi_s$ &\multicolumn{10}{c}{\hrulefill\ $V_{\rm los}$\gm $S_{\rm p}$ \hrulefill}  \\[0.1cm]
           &         &  [Jy]   &[Jy*\kms] &         & \multicolumn{10}{c}{\hrulefill\ [km\,s$^{-1}$, Jy] \hrulefill} \\[0.1cm]
\noalign{\smallskip}\hline\noalign{\smallskip}
   12.06.87&     6959&     3.18&       153&     0.18&  --      &  --      &      28.6&      40.8&  --      &  --      &  --      &  --      &      31.1&      57.8\\
   17.02.90&     7940&     0.99&        56&     0.63&  --      &  --      &      27.9&      17.4&      29.1&      31.5&  --      &  --      &      30.6&      12.9\\
   24.04.90&     8006&     3.29&        62&     0.80&  --      &  --      &      27.9&  $<$19.6 &      28.7&      31.5&  --      &  --      &      30.5&  $$<$$4.0\\
   21.10.90&     8186&     0.16&        67&     0.25&      26.7&       5.3&      27.9&      38.6&      29.2&      18.5&  --      &  --      &      30.6&       2.4\\
   24.10.90&     8189&     2.12&       118&     0.26&      26.7&  $<$8.0  &      27.9&      77.3&      29.1&      36.3&  --      &  --      &  --      &  --      \\
   19.01.91&     8276&     1.90&       106&     0.47&      26.8&  $<$4.1  &      27.9&     102.0&      29.0&      28.9&  --      &  --      &  --      &  --      \\
   01.05.91&     8378&     0.25&       201&     0.73&      26.9&       9.2&      28.0&     199.9&      29.0&      30.1&  --      &  --      &  --      &  --      \\
   06.05.91&     8383&     2.79&        83&     0.74&      26.8&  $<$9.0  &      27.9&      75.0&      29.0&  $<$15.0 &  --      &  --      &  --      &  --      \\
   25.10.91&     8555&     1.30&       614&     0.17&      26.7&     191.7&      27.8&     264.2&      29.1&      26.7&  --      &  --      &      30.3&      18.3\\
   18.01.92&     8640&     0.24&       629&     0.38&      26.9&      77.9&      27.9&     466.3&      29.3&     109.9&  --      &  --      &      30.4&  $<$15.0 \\
   29.02.92&     8682&     0.14&       378&     0.49&      27.1&      29.2&      27.9&     289.5&      29.2&      97.3&  --      &  --      &      30.5&       7.4\\
   19.04.92&     8732&     2.52&       161&     0.61&      26.9&  $<$20.0 &      28.0&     124.6&      29.3&      37.3&  --      &  --      &      30.5&  $<$7.0  \\
   06.07.92&     8810&     0.18&        80&     0.81&      26.7&       9.0&      27.9&      46.2&      29.3&      10.4&  --      &  --      &      30.5&       1.9\\
   02.09.92&     8868&     0.25&       400&     0.95&      27.1&     161.3&      28.2&     133.2&      29.4&      35.7&  --      &  --      &      30.6&      10.0\\
   16.10.92&     8912&     2.00&       464&     0.06&      26.9&     133.2&      28.3&     192.0&      29.3&      67.6&  --      &  --      &      30.2&      21.4\\
   22.12.92&     8979&     0.26&       476&     0.23&      27.3&      39.8&      28.3&     419.0&      29.3&      88.8&  --      &  --      &      30.4&      12.5\\
   26.01.93&     9014&     0.78&       221&     0.32&      27.2&  $<$53.0 &      28.0&     107.8&      29.2&     103.3&  --      &  --      &      30.3&      10.0\\
   21.04.93&     9099&     1.06&        78&     0.53&      27.1&  $<$17.3 &      27.9&      38.5&      29.2&      40.2&  --      &  --      &      30.4&  $<$3.0  \\
   21.04.93&     9099&     0.17&       171&     0.53&      27.1&       9.6&      27.9&      81.1&      29.2&      85.2&  --      &  --      &      30.5&       3.5\\
   02.11.93&     9294&     1.10&       114&     0.02&      27.0&       4.8&      28.1&      57.4&      29.0&      48.9&  --      &  --      &      30.4&       6.7\\
   30.11.93&     9322&     0.74&       126&     0.09&      26.9&       4.9&      28.2&      74.0&      29.2&      42.0&  --      &  --      &      30.3&       8.6\\
   08.03.94&     9420&     0.18&       202&     0.33&      27.0&       6.0&      28.0&     111.8&      29.0&     106.2&      29.8&       9.3&      30.3&       7.9\\
   16.04.94&     9459&     1.45&       147&     0.43&      27.1&  $<$3.0  &      28.0&      71.1&      29.0&      86.6&      29.7&  $<$22.0 &      30.4&  $<$3.0  \\
   18.01.95&     9736&     0.86&       121&     0.12&      27.2&      21.3&      28.1&      45.3&      29.0&      48.4&      29.7&      12.5&      30.5&      30.3\\
   09.03.95&     9787&     0.11&       196&     0.25&      27.2&      22.1&      28.1&      82.6&      29.0&      95.2&      29.7&      23.5&      30.5&      32.2\\
   03.06.95&     9872&     0.18&       248&     0.46&      26.7&     132.0&  --      &  --      &      28.7&     121.4&      29.8&       6.8&      30.4&       7.9\\
   24.06.95&     9893&     0.25&       162&     0.52&      26.8&      31.7&  --      &  --      &      28.7&     106.0&      29.9&       8.3&      30.5&       5.0\\
   03.09.96&    10330&     0.18&        79&     0.61&      27.6&       8.0&  --      &  --      &      28.8&      59.6&  --      &  --      &      30.5&  $<$1.5  \\
   25.06.97&    10625&     0.39&       189&     0.35&      26.9&      40.4&  --      &  --      &      29.1&      42.1&  --      &  --      &      30.5&      16.1\\
   20.12.00&    11899&     1.15&       116&     0.53&      27.0&      10.5&  --      &  --      &      28.6&      94.5&  --      &  --      &      30.2&       6.8\\
   18.09.01&    12171&     0.96&       425&     0.21&      26.8&     189.1&  --      &  --      &      28.6&      88.3&  --      &  --      &      30.6&      22.0\\
   24.10.01&    12207&     1.91&       231&     0.30&      26.9&      97.5&  --      &  --      &      28.6&      59.6&  --      &  --      &      30.5&      10.9\\
   23.03.02&    12357&     1.95&        44&     0.68&      27.1&       9.4&  --      &  --      &      28.6&      29.7&  --      &  --      &      30.5&  $<$4.0  \\          
   26.04.02&    12391&     1.18&        28&     0.76&      27.0&  $<$4.5  &  --      &  --      &      28.6&      17.4&  --      &  --      &      30.5&  $<$2.0  \\          
   02.10.02&    12550&     1.84&       177&     0.16&      26.9&      50.6&  --      &  --      &      28.6&      51.5&  --      &  --      &      30.6&      28.1\\
   29.10.02&    12577&     1.61&       148&     0.23&      27.0&      42.9&  --      &  --      &      28.7&      46.8&  --      &  --      &      30.5&      23.6\\
   19.12.02&    12628&     1.26&       136&     0.35&      27.1&      40.2&  --      &  --      &      28.7&      54.6&  --      &  --      &      30.3&      14.6\\
   14.01.03&    12654&     2.04&        77&     0.42&      27.1&      21.0&  --      &  --      &      28.8&      32.5&  --      &  --      &      30.4&  $<$8.0  \\
   04.04.03&    12734&     1.60&        29&     0.62&      27.3&       6.3&  --      &  --      &      28.8&      18.6&  --      &  --      &      30.4&  $<$3.0  \\
   18.11.03&    12962&     1.35&       100&     0.19&      26.8&      33.2&  --      &  --      &      28.7&      41.0&  --      &  --      &      30.5&      14.4\\
   02.04.04&    13098&     1.27&        57&     0.53&      27.1&      10.3&  --      &  --      &      28.9&      31.5&  --      &  --      &  --      &  --      \\
\noalign{\smallskip}\hline
\end{tabular}
\end{sidewaystable}

\addtocounter{table}{-1}

\newpage

\begin{sidewaystable}
\caption{Maser spectral components of RR\,Aql (continued)}
\label{tab:compRRAql} 
\begin{tabular}{rrrrrrrrrrrrrrr}
           &         &         &          &  $    $ &\multicolumn{2}{c}{A}&\multicolumn{2}{c}{B}&\multicolumn{2}{c}{C}&\multicolumn{2}{c}{D}&\multicolumn{2}{c}{E} \\[0.1cm]
Date & TJD & rms & $S$ & $\varphi_s$ &\multicolumn{10}{c}{\hrulefill\ $V_{\rm los}$\gm $S_{\rm p}$ \hrulefill}  \\[0.1cm]
           &         &  [Jy]   &[Jy*\kms] &         & \multicolumn{10}{c}{\hrulefill\ [km\,s$^{-1}$, Jy] \hrulefill} \\[0.1cm]
\noalign{\smallskip}\hline\noalign{\smallskip}
   13.05.04&    13139&     1.47&        32&     0.63&      27.1&       3.9&  --      &  --      &      28.9&      23.7&  --      &  --      &  --      &  --      \\
   19.06.04&    13176&     2.39&        56&     0.72&      27.1&  $<$5.0  &  --      &  --      &      28.9&      53.5&  --      &  --      &  --      &  --      \\
   15.09.04&    13264&     2.39&       299&     0.94&      27.2&      14.2&  --      &  --      &      28.9&     337.1&  --      &  --      &      30.7&      52.7\\
   20.12.04&    13360&     1.11&       386&     0.18&      27.3&      33.9&  --      &  --      &      28.9&     386.9&  --      &  --      &      30.6&      66.1\\
   12.01.05&    13383&     1.30&       372&     0.24&      27.4&      34.5&  --      &  --      &      28.9&     374.7&  --      &  --      &      30.6&      62.6\\
   15.02.05&    13417&     1.01&       254&     0.32&      27.3&      25.5&  --      &  --      &      28.8&     261.0&  --      &  --      &      30.7&      40.8\\
   14.04.05&    13475&     1.43&       143&     0.47&  --      &  --      &  --      &  --      &      28.9&     162.0&  --      &  --      &  --      &  --      \\
   21.06.05&    13543&     2.34&        68&     0.64&  --      &  --      &  --      &  --      &      28.9&      57.9&  --      &  --      &  --      &  --      \\
   12.07.05&    13564&     3.16&        69&     0.69&  --      &  --      &  --      &  --      &      28.9&      67.8&  --      &  --      &  --      &  --      \\
   22.11.05&    13697&     1.52&       257&     0.02&      27.0&      50.8&  --      &  --      &      28.9&     263.6&  --      &  --      &  --      &  --      \\
   15.02.06&    13782&     2.01&       312&     0.24&      27.1&      60.5&  --      &  --      &      28.9&     282.8&  --      &  --      &  --      &  --      \\
   09.04.06&    13835&     2.21&       174&     0.37&      27.2&      32.0&  --      &  --      &      28.9&     135.7&  --      &  --      &  --      &  --      \\
   07.07.06&    13924&     2.34&        34&     0.59&      27.0&       5.9&  --      &  --      &      29.1&      26.0&  --      &  --      &  --      &  --      \\
   16.10.06&    14025&     2.13&        37&     0.85&      27.3&       9.7&  --      &  --      &      29.0&      24.9&  --      &  --      &  --      &  --      \\
   30.11.06&    14070&     1.16&       119&     0.96&      27.2&      18.4&  --      &  --      &      28.9&      69.4&  --      &  --      &      30.4&      15.4\\
   16.01.07&    14117&     1.26&       236&     0.07&      27.0&      41.7&  --      &  --      &      28.9&     186.1&  --      &  --      &      30.8&      35.4\\
   22.02.07&    14154&     1.36&       236&     0.17&      27.0&      44.6&  --      &  --      &      28.9&     173.8&  --      &  --      &      30.8&      32.5\\
   11.04.07&    14202&     1.48&       167&     0.29&      27.1&      30.8&  --      &  --      &      28.9&     116.1&  --      &  --      &      30.7&      16.9\\
   27.06.07&    14279&     1.77&        66&     0.48&      27.2&      12.2&  --      &  --      &      29.1&      44.7&  --      &  --      &  --      &  --      \\
   24.07.07&    14306&     1.63&        39&     0.55&  --      &  --      &  --      &  --      &      29.2&      23.4&  --      &  --      &  --      &  --      \\
   23.08.07&    14336&     1.73&        29&     0.62&  --      &  --      &  --      &  --      &      29.2&      17.9&  --      &  --      &  --      &  --      \\
   16.10.07&    14390&     1.21&        29&     0.76&  --      &  --      &  --      &  --      &      29.2&      18.1&  --      &  --      &  --      &  --      \\
   26.11.07&    14431&     1.47&        68&     0.86&      27.0&      11.1&  --      &  --      &      29.2&      45.4&  --      &  --      &  --      &  --      \\
   19.12.07&    14454&     1.06&       116&     0.92&      26.9&      24.3&  --      &  --      &      29.1&      71.3&  --      &  --      &  --      &  --      \\
   28.01.08&    14494&     1.40&       212&     0.02&      26.8&      50.7&  --      &  --      &      29.1&     119.9&  --      &  --      &      30.7&      14.4\\
   31.03.08&    14557&     1.33&       215&     0.18&      26.9&      54.2&  --      &  --      &      29.0&     114.6&  --      &  --      &      30.7&      13.8\\
   13.05.08&    14600&     1.56&       184&     0.28&      27.0&      43.5&  --      &  --      &      29.1&      97.0&  --      &  --      &      30.6&       9.4\\
   20.06.08&    14638&     2.04&       113&     0.38&      27.1&      27.8&  --      &  --      &      29.1&      57.3&  --      &  --      &      30.7&       7.0\\
   15.07.08&    14663&     1.32&        60&     0.44&      27.3&      14.4&  --      &  --      &      29.1&      28.9&  --      &  --      &      30.7&  $<$3.0  \\
   12.12.08&    14813&     1.18&        61&     0.82&      26.8&      19.9&  --      &  --      &      29.1&      18.4&  --      &  --      &      30.8&       5.0\\
   03.04.09&    14925&     0.88&       206&     0.09&      26.7&      78.0&  --      &  --      &      28.9&      46.2&  --      &  --      &      30.9&      14.5\\
   15.05.09&    14967&     1.22&       175&     0.20&      26.8&      60.8&  --      &  --      &      28.8&      40.7&  --      &  --      &      30.9&      13.8\\
   22.09.09&    15097&     1.46&        35&     0.52&      27.2&  $<$10.0 &  --      &  --      &      28.5&      13.0&  --      &  --      &  --      &  --      \\
   18.11.09&    15154&     0.91&        15&     0.67&      27.4&       3.3&  --      &  --      &      29.1&       6.7&  --      &  --      &  --      &  --      \\
   20.12.09&    15186&     0.67&        19&     0.75&      27.7&       4.9&  --      &  --      &      29.4&       8.4&  --      &  --      &  --      &  --      \\
   19.01.10&    15216&     1.01&        25&     0.82&      27.2&       7.1&  --      &  --      &      29.2&       8.4&  --      &  --      &  --      &  --      \\
   01.03.10&    15257&     0.69&        54&     0.93&      27.0&      15.1&  --      &  --      &      29.1&      17.9&  --      &  --      &  --      &  --      \\
   07.04.10&    15294&     0.74&        84&     0.02&      26.9&      30.0&  --      &  --      &      29.1&      26.3&  --      &  --      &  --      &  --      \\
   14.05.10&    15331&     0.96&       110&     0.11&      26.9&      39.1&  --      &  --      &      29.0&      32.5&  --      &  --      &  --      &  --      \\
   08.12.10&    15539&     1.43&        16&     0.63&      27.5&  $<$4.0  &  --      &  --      &      28.8&       8.3&  --      &  --      &  --      &  --      \\
   19.01.11&    15581&     1.53&        25&     0.73&      28.2&       4.4&  --      &  --      &      29.2&      10.5&  --      &  --      &  --      &  --      \\
\noalign{\smallskip}\hline
\end{tabular}
\end{sidewaystable}

\addtocounter{table}{-1}

\newpage

\begin{sidewaystable}
\caption{Maser spectral components of RR\,Aql (continued)}
\label{tab:compRRAql} 
\begin{tabular}{rrrrrrrrrrrrrrr}
           &         &         &          &  $    $ &\multicolumn{2}{c}{A}&\multicolumn{2}{c}{B}&\multicolumn{2}{c}{C}&\multicolumn{2}{c}{D}&\multicolumn{2}{c}{E} \\[0.1cm]
Date & TJD & rms & $S$ & $\varphi_s$ &\multicolumn{10}{c}{\hrulefill\ $V_{\rm los}$\gm $S_{\rm p}$ \hrulefill}  \\[0.1cm]
           &         &  [Jy]   &[Jy*\kms] &         & \multicolumn{10}{c}{\hrulefill\ [km\,s$^{-1}$, Jy] \hrulefill} \\[0.1cm]
\noalign{\smallskip}\hline\noalign{\smallskip}
   22.02.11&    15615&     0.64&        29&     0.82&      27.5&       6.7&  --      &  --      &      29.1&      11.7&  --      &  --      &  --      &  --      \\
   20.03.11&    15641&     0.70&        55&     0.89&      27.4&      16.3&  --      &  --      &      29.1&      18.3&  --      &  --      &  --      &  --      \\
   24.02.15&    17078&     0.72&        53&     0.48&  --      &  --      &      28.2&      33.1&  --      &  --      &  --      &  --      &  --      &  --      \\
   28.05.15&    17171&     0.89&        25&     0.71&  --      &  --      &      28.5&      16.9&  --      &  --      &  --      &  --      &  --      &  --      \\
   06.07.15&    17210&     0.77&        42&     0.81&  --      &  --      &      28.5&      23.7&  --      &  --      &  --      &  --      &  --      &  --      \\
   08.09.15&    17274&     0.37&       124&     0.97&  --      &  --      &      28.5&      75.9&  --      &  --      &  --      &  --      &  --      &  --      \\
   12.10.15&    17308&     0.54&       148&     0.05&  --      &  --      &      28.5&      97.0&  --      &  --      &  --      &  --      &  --      &  --      \\
\noalign{\smallskip}\hline\
\end{tabular}
\end{sidewaystable}

\twocolumn

\newpage

\section{all maser spectra}
In this section we show all H$_2$O maser spectra of our targets. There are more spectra here than in the FVt-plots, because in those plots (apart from having averaged spectra taken within 4 days from one another) we have tried to avoid large gaps between observations, in order not to have to interpolate over large time intervals. 

\begin{figure*}
\resizebox{18cm}{!}{
\includegraphics{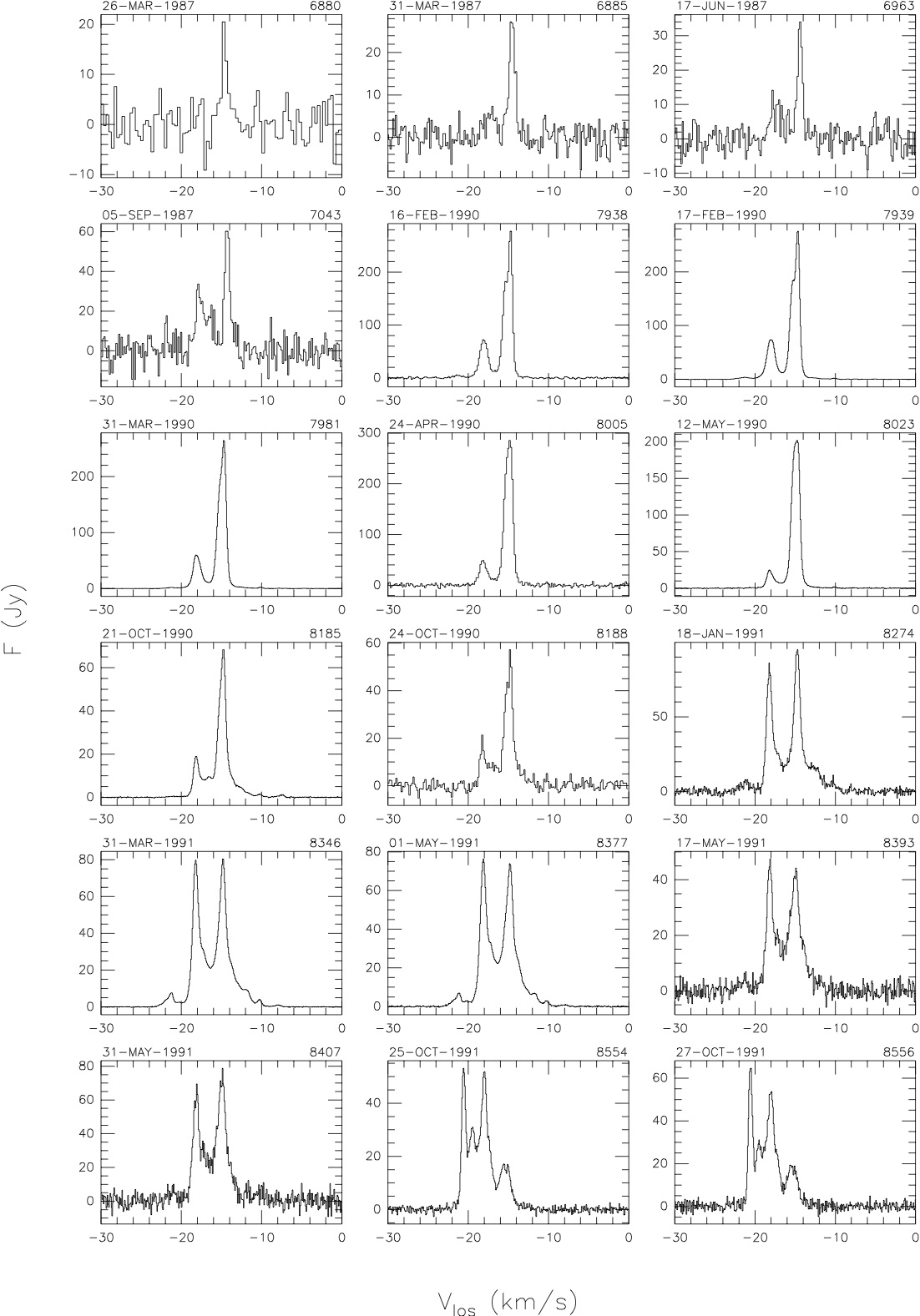}}
\caption{All H$_2$O maser spectra of U~Her. The observing date (top left) and TJD (top right) are indicated for each spectrum.}
\label{fig:uher_all}
\end{figure*}

\addtocounter{figure}{-1}

\begin{figure*}
\resizebox{18cm}{!}{
\includegraphics{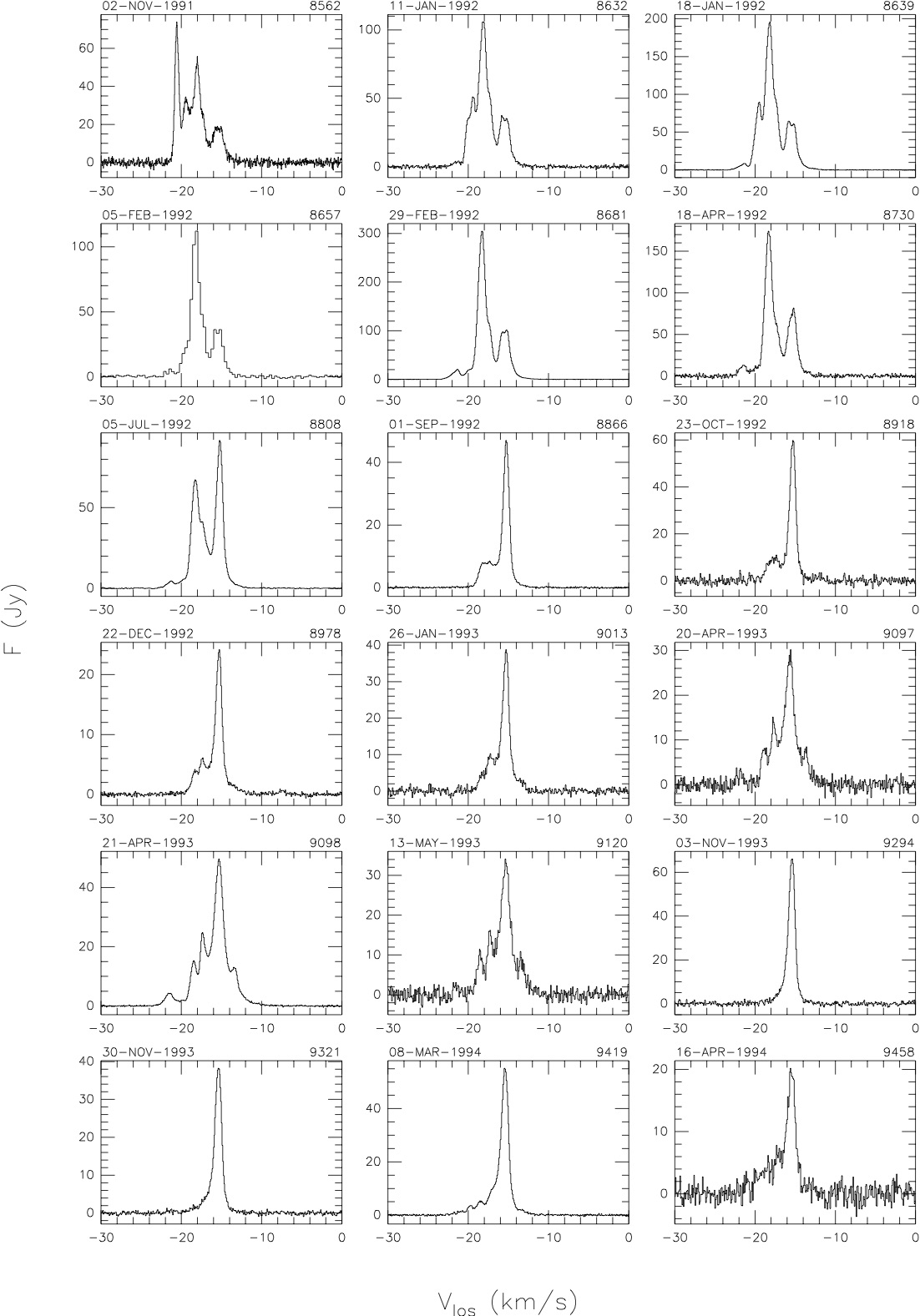}}
\caption{{\it U~Her, continued}}
\label{fig:uher_all}
\end{figure*}

\addtocounter{figure}{-1}

\begin{figure*}
\resizebox{18cm}{!}{
\includegraphics{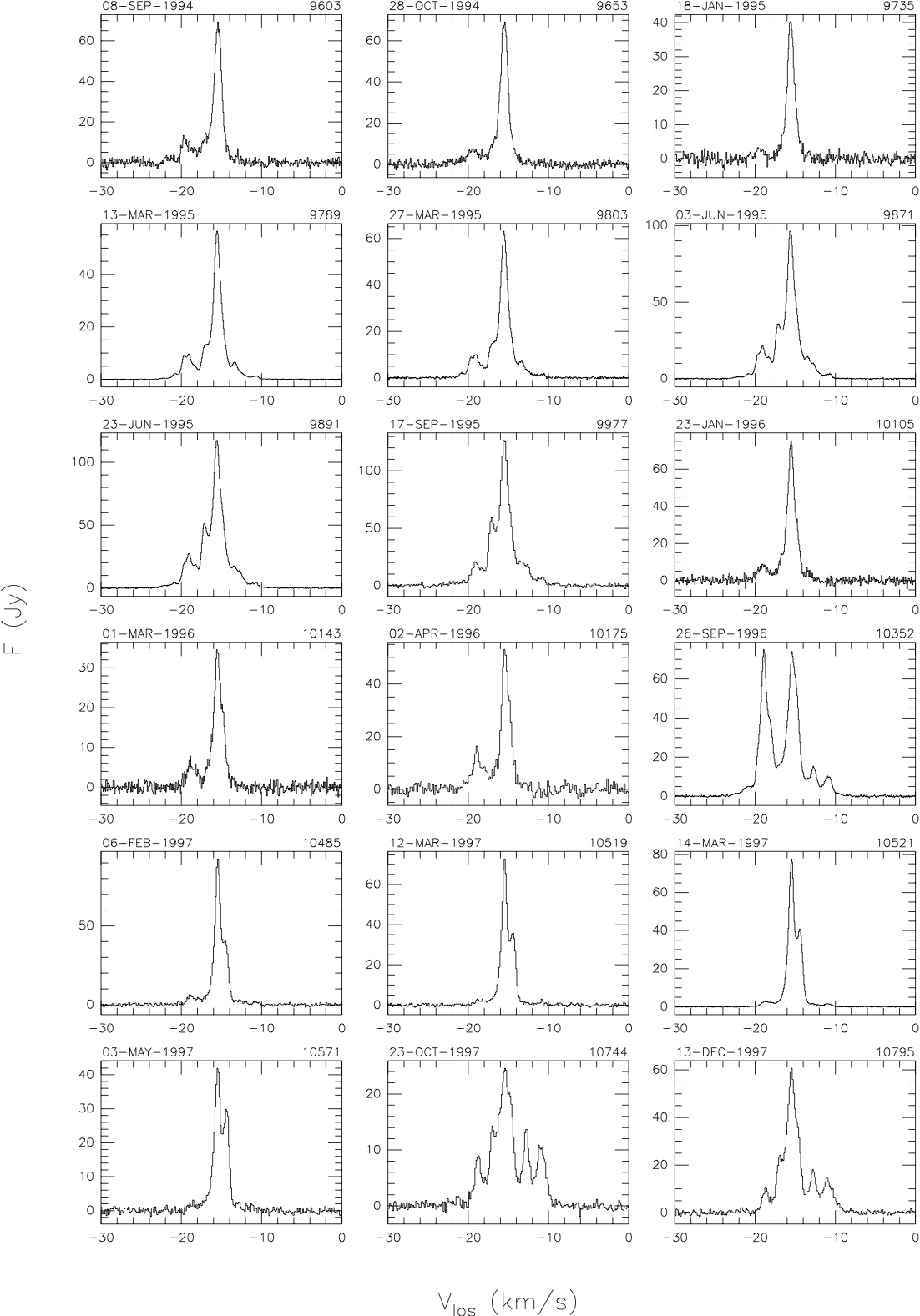}}
\caption{{\it U~Her, continued}}
\label{fig:uher_all}
\end{figure*}

\addtocounter{figure}{-1}

\begin{figure*}
\resizebox{18cm}{!}{
\includegraphics{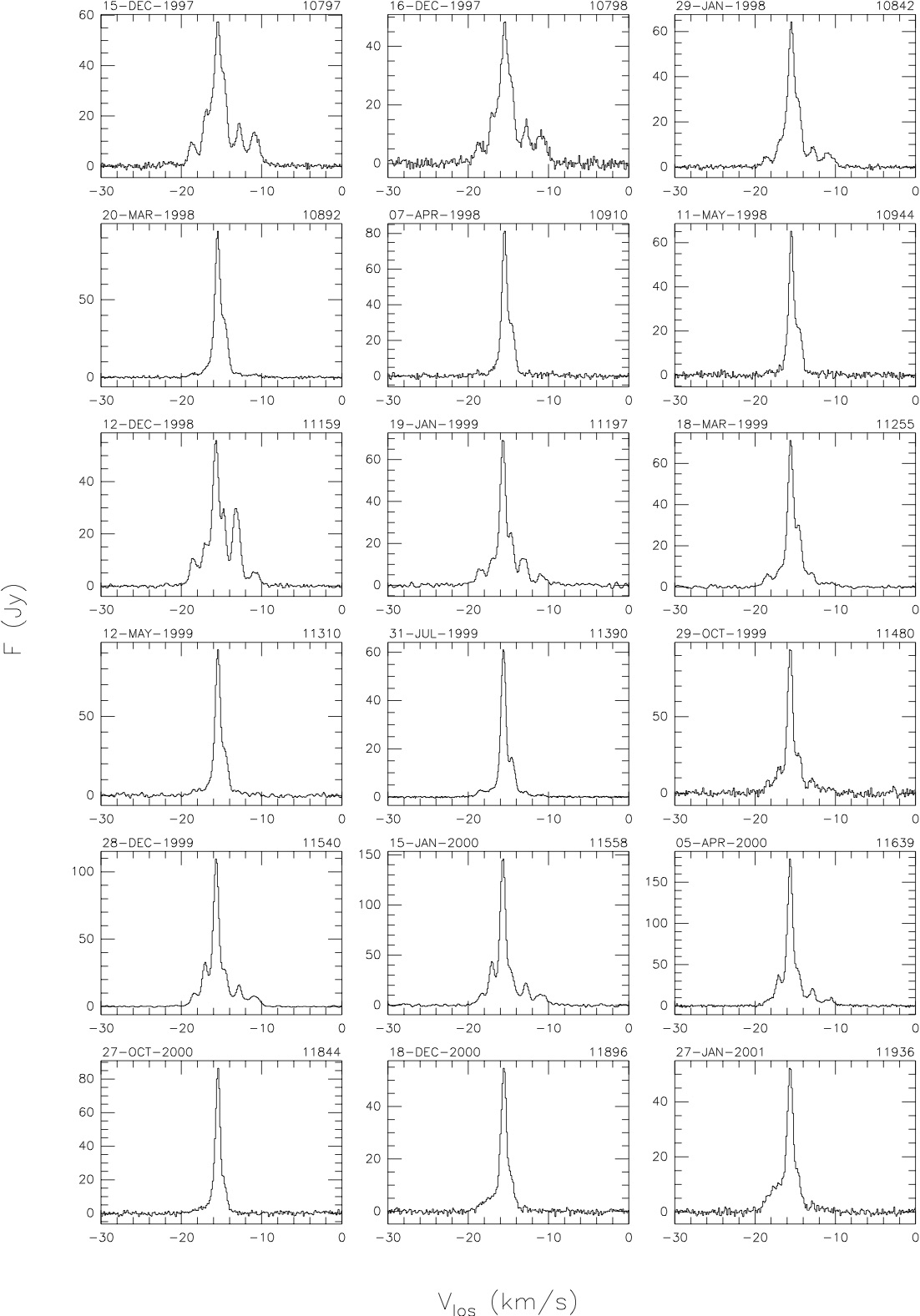}}
\caption{{\it U~Her, continued}}
\label{fig:uher_all}
\end{figure*}

\addtocounter{figure}{-1}

\begin{figure*}
\resizebox{18cm}{!}{
\includegraphics{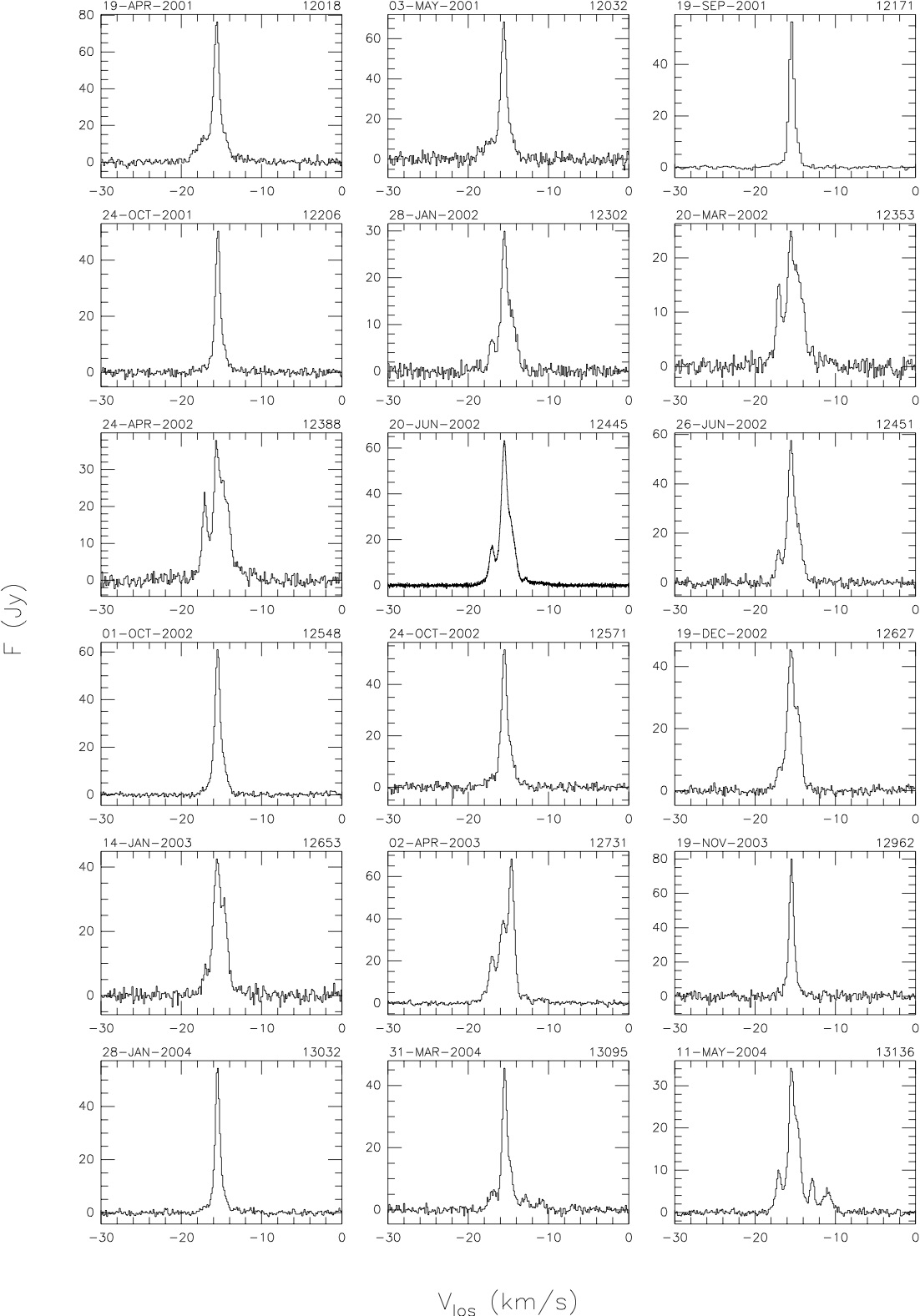}}
\caption{{\it U~Her, continued}}
\label{fig:uher_all}
\end{figure*}

\addtocounter{figure}{-1}

\begin{figure*}
\resizebox{18cm}{!}{
\includegraphics{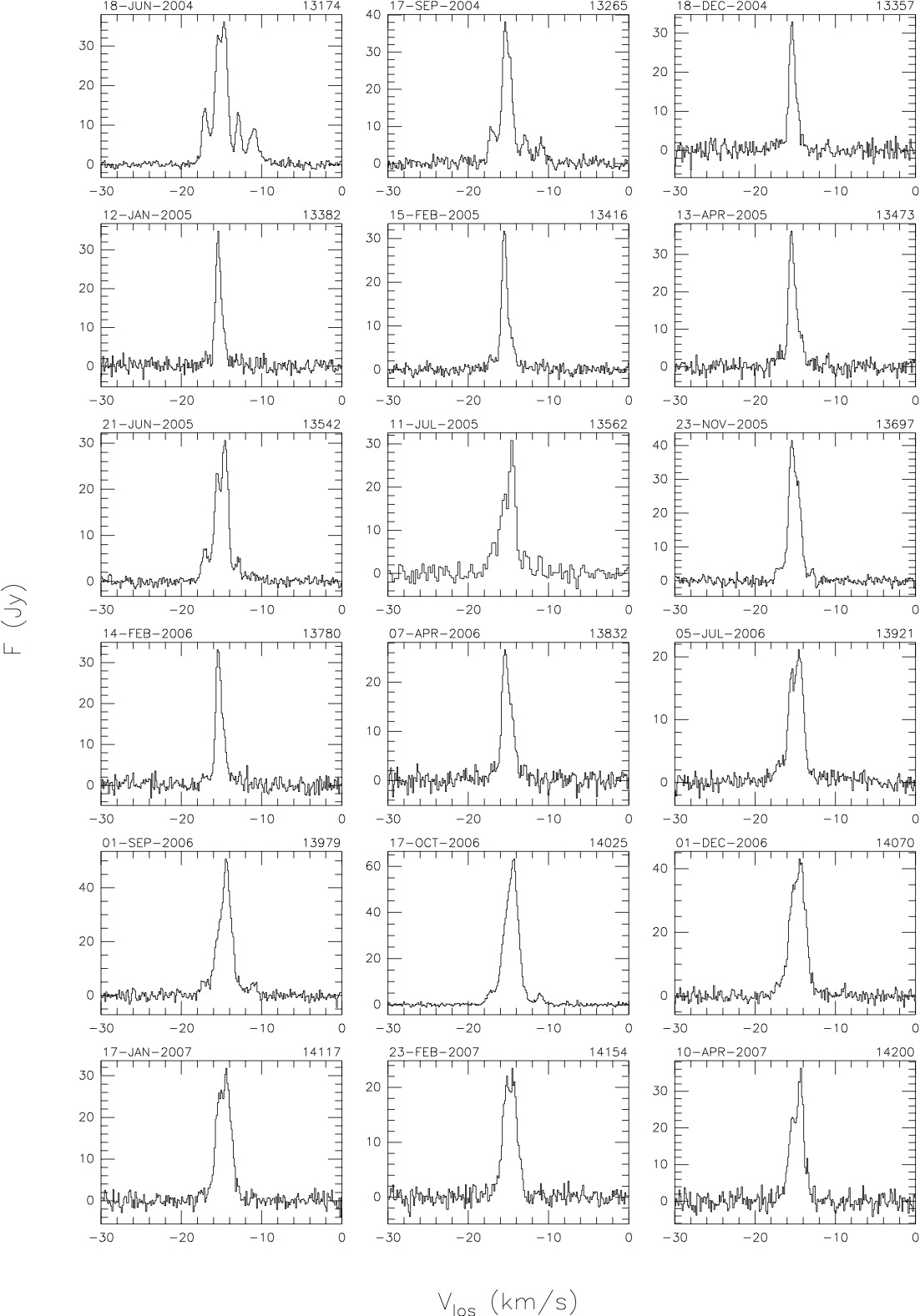}}
\caption{{\it U~Her, continued}}
\label{fig:uher_all}
\end{figure*}

\addtocounter{figure}{-1}

\begin{figure*}
\resizebox{18cm}{!}{
\includegraphics{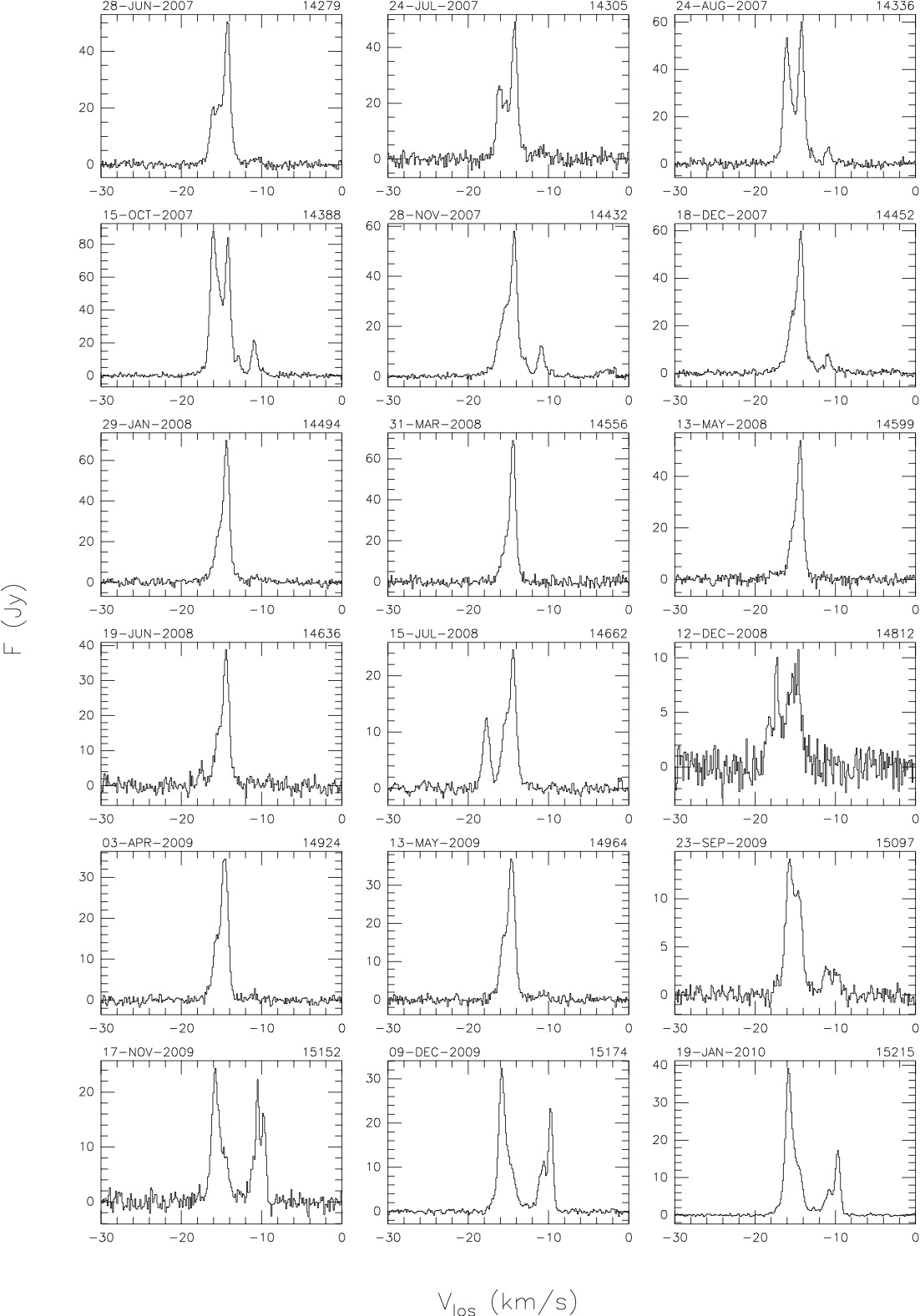}}
\caption{{\it U~Her, continued}}
\label{fig:uher_all}
\end{figure*}

\addtocounter{figure}{-1}

\begin{figure*}
\resizebox{18cm}{!}{
\includegraphics{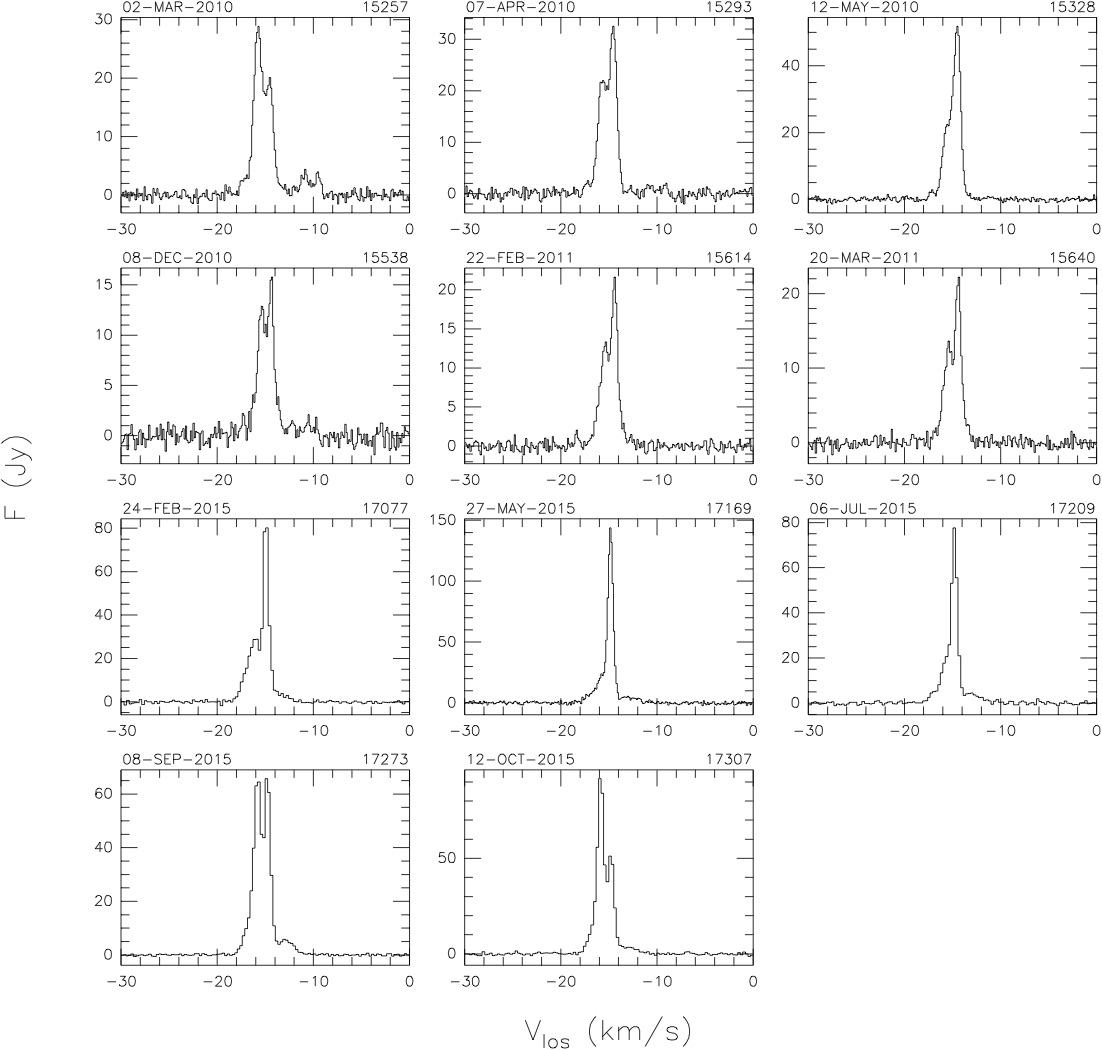}}
\caption{{\it U~Her, continued}}
\label{fig:uher_all}
\end{figure*}

\clearpage

\begin{figure*}
\resizebox{18cm}{!}{
\includegraphics{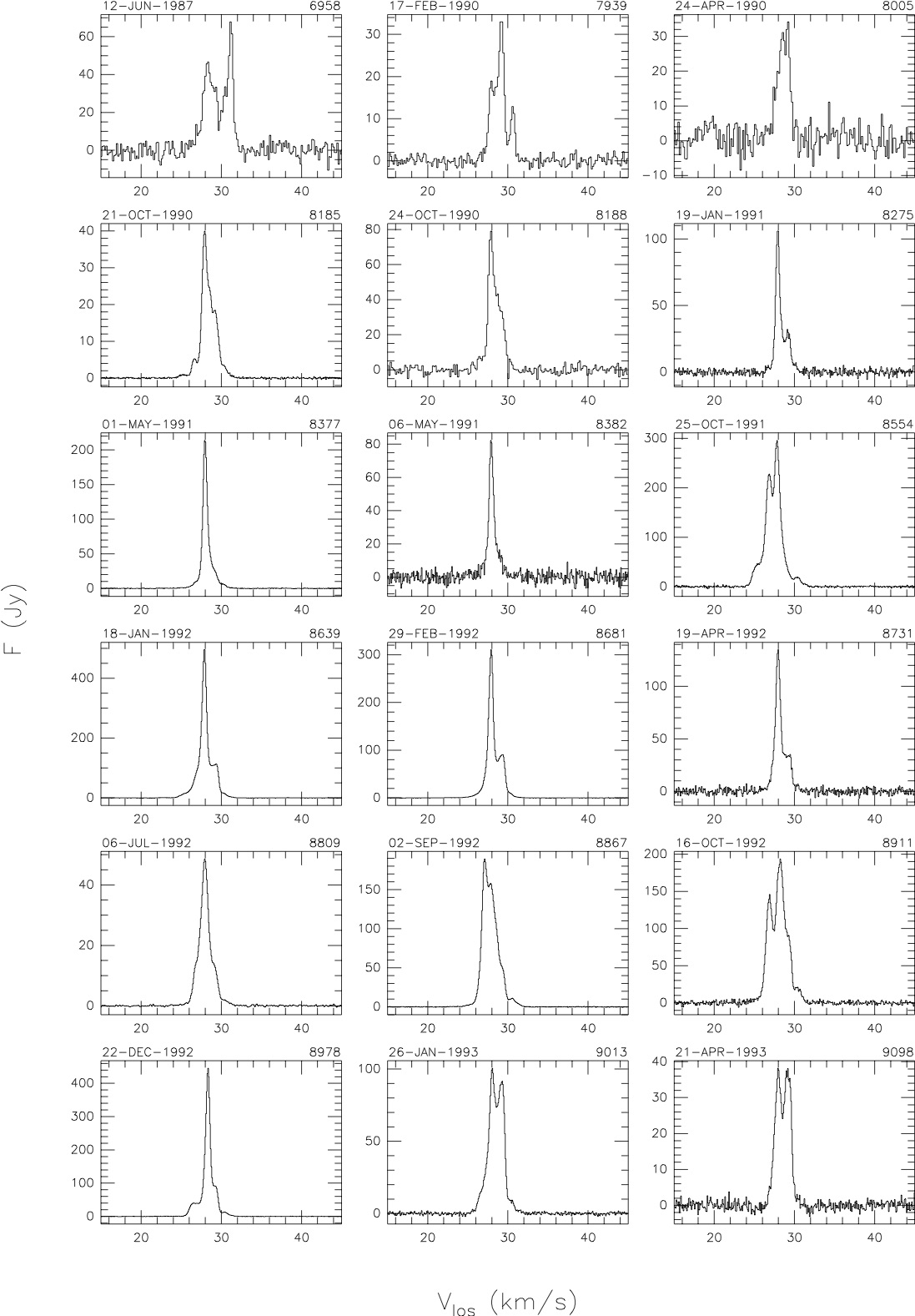}}
\caption{All H$_2$O maser spectra of RR~Aql. The observing date (top left) and TJD (top right) are indicated for each spectrum.}
\label{fig:rraql_all}
\end{figure*}

\addtocounter{figure}{-1}

\begin{figure*}
\resizebox{18cm}{!}{
\includegraphics{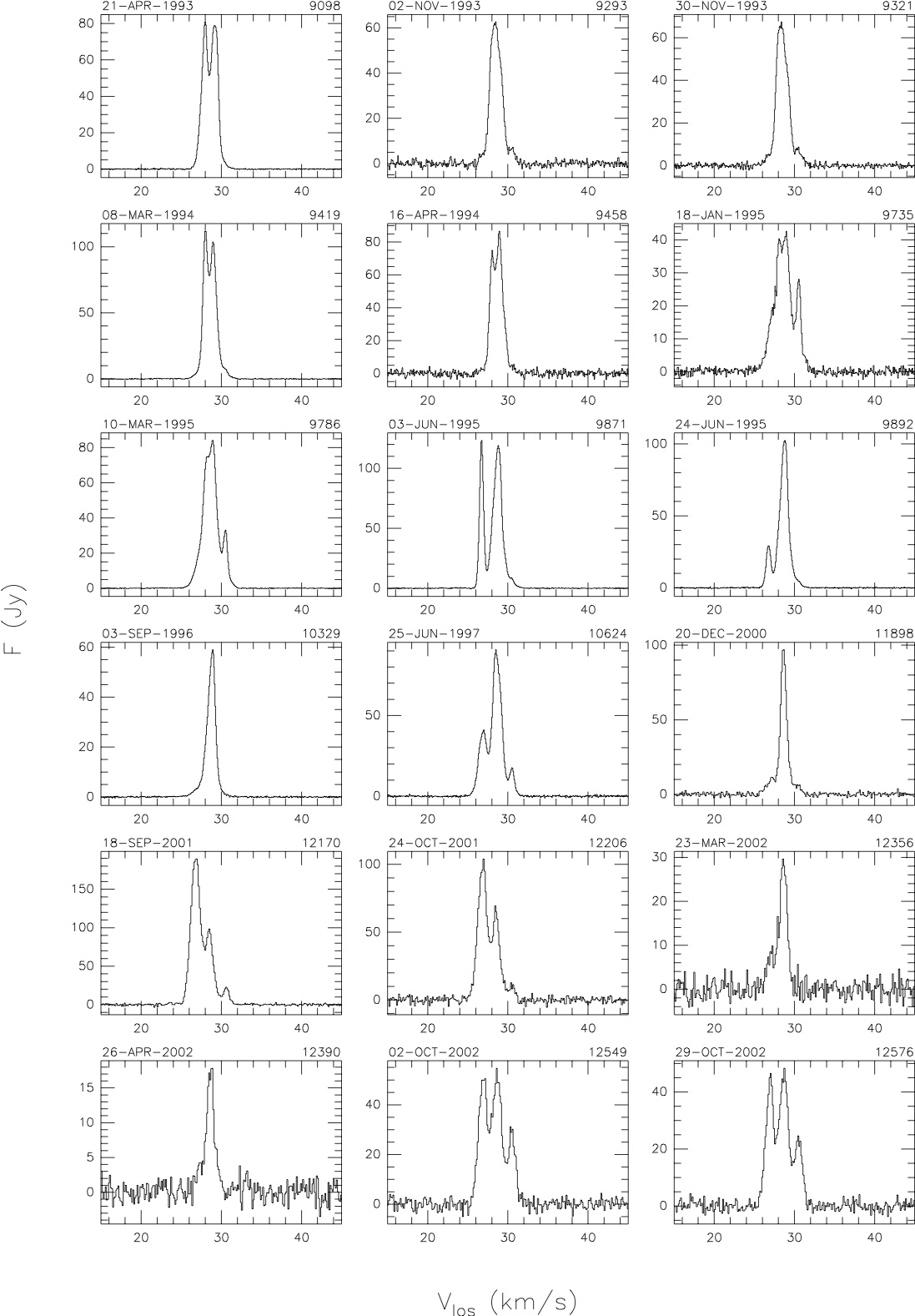}}
\caption{{\it RR~Aql, continued}}
\label{fig:rraql_all}
\end{figure*}

\addtocounter{figure}{-1}

\begin{figure*}
\resizebox{18cm}{!}{
\includegraphics{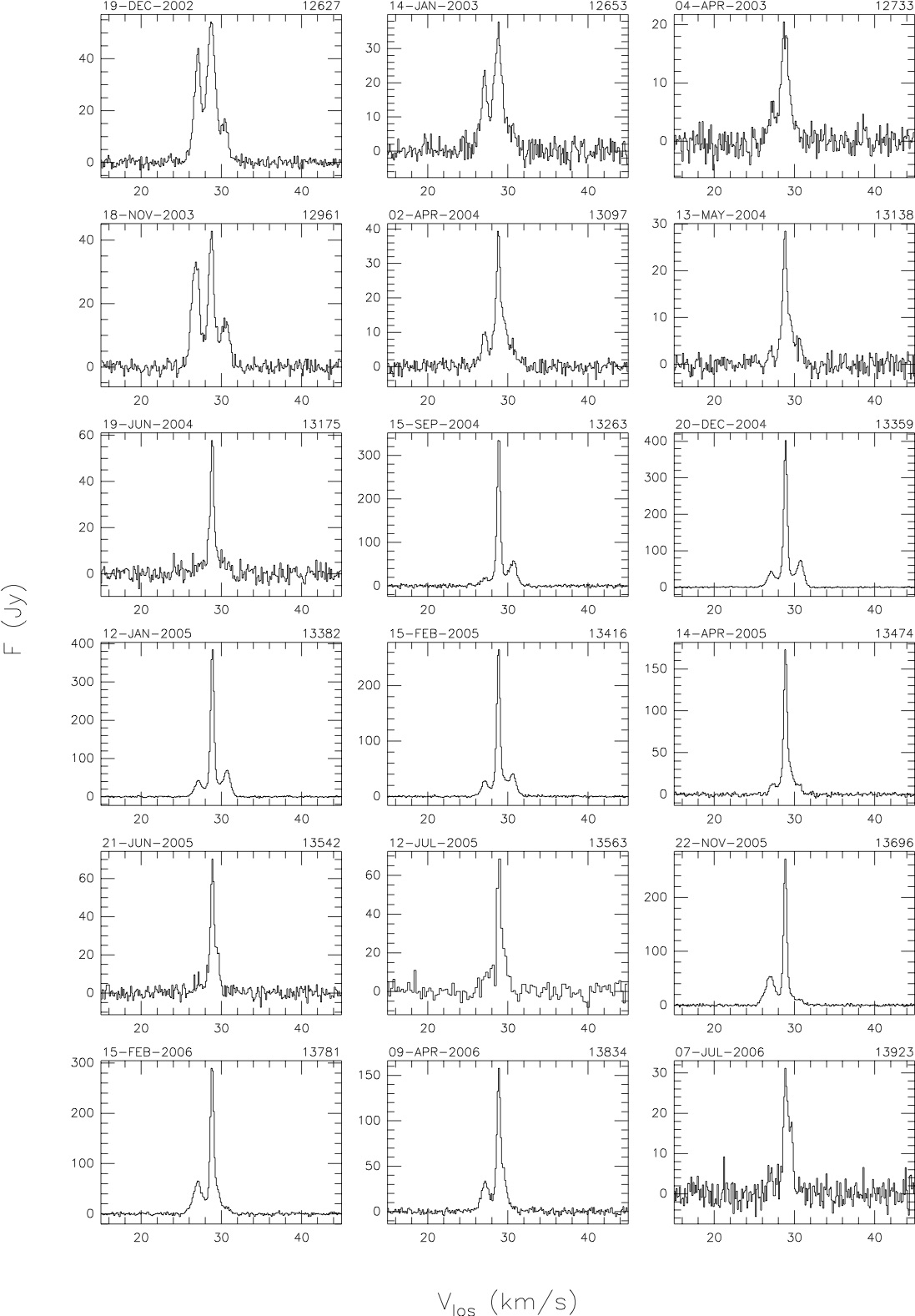}}
\caption{{\it RR~Aql, continued}}
\label{fig:rraql_all}
\end{figure*}

\addtocounter{figure}{-1}

\begin{figure*}
\resizebox{18cm}{!}{
\includegraphics{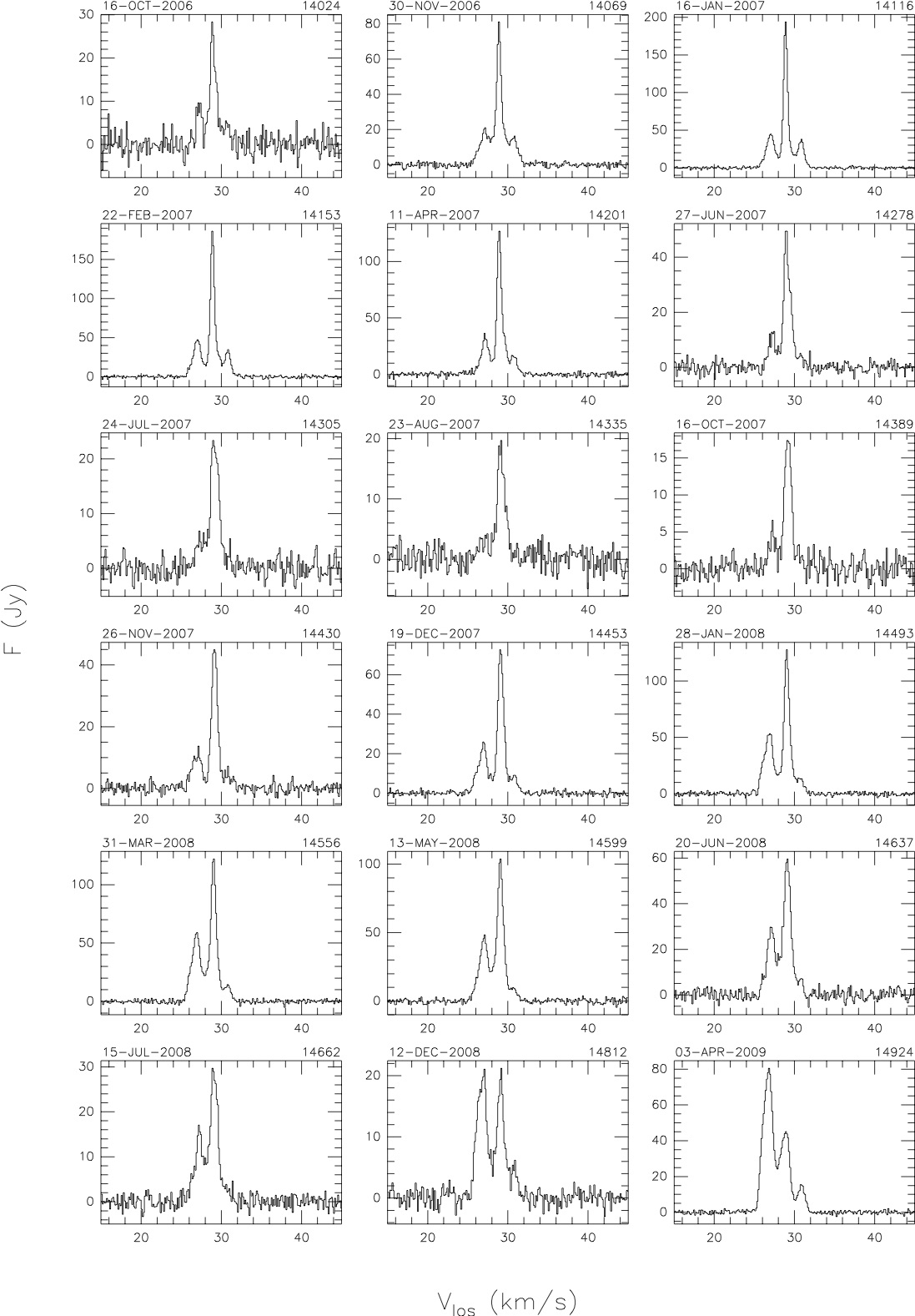}}
\caption{{\it RR~Aql, continued}}
\label{fig:rraql_all}
\end{figure*}

\addtocounter{figure}{-1}

\begin{figure*}
\resizebox{18cm}{!}{
\includegraphics{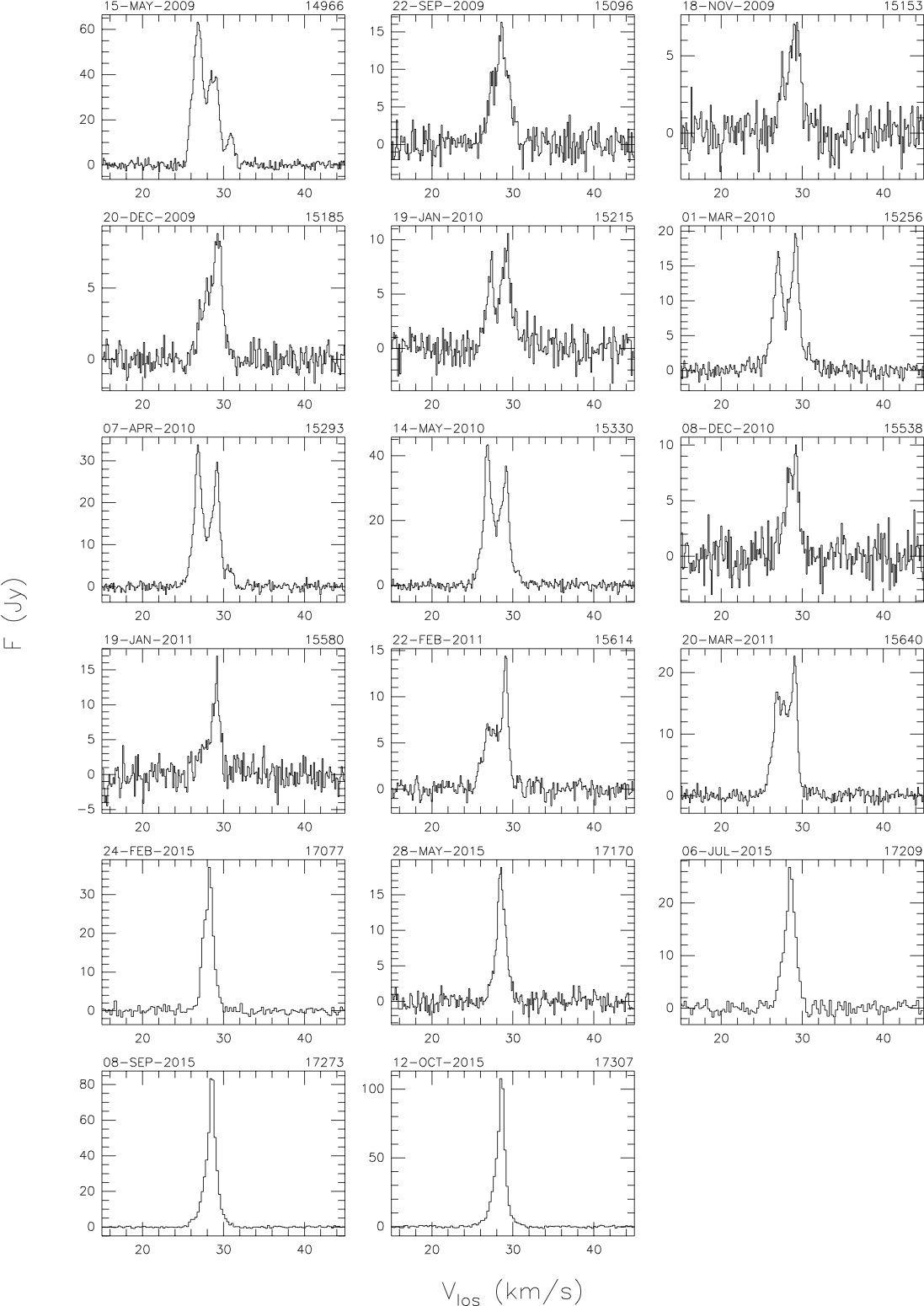}}
\caption{{\it RR~Aql, continued}}
\label{fig:rraql_all}
\end{figure*}